\documentclass[a4paper,11pt]{article}
\pdfoutput=1
\usepackage{jheppub}
\usepackage[T1]{fontenc} 

\usepackage{subfigure}
\usepackage{float}
\def\beq{\begin{equation}}
\def\eeq{\end{equation}}

\usepackage{graphicx}
\usepackage{amsmath}
\usepackage{amssymb}
\usepackage{amsthm}
\usepackage{tikz}
\usepackage{mathrsfs}
\usepackage{multirow}
\usepackage{scalerel}
\usepackage{mathtools}
\usepackage{textcomp}
\usepackage{dsfont}
\usepackage{blkarray}
\usepackage{bm}
\usepackage{todonotes}
\usepackage{hyperref}
\usepackage{pgf-pie}
\usepackage{slashed}
\usepackage[compat=1.1.0]{tikz-feynman}
\allowdisplaybreaks[4]
\graphicspath{{Figures}}
\usepackage{caption}
\usepackage{longtable}

\def\cB{\mathcal{B}}

\def\cL{\mathcal{L}}
\def\cM{\mathcal{M}}
\def\cN{\mathcal{N}}
\def\cO{\mathcal{O}}
\def\cS{\mathcal{S}}

\def\cP{\mathcal{P}}

\def\re{\text{Re}}
\def\im{\text{Im}}
\def\Li{\text{Li}}

\newcommand{\Sl}[1]{\slashed{#1}}
\newcommand{\nn}{\nonumber}
\newcommand{\df}{\mathrm{d}}

\definecolor{darkyellow}{rgb}{0.5, 0.5, 0.0}
\definecolor{darkpurple}{rgb}{0.5, 0.2, 0.8}
\definecolor{darkblue}{rgb}{0.0, 0.0, 0.8}
\definecolor{darkgreen}{rgb}{0.0, 0.4, 0.0}
\definecolor{darkred}{rgb}{0.5, 0.0, 0.0}

\usepackage[normalem]{ulem}
\usepackage{marginnote}

\title{\boldmath NNLL Resummation for Projected Three-Point Energy Correlator}

\author[a]{Wen Chen,}
\author[b]{Jun Gao,}
\author[a]{Yibei Li,}
\author[a]{Zhen Xu,}
\author[c]{Xiaoyuan Zhang,}
\author[a]{Hua Xing Zhu\footnote{Current address: School of Physics, Peking University, Beijing 100871, China}}

\affiliation[a]{Zhejiang Institute of Modern Physics, Department of Physics, Zhejiang University, Hangzhou, 310027, China}
\affiliation[b]{INPAC, Shanghai Key Laboratory for Particle Physics and Cosmology, School of
Physics and Astronomy, Shanghai Jiao-Tong University, Shanghai 200240, China}
\affiliation[c]{Department of Physics, Harvard University, Cambridge, MA 02138, USA}

\emailAdd{chenwenphy@zju.edu.cn}
\emailAdd{jung49@sjtu.edu.cn}
\emailAdd{yblee777@zju.edu.cn}
\emailAdd{zhen.xu@zju.edu.cn}
\emailAdd{xiaoyuanzhang@g.harvard.edu}
\emailAdd{zhuhx@pku.edu.cn}

\abstract{The projected energy correlator measures the energy deposited in multiple detectors as a function of the largest angular distance $x_L = (1 - \cos\chi_L)/2$ between detectors.
The collinear limit $x_L\to 0$ of the projected energy correlator is particularly interesting for understanding the jet-substructures, while the large logarithms of $x_L$
could potentially spoil the perturbation theory and must be resummed. As a necessary ingredient for its resummation at next-to-next-to-leading logarithmic (NNLL) accuracy, we calculate the two-loop jet functions for the projected three-point energy correlator (E3C), using direct integration method and the parameter space Integration-by-Part (IBP) method. We then present the NNLL resummation for $e^+e^-$ annihilation and an approximate NNLL resummation for $pp\rightarrow jj$ process, where the two-loop hard constant is estimated in the latter case. The convergence is improved and the hadronization effect in the collinear limit is suppressed when considering the ratio of E3C distribution to two-point energy-energy correlator (EEC). Our results show  potential in precision determination of strong coupling constant using energy correlators from both $e^+e^-$ data and $pp$ data.

}

\begin{document} 
\maketitle
\pagebreak

\newpage

\section{Introduction}
\label{sec:intro}

Energy correlators are a class of multi-particle angle correlation functions, weighted by the particle energy. Thanks to the energy weighting, they are infrared and collinear safe observables and can be calculated in perturbation theory. The simplest energy correlator is a two-point energy correlator, or Energy-Energy Correlation function~(EEC).
Proposed in 1970s \cite{Basham:1978zq,Basham:1978bw}, EEC measures the correlation of energy deposited in two detectors as a function of the angle $\chi$ between them. In perturbation theory, the definition of EEC reads
\begin{equation}
    \frac{d\sigma^{[2]}}{d\cos\chi} \equiv \sum_{i,j}\int d \sigma  \frac{E_i E_j}{Q^2} \delta\left(\vec n_i\cdot \vec n_j-\cos\chi\right)\,,
\end{equation}
where $i,\,j$ run over all the final state particles, $\vec{n}_i$ and $\vec{n}_j$ are unit three-vectors that define the directions of the particles, and $Q$ is the total energy in the center-of-mass frame. Compared with other event shape variables studied at Large Electron–Positron Collider (LEP),  one advantage of EEC is its simple analytic properties. As far as we are aware of, EEC is the only event shape that can be calculated analytically beyond leading order, e.g. it's now known analytically through to next-to-next-to-leading order (NNLO) \cite{Belitsky:2013ofa,Henn:2019gkr} in $\cN=4$ super Yang-Mills (SYM) theory and through to NLO in QCD \cite{Dixon:2018qgp,Luo:2019nig,Gao:2020vyx}.

In recent years, increasing attention has been paid to generalization of EEC to $N$-point energy correlators, which measure the energies of the outgoing particles with $N$ detectors at colliders and turn out to be a function of ${N(N-1)}/{2}$ angles among these detectors \cite{Hofman:2008ar,Chen:2019bpb,Chen:2020adz,Chen:2022jhb,Chen:2022swd,Chang:2022ryc,Yang:2022tgm}. For example, the three-point energy correlator (EEEC) is defined as
\begin{multline}
\label{eq:eeecdef}
\frac{d^3\sigma}{dx_1dx_2dx_3}\equiv \sum_{i,j,k}\int d\sigma \frac{E_iE_jE_k}{Q^3}\\
\times\delta\left(x_1-\frac{1-\cos\theta_{jk}}{2}\right) \delta\left(x_2-\frac{1-\cos\theta_{ik}}{2}\right) \delta\left(x_3-\frac{1-\cos\theta_{ij}}{2}\right)\,,
\end{multline}
which gives rise to rich functional dependence on the angles and can be used to probe various properties of perturbative QCD.
The LO EEEC was first computed in the triple collinear limit in Ref.~\cite{Chen:2019bpb}, later genelarized to arbitrary angle dependence in both $\cN=4$ SYM \cite{Yan:2022cye} and QCD \cite{Yang:2022tgm}.
To reduce the dimension of the kinematic space of the measured angles without losing too much useful information, one can project the kinematic dependence into a 1D subspace, which leads to the so-called \textit{projected energy correlator} \cite{Chen:2020vvp}. In momentum space, projected $N$-point energy correlator (ENC) is given by restricting the maximum angular distance to be $x_L$:
\begin{equation}\label{eq:pecdef}
\frac{d\sigma^{[N]}}{dx_L}\equiv\sum_n\sum_{1\leq i_1,\cdots i_N\leq n}\int d\sigma \frac{\prod_{a=1}^N E_{i_a}}{Q^N}\delta(x_L-\text{max}\{x_{i_1,i_2},x_{i_1,i_3}, \cdots x_{i_{N-1}, i_N}\})\,,
\end{equation}
and for example, EEEC is then reduced to the projected three-point correlator (E3C). In this work we are mainly interested in the small angle, or collinear limit of E3C, namely $x_L \to 0$. 

It is well-known in the boundary of phase space, incomplete cancellation of infrared divergences can lead to large logarithms that could possibly spoil the convergence of the perturbation theory and thus it is essential to resum these large logarithms to all orders. 
EEC is special as it exhibits both large logarithms in collinear limit and back-to-back limit.
In this work we are interested in the large logarithms in the collinear limit, for which the most singular terms behave as $\alpha_s^n \ln^n x_L$ at $n$ loops.
In the collinear region, EEC can be factorized into a hard function and a jet function, both of which live in the flavor space. The resummation of collinear EEC has been performed up to NNLL accuracy in both QCD \cite{Dixon:2019uzg} and $\cN=4$ SYM \cite{Dixon:2019uzg,Kologlu:2019mfz,Korchemsky:2019nzm}. More interestingly, the collinear factorization can be easily generalized to three-point energy correlator \cite{Chen:2019bpb} and even the projected $N$-point energy correlator \cite{Chen:2020vvp}.
Previously, LL and NLL resummation has been performed in \cite{Chen:2020vvp,Komiske:2022enw,Lee:2022ige}. To improve upon those results, it is necessary to compute the relevant jet and hard function to higher order.
While the hard function is universal for them, the jet functions differ by the measurement function. One of the key new results in this paper is the calculation of two-loop jet function for projected three-point energy correlator, which is the last missing ingredient for NNLL resummation of projected three-point energy correlator in $e^+e^-$ collider. 

One of the main motivations for improving the theoretical accuracy of projected energy correlators comes from the possibility of determining the strong coupling constant $\alpha_s$ by measuring the ratio of projected energy correlators~\cite{Chen:2020vvp}. 
 Measurements of strong coupling constant using classical QCD event shape observable has been actively studied for a long time, e.g.  \cite{L3:1992nwf,SLD:1994idb,DELPHI:1996oqw,ALEPH:2003obs,DELPHI:2004omy,OPAL:2004wof,Dissertori:2007xa,Davison:2009wzs,Bethke:2009ehn,Dissertori:2009ik,Abbate:2010xh,Abbate:2012jh,Hoang:2014wka,Hoang:2015hka,Becher:2008cf,Chien:2010kc,Luisoni:2020efy,Bhattacharya:2022dtm,Bhattacharya:2023qet}. In recent years, there has been increasing attention to using jet substructure observables to extract $\alpha_s$, such as soft-drop thrust and jet mass \cite{Marzani:2019evv,Hannesdottir:2022rsl}, see also \cite{LeBlanc:2022bwd} for $\alpha_s$ determination from jet substructure by demixing quark and gluon jets. 
Since we are mainly concerned with the collinear limit of projected energy correlators in this paper, our results naturally provide theory input for measuring projected energy correlator within a jet, treating it as a jet substructure observable. We will show that considering the ratio of E3C and EEC  can significantly reduce  scale uncertainties and hadronization corrections, which makes it a good candidate for precision determination of $\alpha_s$ using jet substructure. We also note that energy correlators have the advantage that they can be defined and calculated using charged hadrons only~\cite{Chen:2020vvp,Li:2021zcf}. Using the track function formalism~\cite{Chang:2013rca,Jaarsma:2022kdd}, it is possible to perform precision calculation for projected energy correlators on tracks in the future.  

The outline of this paper is as follows. In Sec.~\ref{sec:formalism}, we present the factorization theorem for ENC in the collinear limit and the RG evolution for both hard function and jet function. The desired orders required for all the ingredients to achieve NNLL resummation are briefly summarized there.
In Sec.~\ref{sec:calculate_jet_fn}, we calculate the two-loop E3C jet function. Modern multiloop techniques like IBP and differential equation (DE) are applied for both finite and contact terms. Combining all together, we are able to extract the two-loop E3C jet constants, which is the last missing piece of the NNLL resummation for collinear E3C in $e^+e^-$ collision. 
In Sec.~\ref{sec:resummation_ee}, we present the matched NNLL results for both E3C and the ratio of E3C to EEC in $e^+e^-$ collision. A qualitative analysis is performed to estimate the leading hadronization correction. 
The resummation procedure is extended to the case of $pp$ collision, in particular, the $pp\rightarrow \text{dijet}$ process in Sec.~\ref{sec:resummation_pp}. We present the highest perturbative prediction given the available ingredients, the approximate NNLL, with the missing two-loop hard function constants estimated and included as an additional uncertainty. 
We summarize and conclude in Sec.~\ref{sec:conclusion}.


\section{Resummation formalism}
\label{sec:formalism}

\subsection{Factorization theorem}
In this subsection, we summarize the factorization theorem for the projected  $N$-correlator in the collinear limit and describe the necessary ingredients for NNLL resummation \cite{Chen:2020vvp}.  Similar to EEC, $N$-point energy correlator~(ENC) in this limit is dominated by the logarithmic series of the largest angular distance $x_L$
\begin{align}
  \frac{d\sigma^{[N]}}{dx_L}= \sum_{L=1}^\infty \sum_{j=-1}^{L-1} \left(\frac{\alpha_s(\mu)}{4 \pi} \right)^L c_{L,j} \cL^j (x_L) +\ldots \,,  
\end{align}
where $\cL^{-1}(x_L)=\delta(x_L)$ and $\cL^{j}(x_L)= \left[\ln^j(x_L)/x_L \right]_+$ for $j\geq 0$, with standard plus distribution. 
We do the logarithm counting in the projected  $N$-point 
 energy correlator cumulant, defined as
\begin{align}
  \label{eq:cumulant}
  \Sigma^{[N]}\left(x_L, \ln\frac{Q^2}{\mu^2}\right) =  \frac{1}{\sigma_{\text{tot}}} \int_{0}^{x_L}	dx_L^{\prime} \, \frac{ d\sigma^{[N]}}{d x_L^{\prime}}\left(x_L^{\prime}, \ln\frac{Q^2}{\mu^2}\right) \, , 
  \end{align}
which maps $[\ln^j(x_L)/x_L ]_+ \to 1/(j+1)\times\ln^{j+1}(x_L)$ and $\delta(x_L)\to 1$. Then $\text{N}^{k}\text{LL}$ accuracy refers to the logarithmic series $\sum_{i=0}^{\infty}\sum_{j=\text{max}\{0, i-k\}}^{i} \left(\frac{\alpha_s(\mu)}{4 \pi} \right)^i d_{i,j}\ln^j x_L$ in the cumulant $\Sigma^{[N]}$.

At leading power, the $e^+e^-$ cumulant $\Sigma^{[N]}$ can be written in terms of a modified factorization formula in the collinear limit $x_L \to 0$ \cite{Chen:2020vvp}:
\begin{align}
  \label{eq:fac_nu}
  \Sigma_{ee}^{[N]}\left(x_L, \ln\frac{Q^2}{\mu^2}\right) = \int_0^1  d x\,
x^N \vec{J}^{[N]}\left(\ln \frac{x_L x^2 Q^2}{\mu^2}\right) \cdot \vec{H}_{ee}\left(x, \ln \frac{Q^2}{\mu^2}\right)\, ,
\end{align}
where the hard function $\vec{H}_{ee}^{[N]}$ encodes the production of a parent parton with energy fraction $x$ with respect to the center of mass energy, and the jet function $\vec{J}^{[N]}$ encodes the evolution of the parent parton into a number of collinear partons which contribute to the observable.
Similar factorization formula for EEC was first obtained in \cite{Dixon:2019uzg}, and checked explicitly with known NLO results in QCD \cite{Dixon:2018qgp,Luo:2019nig} and ${\cal N} = 4$ SYM ~\cite{Belitsky:2013ofa,Henn:2019gkr}. We note the explicit dependence on the variable $x$ in both the jet function  and the hard function.
Ignoring the dependence on different quark flavor, both jet and hard functions are two-component vectors living in the flavor space, i.e. $\vec{J}^{[N]}=\{J_q^{[N]}, J_g^{[N]}\}$, $\vec{H}_{ee}=\{H_{ee,q},H_{ee,g}\}$. We will describe their definition for both $e^+e^-$ annihilation and $pp$ collision in detail in the following subsections. We also emphasize that the factorization theorem holds for any $N$ at leading power, though we only calculate the $N=3$ case in this paper. Finally the energy weights in the distribution makes projected $N$-point energy correlator insensitive to the soft radiations and non-global logarithms.

In hadron colliders, the largest angular distance $x_L$ is replaced by the rapidity-azimuth distance $R_L = \max_{i,j 
\in X_E} \sqrt{\Delta \eta_{ij}^2 + \Delta \phi_{ij}^2}$, where $X_E$ is the set of particles that contributes to the energy weight. When the projected energy correlators are measured within a jet, as is typical for jet substructure observable, the cumulant $\Sigma^{[N]}_{\mathrm{had}}$ also depends on the jet radius $R_0$ parameter. In the limit of $R_L \ll R_0$, the modified factorization formula can be written as 
\begin{align}
  \label{eq:fac_nu_pp}
  \Sigma^{[N]}_{\mathrm{had}}\left(R_0, R_L, \ln\frac{p_T^2}{\mu^2}\right) = \int_0^1  d x\,
x^N \vec{J}^{[N]}\left(\ln \frac{R_L^2 x^2 p_T^2}{\mu^2}\right) \cdot \vec{H}_{\mathrm{had}}\left(R_0, x, \ln \frac{p_T^2}{\mu^2}\right)\, ,
\end{align}
where $p_T$ is the jet transverse momentum. 
Around $R_L \sim R_0$, the jet function can also depend on $R_0$. However, there is no large logarithms associated with $R_0$, and its dependence can be obtained from fixed-order matching. For simplicity, 
we will ignore the $R_0$ dependence in the jet function. In that case the jet function become universal between $e^+e^-$ and $pp$ collision. For $pp$ collision, the hard function depends on the partonic scattering process, as well as parton distribution functions~(PDFs). 

\subsection{Hard functions}
\label{subsec:hard}

\subsubsection{$e^+e^-$ annihilation} 

For $e^+e^-$, the hard function is simply the semi-inclusive hadron fragmentation function \cite{Mitov:2006ic}, which depends on the parton flavor and parton energy fraction $x=\frac{2p\cdot q}{Q^2}$,
where $q$ is the total momentum and $p$ is the parton momentum.
 The leading order hard function follows from the born process $e^+e^-\to q\bar q$, $\vec{H}_{ee}^{(0)} (x) = \{2\delta(1-x), 0\}$. At one-loop, we find
\begin{align}
\frac{1}{2}  H_{ee,q}^{(1)}(x) = &\  \frac{\alpha_s}{4 \pi} C_F 
\Bigg[
\left(\frac{4 \pi ^2}{3}-9\right) \delta(1-x) +4 \left[\frac{\ln(1-x)}{1-x} \right]_+
\nn\\
&\
+\left(4 \ln
   (x)-\frac{3}{2}\right)\left(2 \frac{1}{[1-x]}_+-x-1\right)-\frac{9 x}{2}-2 (x+1) \ln
   (1-x)+\frac{7}{2}
\Bigg] \,,
\nn\\
H_{ee,g}^{(1)}(x) = &\, \frac{\alpha_s}{4 \pi}C_F \Bigg[
\frac{4 \left(x^2-2 x+2\right) \ln (1-x)}{x}+\frac{8 \left(x^2-2 x+2\right) \ln   (x)}{x}
\Bigg] \,.
\end{align}
The factor $1/2$ in front of the quark channel indicates for identical contribution from anti-quark, since we do not dinstinguish quark and anti-quark flavor. At two-loop, the hard function can be found from the coefficient functions in \cite{Mitov:2006ic}. 
Similar to the hadron fragmentation function, the renormalization group evolution (RGE) for the hard function $\vec{H}$ is simply the DGLAP equation,
\begin{equation}
  \label{eq:hard_evo}
  \frac{d \vec{H}(x, \ln\frac{Q^2}{\mu^2})}{d \ln \mu^2}
= 
- \int_x^1\! \frac{dy}{y} \widehat{P}(y) \cdot \vec{H} \left( \frac{x}{y},
\ln\frac{Q^2}{\mu^2}\right) \,,
\end{equation}
with $\widehat{P}(y)$ being the singlet timelike splitting matrix, which is now known to three loops~\cite{Chen:2020uvt,Almasy:2011eq}. While it is very difficult to derive an analytic solution for DGLAP to all orders in $\alpha_s$, as we will see below, our resummation only uses a $\alpha_s$-expanded solution (which turns out to be a very good approximation) and only requires certain moments of the hard function.
Explicitly, we will only need the regular and logarithmic moments for the hard function defined as the following~\cite{Dixon:2019uzg}, 
\begin{align}
\int_0^1 dx \, x^N \, H_{q,g}(x,\mu=Q)\ &=\ \sum\limits_{L=0}^\infty
\left( \frac{\alpha_s}{4\pi} \right)^{L} h_L^{q,g}(N) \,, \nn\\
\int_0^1 dx \, x^N \, \ln x \, H_{q,g}(x,\mu=Q)\ &=\ \sum\limits_{L=1}^\infty
\left( \frac{\alpha_s}{4\pi} \right)^{L} \dot{h}_L^{q,g}(N) \,,  \nn\\
\int_0^1 dx \, x^N \, \ln^2 x \, H_{q,g}(x,\mu=Q)\ &=\ \sum\limits_{L=1}^\infty
\left( \frac{\alpha_s}{4\pi} \right)^{L} \ddot{h}_L^{q,g}(N) \,.
\end{align}

Here we use $x^N \ln x =\partial_N x^N$ and the dot on the RHS stands for the derivative. The expressions of needed hard function moments can be found in Appendix~\ref{sec:hard_and_jet}.

\subsubsection{$pp$ collision}

In hadronic collisions, we mainly focus on the dijet production $pp\rightarrow jj$, which has a relatively large cross section at the LHC. Different from $e^+e^-$ collider, this hard function incorporates the partonic scattering cross sections, the contribution from parton distribution functions (PDFs) and the jet algorithms for clustering the particles.
Currently, to the best of our knowledge, the hard function is not know at two-loop. However, important progress are being made to compute those hard functions, e.g. \cite{Gehrmann:2022cih}.
Similar to the $e^+e^-$ case, our resummation will only need the hard function moments.

In this work we evaluate the needed moments of the hard function numerically in \textsc{Madgraph5}~\cite{Alwall:2011uj,Alwall:2014hca}. To investigate the sensitivity of the result to the values of $\alpha_s$, we used three different PDF sets: {\texttt{NNPDF31\_nnlo\_as\_0112}}, {\texttt{NNPDF31\_nnlo\_as\_0118}} and {\texttt{NNPDF31\_nnlo\_as\_0124}} through \textsc{Lhapdf}~\cite{Buckley:2014ana}. Each PDF set fixes also the value of $\alpha_s(m_Z)$ and the corresponding evolution in \textsc{Madgraph5}. 
To address the fact that the hard function contains collinear divergence when resolving the energy fraction of the quarks and gluons, we use the one cut-off phase space slicing to regularize the collinear singularity, as implemented in \cite{Liu:2023fsq}. With the collinear divergent contribution singled out and calculated analytically, the remaining contributions can be evaluated numerically. The detailed discussion can be found in Appendix~\ref{sec:hard_and_jet}.

For $pp\rightarrow jj$, we adopt the anti-$k_t$ algorithm~\cite{Cacciari:2008gp} for jet detection and use the following parameters in the calculation
\begin{equation}
	\label{eq:jj_kin1}
R_0=0.4,\qquad p_T > 15\, {\rm GeV}, \qquad |\eta|<1.5 \,. 
\end{equation}
The two leading jets are further subject to the following cuts
\begin{equation}
	\label{eq:jj_kin2}
 |\Delta \phi(j_1, j_2)| >2 , \qquad |p_T^1 - p_T^2|/(p_T^1 + p_T^2) < 0.5\,,
 \end{equation}
and cast to the corresponding $p_t$ bins for the analysis. 
The calculated moments need to be normalized with the cross section $\sigma_J$ of jet production within specific $p_t$ range. In particular, we expand $H_{\text{had}}/\sigma_J$ to NLO in $a_s$, and take the $\mathcal{O}(a_s^0)$ and $\mathcal{O}(a_s^1)$ as the leading and next-to-leading order results. For the purpose of phenomenological studies, we will focus on two different $p_t$ ranges: $[300,350]$ GeV and $[500,550]$ GeV. The hard function moments needed for NNLL are also summarized in Appendix~\ref{sec:hard_and_jet}.

\subsection{Jet functions}

The E3C jet function, on the other hand, encodes the measurement information. From RG invariance of the modified factorization formula \eqref{eq:fac_nu}, the jet function satisfies a modified timelike DGLAP evolution equation
\begin{align}
  \label{eq:jet_evo}
  \frac{d \vec{J}^{[N]}( \ln \frac{x_L Q^2}{\mu^2})}{d \ln\mu^2} = \int_0^1  dy\, y^N \vec{J}^{[N]}\left( \ln \frac{x_L y^2 Q^2}{\mu^2}\right)
\cdot 
\widehat{P}(y) \,. 
\end{align}

In order to write down an operator description of the E3C jet function, we first recall the collinear EEEC jet function from \cite{Chen:2019bpb}:
\begin{align}\label{eq:jet_func_quark}
J_q(x_1,x_2,x_3,Q,\mu^2)&=\notag\\
&\hspace{-0.1cm}\int \frac{dl^+}{2\pi}\frac{1}{2N_C} \text{Tr} \int d^4x e^{i l\cdot x} \langle 0 | \frac{\Sl{\bar n}}{2} \chi_n(x) \widehat\cM_{\text{EEEC}} ~ \delta (Q+\bar n \cdot \cP) \delta^2(\cP_\perp) \bar \chi_n(0) |0\rangle\notag\\
J_g(x_1,x_2,x_3,Q,\mu^2)&=\notag\\
&\hspace{-1.5cm}\int \frac{dl^+}{2\pi}\frac{1}{2 (N^2_C-1)} \text{Tr} \int d^4x e^{i l\cdot x} \langle 0 |  \cB^{a,\mu}_{n,\perp}(x) \widehat\cM_{\text{EEEC}} ~ \delta (Q+\bar n \cdot \cP) \delta^2(\cP_\perp) \cB^{a,\mu}_{n,\perp}(0) |0\rangle \,,
\end{align}
where $\chi_n\equiv W_n^\dagger \xi_n$ is  the collinear quark and $\cB^{\mu}_{n,\perp}\equiv \frac{1}{g}\left[\frac{1}{\bar n\cdot \cP}W_n^\dagger [i\bar n\cdot D_n, iD_{n\perp}^\mu] W_n\right]$ is the collinear gluon, and $\cP_{n\perp}^{\mu}$ form a complete set of collinear gauge invariant building blocks \cite{Marcantonini:2008qn} in SCET~\cite{Bauer:2000yr,Bauer:2000ew,Bauer:2001yt,Bauer:2001ct,Beneke:2002ph}. The triple collinear measurement function $\widehat\cM_{\text{EEEC}}$ is defined as
\begin{equation}\label{eq:measure_EEEC}
    \widehat{\mathcal{M}}_\text{EEEC}(x_1,x_2,x_3)=\sum_{i,j,k}\frac{E_i E_j E_k}{Q^3}\delta\left(x_1-\frac{\theta_{ij}^2}{4}\right)\delta\left(x_2-\frac{\theta_{jk}^2}{4}\right)\delta\left(x_3-\frac{\theta_{ki}^2}{4}\right) \, ,
\end{equation}
with $\theta_{ij}$ being the angle between parton $i$ and $j$. Then our E3C jet function has the same form as EEEC jet function, with a replacement of the measurement function:
\begin{align}
    \widehat{\mathcal{M}}_\text{EEEC}\Rightarrow \widehat{\mathcal{M}}_\text{E3C}(x_L)&=\int_{0}^{x_L} dx_L^\prime\int_{K} dx_1 dx_2 dx_3 \,\widehat{\mathcal{M}}_\text{EEEC}\,\delta\left(x_L^\prime-\text{max}(x_1,x_2,x_3)\right)\notag\\
    &=\int_{K} dx_1 dx_2 dx_3 \,\widehat{\mathcal{M}}_\text{EEEC}\,\theta\left(x_L-\text{max}(x_1,x_2,x_3)\right) \, .
    \label{eq:measure_E3C}
\end{align}
There are two folds integration in the first line. The first one is performed in the allowed kinematic space $\{x_1,x_2,x_3\}\in K$ that will be discussed below, projecting the shape-dependent EEEC jet function into a single-scale jet function. The second integration brings the differential measurement to the cumulant level. For $N>3$, the measurement function takes a similar structure, with more $\delta$ functions and integrations.
Perturbatively, the E3C jet function can be written as $J_{q,g}=\sum_L (\alpha_s/4\pi)^L J^{(L)}_{q,g}$, and we use the normalization condition $2^3 \cdot J^{(0)}_{q}=2^3 \cdot J^{(0)}_{g}=1$ as in Ref.~\cite{Chen:2020vvp}. The one-loop correction can be calculated from the QCD $1 \to 2$ timelike splitting kernel and is given by 
\begin{align}
  \label{eq:jet_one_loop}
 2^3 J^{(1)}_{q} = &\ \frac{9C_F}{2} \ln \frac{x_L Q^2}{\mu^2} -\frac{37 C_F}{2}   \,,
\nn\\
 2^3 J^{(1)}_{g} = &\ \left(\frac{21 C_A}{5 }+\frac{3 n_f}{10}\right)  \ln \frac{x_L Q^2}{\mu^2}
-\frac{449 C_A}{25}-\frac{21 n_f}{25}
\,. 
\end{align}
Note that the $\mu$-dependent terms are precisely captured by the jet RGE, while the remaining constants have to come from the fixed-order calculation. One of the main result in this paper is to calculate the two-loop constants described below.

\subsection{Two-loop calculation for the E3C jet function}
\label{sec:calculate_jet_fn}

In this subsection, we present the two-loop calculation of the E3C jet functions for both quark jets and gluon jets. Since they are universal in the small angle limit, they can be used in both $e^+e^-$ collision and $pp$ collision.

We start from recalling the definition of E3C at finite angle before taking the small angle limit.
At two loops, E3C receives contributions from double-real (RR) and real-virtual (RV) as well as double-virtual (VV) corrections to $q\to q$, from which the quark jet function can be extracted by matching to the factorization formula, \eqref{eq:fac_nu}. Similarly, the gluon jet function can be extracted from the NLO E3C distribution of Higgs gluonic decay $H\to gg$. To organize the calculation, we rewrite the definition of E3C in Eq.~\eqref{eq:pecdef} with the number of energy weight:
\begin{align}
    \label{eq:PEEECtoals2}
    \frac{1}{\sigma_{0}}\frac{d\sigma^{[3]}}{dx_L}&=\sum_{1\leq i_1\neq i_2\neq i_3\leq 4}\int \df\text{LIPS}_4\,|\mathcal{M}_{4}|^2 \frac{E_{i_1} E_{i_2} E_{i_3}}{Q^3}\delta(x_L-\max\{x_{i_1,i_2},x_{i_1,i_3},x_{i_2,i_3}\})\notag\\
    &+
    \sum_{n\in\{3,4\}}\sum_{1\leq i_1\neq i_2\leq n}\int \df\text{LIPS}_n\,|\mathcal{M}_{n}|^2 \frac{E^2_{i_1} E_{i_2}}{Q^3}\delta(x_L-x_{i_1,i_2})\notag\\
    &+
    \sum_{n\in\{2,3,4\}}\sum_{1\leq i_1\leq n}\int \df\text{LIPS}_n\,|\mathcal{M}_{n}|^2 \frac{E^3_{i_1}}{Q^3}\delta(x_L) \,,
\end{align}
where we normalize the distribution to the born cross-section in $d$ dimension.
The first line represents the contribution from nonidentical energy weights measurement and the other lines are called contact terms. If we define $x_1=x_L z \bar z$, $x_2=x_L (1-z)(1-\bar z)$ and $x_3=x_L$, then in the collinear limits, they are the contact terms for $\delta(z\bar z)$ that captures the strict squeeze limit and $\delta(x_L)$ that captures the strict triple collinear limit. The main goal of this section is to compute the collinear limit of Eq.~\eqref{eq:PEEECtoals2} and extract the corresponding two-loop constants.

The lowest regular distribution of the E3C quark jet function comes from tree-level process $\gamma^{*}\to \text{4 partons}$ in electron-positron annihilation, which under the triple collinear limit, factorizes into the born process $\gamma^*\to q\bar q$ and the $1\to 3$ splitting functions, and we will call it nonidentical energy weight term. Below we will introduce two different methods to compute this part. The traditional method is to calculate the EEEC jet function to order $\cO(\epsilon)$ and to integrate two angular distances $x_2,\, x_3$ numerically by the interpolation method. The OPE singularities (sometimes called squeezed singularities) of EEEC are subtracted and integrated in $d$ dimension separately. The second approach benefits from the parameter space IBP method \cite{Chen:2019mqc,Chen:2019fzm,Chen:2020wsh} developed very recently. Only 7 master integrals are needed to express EEEC, allowing the precise calculation of the remaining two-fold integral. 

The other two parts contribute to the contact terms and cancel the infrared divergence, which is guaranteed by the Kinoshita-Lee-Nauenberg (KLN) theorem \cite{PhysRev.133.B1549,Kinoshita:1962ur}. Similar to EEC at NLO, the measurement function in the contact terms can be treated as a non-standard cut propagators, which allows for a generalized IBP reduction in \textsc{Litered}~\cite{Lee:2012cn,Lee:2013mka} and \textsc{Fire6}~\cite{Smirnov:2019qkx}. The master integrals then can be calculated in packages like \textsc{Canonica}~\cite{Meyer:2017joq} or \textsc{Libra}~\cite{Lee:2014ioa,Lee:2020zfb} with the differential equation method implemented.

\subsubsection{Nonidentical energy-weight terms}\label{sec:3_1_nonidentical}

We start by computing the nonidentical energy-weight contribution in the traditional approach. As discussed in Ref.~\cite{Chen:2019bpb}, the inclusive jet function $J_{\widehat{ijk}}$ is related to the $1\to 3$ splitting function $P_{ijk}$ \cite{Campbell:1997hg,Catani:1998nv,Ritzmann:2014mka} through 
\begin{equation}
\label{eq:jetR_def}
    J^{\text{nonid}}\equiv J_{\widehat{ijk}}=\int\mathrm{d}\Phi^{(3)}_c\left(\frac{\mu^2e^{\gamma_E}}{4\pi}\right)^{2 \epsilon}\frac{4g^4}{s_{123}^2}\sum_{i,j,k}P_{ijk}\widehat{\mathcal{M}}_\text{EEEC} \,,
\end{equation}
where $\mathrm{d}\Phi^{(3)}_c$ is the triple collinear phase space \cite{Gehrmann-DeRidder:1997fom,Ritzmann:2014mka}, and $i,j,k$ run over all final-state particles. The fully differential distribution with respect to all angular distances $\{x_1,x_2,x_3\}$ in $d=4-2\epsilon$ dimension is then written as
\begin{multline}
	\label{eq:eeec_jet_expanded}
    \frac{d J^{\text{nonid}}}{dx_L d\text{Re}(z) d\text{Im}(z)}=\left(\frac{\mu^2}{Q^2}\right)^{2\epsilon}\frac{\alpha_s^2}{\pi^3}\frac{e^{2\epsilon \gamma_E}}{\Gamma(1-2\epsilon)}\frac{1}{x_L^{1+2\epsilon}}\frac{1}{ (2\text{Im}(z))^{2\epsilon}}\\
    \times\left[G(z)+\epsilon F(z)+\epsilon^2 H(z)+\cO(\epsilon^3)\right] \,,
\end{multline}
where $G(z),F(z),H(z),\cdots$ the shape function in $\epsilon$ expansion. The order $\cO(1)$ part $G(z)$ is computed analytically in \cite{Chen:2019bpb} and following the same approach, we also calculate the complete result for $F(z)$ and the $z\to 1$ limit of $H(z)$. We will see that these are all the needed ingredients for nonidentical part. Note that the $x_L$ dependence is defined by plus distribution, where
\begin{equation}
\label{eq:plusdistdef}
    \frac{1}{x_L^{1+2\epsilon}}=-\frac{\delta(x_L)}{2\epsilon}+\left(\frac{1}{x_L}\right)_{+}-2\epsilon \left(\frac{\ln x_L}{x_L}\right)_{+}+\cdots \,.
\end{equation}

\begin{figure}[ht]
\centering 
\includegraphics[scale=0.4]{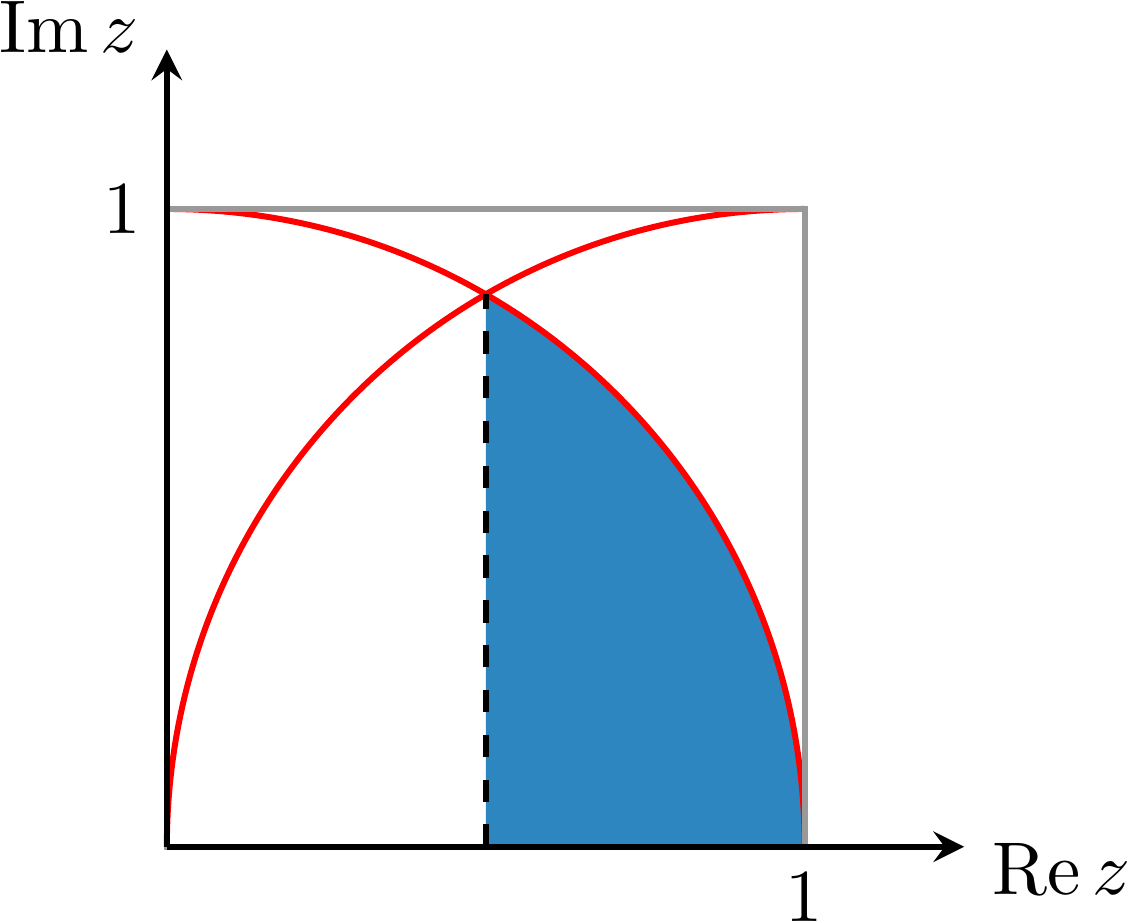}
\caption{The integration region $\cS$ for E3C jet function. The $S_3$ symmetry is applied to reduce the entire region of $z$ into $6$ times the blue region. The integration range for $z$ then becomes $\int_0^{\frac{\sqrt{3}}{2}}d\im (z)\int_\frac{1}{2}^{\sqrt{1-(\im (z))^2}}d\re (z)$.}
\label{fig:eeecregion}
\end{figure}

In order to perform the integral over $z$, we need to figure out the integration region first. Compared with the first line in Eq.~\eqref{eq:PEEECtoals2}, it is straightforward to show that 
\begin{equation}
\label{eq:e3cint}
    \frac{d J^{\text{nonid}}}{dx_L}=\left(\frac{\mu^2}{Q^2}\right)^{2\epsilon}\frac{\alpha_s^2}{\pi^3}\frac{e^{2\epsilon \gamma_E}}{\Gamma(1-2\epsilon)}\frac{6}{x_L^{1+2\epsilon}}\underbrace{ \int_{\cS}\frac{d\re (z)d\im (z)}{(2\im (z))^{2\epsilon}}\left[G(z)+\epsilon F(z)+\epsilon^2 H(z)+\cO(\epsilon^3)\right]}_{\equiv A(\epsilon)} \,,
\end{equation}
where the constant factor 6 comes from the $S_3$ permutation symmetry and the integration region $\cS$ is given in Fig.~\ref{fig:eeecregion}. To calculate $A(\epsilon)$ numerically, we also need to subtract the OPE singularities around $z\rightarrow 1$ at the integrand level, and evaluate its $z$ integration analytically in $d$ dimension.
The full asymptotic expansion of $z\to 1$ is given in the appendix~\ref{sec:eeec_squeeze}. The most singular term is proportional to $\frac{1}{(1-z)(1-\bar z)}$, which gives rise to 
\begin{multline}
\label{eq:example_int}
    \int_0^{\frac{\sqrt{3}}{2}}d\im (z)\int_\frac{1}{2}^{\sqrt{1-(\im (z))^2}}d\re (z)\frac{1}{(2\im (z))^{2\epsilon}}\frac{1}{(1-z)(1-\bar z)} \\
    =-\frac{\pi}{4\epsilon}-\kappa
    +\epsilon\left(-\frac{167}{1080}\pi^3-\frac{1}{20}\pi\ln^2 3+\kappa \ln 3+\frac{12}{5} \eta\right)+\cO(\epsilon^2) \,.
\end{multline}
Here $\kappa=\im \Li_2 e^{i\frac{\pi}{3}}$ is the Gieseking's constant living in the transcendentality-two family and $\eta=\im \Li_3\left(\frac{i}{\sqrt{3}}\right)$ is a parity-odd transcendentality-three constant. These constants are typical numbers in loop integrals, especially in trijet observable calculations.

With subtraction terms, the integral $A$ in Eq.~\eqref{eq:e3cint} up to order $\cO(\epsilon)$ is then written as 
\begin{align}
    A&=\int_0^{\frac{\sqrt{3}}{2}}d\im (z)\int_\frac{1}{2}^{\sqrt{1-(\im (z))^2}}d\re (z)\frac{1}{(2\im (z))^{2\epsilon}}\left[G(z\to 1)+\epsilon F(z\to 1)+\epsilon^2 H(z\to 1)\right]\notag\\
    & \hspace{-0.4cm}+\int_0^{\frac{\sqrt{3}}{2}}d\im (z)\int_\frac{1}{2}^{\sqrt{1-(\im (z))^2}}d\re (z)\frac{1}{(2\im (z))^{2\epsilon}}\left[\left(G(z)-G(z\to 1)\right)+\epsilon \left(F(z)-F(z\to 1)\right)\right] .
\end{align}
The first term is proportional to Eq.~\eqref{eq:example_int} and it is straightforward to compute it to $\cO(\epsilon)$. For the second integral, we have to expand in $\epsilon$ and evaluate it numerically. To implement the interpolation method, we first change the integration variables via $v_1=\frac{2}{\sqrt{3}}\im (z)$ and $v_2=\frac{\re (z)-\frac{1}{2}}{\sqrt{1-(\im(z))^2}-\frac{1}{2}}$, such that both $v_{1,2}$ range from 0 to 1. Then we can build a 2D lattice by discretizing $v_{1,2}$ and approximate our integrand with polynomials. This allows one to perform the two-fold numerical integral directly in \textsc{Mathematica}. To check the stability of the integration and estimate the statistical error, we vary the lattice size and the order of polynomials and see which significant figure remains unchanged. Eventually we obtain both $\delta(x_L)$ contact term and $\frac{1}{x_L}$ finite term for the nonidentical energy weight contribution. The explicit expression for both quark and gluon jet function can be found in Eq.~\eqref{eq:E3C_quark_jet_1}-\eqref{eq:E3C_gluon_jet_2} in the appendix.

Alternatively, benefiting from the recent development of the IBP method in the Feynman parameter space, we can simplify the whole jet function calculation with integral reduction.
First of all, recall that Eq.~\eqref{eq:jetR_def} takes the form 
\begin{equation}
\label{eq:JetFunct}
    J^{\text{nonid}}\equiv \int \mathrm{d}x_1\mathrm{d}x_2\mathrm{d}x_3 \frac{d J^{R}}{dx_1dx_2dx_3}\propto \int\mathrm{d}x_1\mathrm{d}x_2\mathrm{d}x_3\mathrm{d}\omega_1\mathrm{d}\omega_2\mathrm{d}\omega_3\delta(1-\omega_1-\omega_2-\omega_3)\hat{P}_{ijk}  \,.
\end{equation}
Here $\hat{P}$ is a homogeneous function of the energy fraction $\omega_i$ of the final-state particles. Explicitly, it is of the form
\begin{equation}
    \hat{P}_{ijk}=\frac{\omega_1^{\alpha_1}\omega_2^{\alpha_2}\omega_3^{\alpha_3}}{f_1^{\beta_1}f_2^{\beta_2}} \,,
\end{equation}
with $f_1$ linear in $\omega_i$, and $f_2$ a polynomial of $\omega_i$ of degree $2$. Following the idea in Ref~\cite{Chen:2019bpb}, the integral $\frac{\mathrm{d}^3J^R}{\mathrm{d}x_1\mathrm{d}x_2\mathrm{d}x_3}$ in Eq.~\eqref{eq:JetFunct} can be related to a Feynman parameter integral through\footnote{In the special cases where $\beta_1=0$ or $f_1=U$, we don't need to introduce the parameter $\omega_4$.}
\begin{align}
\label{eq:EEECtoFeyint}
    \frac{d^3J^{\text{nonid}}}{dx_1dx_2dx_3}=&\frac{\Gamma(\beta_1+\beta_2)}{\Gamma(\beta_1)\Gamma(\beta_2)}\int\mathrm{d}\omega_1\mathrm{d}\omega_2\mathrm{d}\omega_3\mathrm{d}\omega_4\delta(1-\omega_1-\omega_2-\omega_3)\frac{\omega_1^{\alpha_1}\omega_2^{\alpha_2}\omega_3^{\alpha_3}\omega_4^{\beta_1-1}}{(f_2+f_1\omega_4)^{\beta_1+\beta_2}}\notag\\
    =&\frac{\Gamma(\beta_1+\beta_2)}{\Gamma(\beta_1)\Gamma(\beta_2)}\int\mathrm{d}\omega_1\mathrm{d}\omega_2\mathrm{d}\omega_3\mathrm{d}\omega_4\delta(1-\omega_1-\omega_2-\omega_3)\frac{\omega_1^{\alpha_1}\omega_2^{\alpha_2}\omega_3^{\alpha_3}\omega_4^{\beta_1-1}}{(f_2+f_1\omega_4)^{\beta_1+\beta_2}}\notag\\
    =&\frac{\Gamma(\beta_1+\beta_2)}{\Gamma(\beta_1)\Gamma(\beta_2)}\int\mathrm{d}\omega_1\mathrm{d}\omega_2\mathrm{d}\omega_3\mathrm{d}\omega_4\delta(1-U)\frac{\omega_1^{\alpha_1}\omega_2^{\alpha_2}\omega_3^{\alpha_3}\omega_4^{\alpha_4}}{U^{\lambda_1}F^{\lambda_2}}\notag\\
    \equiv&\frac{\Gamma(\alpha_1)\Gamma(\alpha_2)\Gamma(\alpha_3)}{\Gamma(\beta_1)\Gamma(\beta_2)}I(\alpha_0,\alpha_1,\alpha_2,\alpha_3,\alpha_4) \,,
\end{align}
where $U=\omega_1+\omega_2+\omega_3$, $F=f_2+f_1\omega_4$, $\lambda_1=\alpha_1+\alpha_2+\alpha_3-\beta_1-2\beta_2+3$, $\lambda_2=\beta_1+\beta_2$, and $\alpha_0=-\beta_1-\beta_2$. The integral in the last line is a standard parametric Feynman integral, which can be reduced with IBP reduction~\cite{Tkachov:1981wb,Chetyrkin:1981qh} in the parametric representation~\cite{Lee:2014tja,Chen:2019mqc,Chen:2019fzm,Chen:2020wsh}\footnote{The algorithms described in ref.~\cite{Chen:2019fzm} to generate symbolic rules work only when all the indices are nonnegative. Thus, here we carry out the reduction by merely solving IBP identities using \textsc{Kira}~\cite{Maierhofer:2017gsa,Klappert:2019emp,Klappert:2020aqs,Klappert:2020nbg}.}. The master integrals are
\begin{align}
&\mathcal{I}_{1}=I_1\left(\alpha _0,-2 \epsilon ,1-2 \epsilon ,-2 \epsilon \right),
&\mathcal{I}_{2}=I_1\left(\alpha _0,1-2 \epsilon ,-2 \epsilon ,-2 \epsilon \right),\notag\\
&\mathcal{I}_{3}=I_1\left(\alpha _0,-2 \epsilon ,-2 \epsilon ,1-2 \epsilon \right),
&\mathcal{I}_{4}=I_1\left(\alpha _0,-2 \epsilon ,-2 \epsilon ,-2 \epsilon \right),\notag\\
&\mathcal{I}_{5}=I_2\left(\alpha _0,-2 \epsilon ,-2 \epsilon ,-2 \epsilon ,0\right),
&\mathcal{I}_{6}=I_3\left(\alpha _0,-2 \epsilon ,-2 \epsilon ,-2 \epsilon ,0\right),\notag\\
&\mathcal{I}_{7}=I_4\left(\alpha _0,-2 \epsilon ,-2 \epsilon ,-2 \epsilon ,0\right) \,,
\end{align}
with the integrals $I_i$ defined by the $F$ polynomials
\begin{align}
&F_1=x_1\omega_2\omega_3+x_2\omega_1\omega_3+x_3\omega_1\omega_2\,,\quad 
&F_2=F_1+(\omega_1+\omega_2)\omega_4 \,,\notag\\
&F_3=F_1+(\omega_1+\omega_3)\omega_4\,,\quad
&F_4=F_1+(\omega_2+\omega_3)\omega_4  \,,
\end{align}
and $\alpha_0=6\epsilon-2$\footnote{Notice that though here $\alpha_0$ and $\epsilon$ are not independent, we should treat them as independent parameters during the IBP reduction, because otherwise some integrals may be ill-defined.}. The master integrals can be evaluated using the differential equation technique~\cite{Kotikov:1990kg,Remiddi:1997ny}. For simplicity, we set $\mu=x_3=1$, and introduce $u$ and $v$ following $z=u(1+iv)$. Then we construct the differential-equation system with respect to $u$, and derive the canonical basis~\cite{Henn:2013pwa} using \textsc{Libra}~\cite{Lee:2014ioa,Lee:2020zfb}
\begin{align}
\mathcal{I}_1^\prime=&6 u (v-1) \mathcal{I}_4+\frac{x_1 (1-2 \epsilon )}{\epsilon }\mathcal{I}_1\, ,\notag\\
\mathcal{I}_2^\prime=&6 (u-1) \mathcal{I}_4+\frac{x_2 (1-2 \epsilon )}{\epsilon }\mathcal{I}_2\, ,\notag\\
\mathcal{I}_3^\prime=&6 \left(u v+u-x_1\right) \mathcal{I}_4+\frac{x_1 x_2 (1-2 \epsilon )}{\epsilon }\mathcal{I}_3\, ,\notag\\
\mathcal{I}_4^\prime=&6 u v \mathcal{I}_4\, ,\notag\\
\mathcal{I}_5^\prime=& \left(x_1-x_2\right)\mathcal{I}_5\, ,\notag\\
\mathcal{I}_6^\prime=& \left(x_3-x_1\right)\mathcal{I}_6\, ,\notag\\
\mathcal{I}_7^\prime=& \left(x_2-x_3\right)\mathcal{I}_7 \, ,
\end{align}
with the corresponding alphabet $\{u,\, 2u-1,\, x_2,\, x_2-1\}$. By solving the differential-equation system, we can express the master integrals via Goncharov polylogarithms (GPLs) \cite{goncharov1mpl,Goncharov:1998kja,Borwein:1999js}. The GPL is defined iteratively by 
\begin{equation}\label{eq:gpl_def}
G(a_1,\cdots a_n; x)\equiv\int_0^x \frac{dt}{t-a_1} G(a_2,\cdots a_n; t)\,,
\end{equation}
with
\begin{equation}
G(;x)\equiv1,\quad G(\vec 0_n;x)\equiv\frac{1}{n!}\ln^n (x)\,.
\end{equation}

After finishing the simplified calculation of EEEC in the collinear limit, we still need to integrate two angular distances for the projected EEEC as the previous approach. By virtue of the $S_3$ permutation symmetry, this amount to consider
\begin{align}
    \frac{dJ^{\text{nonid}}}{dx_L}=&6\int\mathrm{d}x_1\mathrm{d}x_2~\Theta(x_1,x_2)\frac{d^3J}{dx_1dx_2dx_3}\notag\\
    =&24\int\mathrm{d}u\mathrm{d}v~\Theta(x_1,x_2)u^2v\frac{d^3J}{dx_1dx_2dx_3}\notag\\
    \equiv&\int\mathrm{d}u\mathrm{d}v~\Theta(x_1,x_2)\Tilde{J}(u,v)\,,
\end{align}
where $\Theta(x_1,x_2)\equiv\theta\left(1-\sqrt{x_2}\right)\theta\left(\sqrt{x_2}-\sqrt{x_1}\right)\theta\left(\sqrt{x_2}+\sqrt{x_1}-1\right)$. Now the OPE singularity corresponds to $u\to 0$ limit, and similarly, we need to subtract the singular behavior and do the integration separately:
\begin{equation}\label{eq:subtractedjet}
    \frac{dJ^{\text{nonid}}}{dx_L}=\int\mathrm{d}u\mathrm{d}v~\Theta(x_1,x_2)\tilde{J}(u\to0)+\int\mathrm{d}u\mathrm{d}v~\Theta(x_1,x_2)\left[\Tilde{J}(u,v)-\tilde{J}(u\to0)\right] \,,
\end{equation}
where again we can evaluate the first integral in $d$ dimension and expand the integrand of the second one in $\epsilon$.

To calculate the $\tilde{J}(u\to0)$, now we can directly extract the asymptotic expansion of the integral $I$ in Eq.~\eqref{eq:EEECtoFeyint} from DE, in which we identify two expansion regions:
\begin{align}
    \text{hard region:}\quad &\omega_1\sim\omega_2\sim\omega_3\sim 1 \,,\notag\\
    \text{small region:}\quad &\omega_2\sim\omega_3\sim 1,~\omega_1\sim u^2  \,.
\end{align}
Evantually we only need to integrate the reduced master integrals in $d$ dimension.

Regarding the second integral in Eq.~\eqref{eq:subtractedjet}, the $u$ integral is straightforward since $\tilde{J}(u,v)$ is expressed in terms of GPLs of the form $G(\dots,u)$. However, the $v$ integral becomes unstable in two regions $v\to0$ and $v\to\infty$. To resolve this problem, we decompose the $v\in [0,\infty]$ integration into three parts: $[0,~\frac{1}{C}]$, $[\frac{1}{C},~C]$, and $[C,~\infty]$, with a arbitrary cut parameter $C>1$. In the region $(\frac{1}{C},~C)$, we carry out the integration numerically, with the GPLs numerically using \textsc{Handyg}~\cite{Naterop:2019xaf}. The other two regions require expanding the integrand in $v$ (or $\frac{1}{v}$) to $\mathcal{O}(v^{100})$ (or $\mathcal{O}(v^{-100})$) and performing the integration analytically. This expansion can easily be done by asymptotically solving the differential equations satisfied by the GPLs. Eventually, we find the same result as in Eq.~\eqref{eq:E3C_quark_jet_1}-\eqref{eq:E3C_gluon_jet_2}.

\subsubsection{Contact terms}\label{sec:3_2_contact}

While it is convenient to calculate the nonidentical $E_{i_1}E_{i_2}E_{i_3}$ part starting with the splitting functions, it is preferable to compute the full angular dependence on $x_L$ for corresponding processes (namely $e^+e^-$ annihilation and gluonic Higgs decay) with energy weights $E^2_{i_1}E_{i_2}$ ($i_1\neq i_2$) and $E^3_{i_1}$, and extract the contact term from the collinear limit $x_L\to 0$. In other words, we will adopt the full matrix elements squared and compute the full phase space integral using modern multi-loop techniques, with which the collinear expansion gives $\df \sigma^{[3]}_{\text{E}^2\text{EC}}(x_L)/\df x_L$ (the $E^2_{i_1}E_{i_2}$ ($i_1\neq i_2$) part) and $\df \sigma^{[3]}_{\text{E}^3\text{C}}(x_L)/\df x_L$ (the $E^3_{i_1}$ part) in the $x_L\to 0$ limit. 

We start with the relevant processes in perturbation theory for two-loop jet functions,
\begin{align}
{\bf e^+e^- \textbf{annihilation }}\qquad\qquad& \textbf{Higgs decays}\nonumber\\
\gamma^*\rightarrow q\bar{q}+VV \qquad\qquad& H\rightarrow gg+VV\nonumber\\
\gamma^*\rightarrow q\bar{q}g+V \qquad\qquad& H\rightarrow ggg+V\nonumber\\
& H\rightarrow q\bar{q}g+V\nonumber\\
\gamma^*\rightarrow q\bar{q}gg \qquad\qquad& H\rightarrow gggg\nonumber\\
\gamma^*\rightarrow q\bar{q}q\bar{q} \qquad\qquad& H\rightarrow q\bar{q}gg\nonumber\\
\gamma^*\rightarrow q\bar{q}q'\bar{q}' \qquad\qquad& H\rightarrow q\bar{q}q\bar{q}\nonumber\\
& H\rightarrow q\bar{q}q'\bar{q}'
\end{align}
where $V$ and $VV$ denotes one-loop and two-loop correction respectively. In particular, in the $x_L\to 0$ limit, $1\to2$ processes only contribute to $\delta(x_L)$-terms (i.e., $\df\sigma^{[3]}_{\text{E}^3\text{C}}(x_L)/\df x_L$).

The calculation setup of $\df\sigma^{[3]}_{\text{E}^2\text{EC}}(x_L,\epsilon)/\df x_L$ shares the same structure as the original EEC, which basically follows the approach described in Ref.~\cite{Dixon:2018qgp} and more detail in \cite{Luo:2019nig}. Briefly speaking, using the Cutkosky rules \cite{Cutkosky:1960sp, Anastasiou:2002yz}, we can replace the phase-space on-shell delta functions with the cut propagators
\begin{equation}
\delta(p^2)=\frac{1}{2\pi \mathrm{i} }\left(\frac{1}{p^2-\mathrm{i}0}-\frac{1}{p^2+\mathrm{i}0}\right)  \,,
\end{equation}
and also the EEC measurement function $\delta(x_L-x_{i,j})$ with
\begin{align}
&\delta \left(x_L-\frac{1-\cos\theta_{ij}}{2}\right)=\frac{(p_i\cdot p_j)}{x_L}\delta\left[2x_L(p_i\cdot Q)(p_j\cdot Q)-p_i\cdot p_j\right]\nn\\
= &\frac{1}{2\pi\mathrm{i}}\frac{(p_i\cdot p_j)}{x_L}\left\{\frac{1}{\left[2x_L(p_i\cdot Q)(p_j\cdot Q)-p_i\cdot p_j\right]-\mathrm{i} 0}-\frac{1}{\left[2x_L(p_i\cdot Q)(p_j\cdot Q)-p_i\cdot p_j\right]+\mathrm{i} 0}\right\}  \,,
\end{align}
where we set the center-of-mass energy $Q=1$ for simplicity. After topology classification and identification as described in Ref.~\cite{Luo:2019nig}, the  E${}^2$EC integral can be reduced to a set of master integrals $\widetilde{\mathcal{I}}_k(x_L,\epsilon)$ using IBP reduction and E${}^2$EC distribution can be written as a linear combination of the master integrals, 
\begin{equation}
\label{eq:E2EC_MIs}
    \frac{\df}{\df x_L}\sigma^{[3]}_{\text{E}^2\text{EC}}(x_L,\epsilon)=\sum_k \mathcal{C}_k(x_L, \epsilon)\widetilde{\mathcal{I}}_k(x_L,\epsilon)  \,.
\end{equation}
Specifically, we generate the standard IBP equations using \textsc{Litered}~\cite{Lee:2012cn,Lee:2013mka}, add the missing one that is associated with the EEC measurement function by hand, and do the reduction in \textsc{Fire6}~\cite{Smirnov:2019qkx}. The master integrals turn out to be the same as in NLO EEC calculation for both $e^+e^-$ annihilation and gluonic Higgs decays, which can be converted into the canonical basis using the DE package \textsc{Canonica}~\cite{Meyer:2017joq}. 

In order to obtain the collinear $\df\sigma^{[3]}_{\text{E}^2\text{EC}}(x_L,\epsilon)/\df x_L$, one could surely expand the differential equation asymptotically and derive the analytical expression of the master integrals in that limit. However, the fact that the most singular power of  $\mathcal{C}_k$'s is $x_L^{-8}$ requires us to compute the master integrals up to $\cO(x_L^7)$ order, which turns out to be expensive and time-consuming. This becomes worse in the higher-point energy correlator since the singular power increases as well.
One antidote is to reconstruct the coefficients from DE following an ansatz on the structure of asymptotic expansion. 
In fact, the pattern turns out to be
$x_L^{-\epsilon}U^{(1)}_1(x_L,\epsilon)$ at $\cO(\alpha_s)$ and $x_L^{-\epsilon}U^{(2)}_1(x_L,\epsilon)+x_L^{-2\epsilon}U^{(2)}_2(x_L,\epsilon)$ at $\cO(\alpha_s^2)$, where $U$ denotes a series in $x_L$ with rational fractions of $\epsilon$ as the coefficients. 

Therefore, we perform the asymptotic expansion in the following way. First of all, we solve the canonical DE at $0<x_L<1$ to transcendental-weight 5, which can be used to obtain the finite part of the contact term via Eq.~\eqref{eq:E2EC_MIs}. The result can be converted to Harmonic polylogarithms (HPLs) with the package \textsc{Hpl}~\cite{Maitre:2005uu} or even classical polylogarithms. Then we can extract the leading power $x_L^{-1}$ and match it to a resummed ansatz 
\begin{equation}\label{eq:resum_E2EC}
x_L^{-1-\epsilon}C_1(\epsilon)+x_L^{-1-2\epsilon}C_2(\epsilon)  \,,
\end{equation}
with unknown $\epsilon$-series $C_1(\epsilon)$ and $C_2(\epsilon)$. 
The matching between fixed order calculation and the resummed structure in $\epsilon$ leads to the solution of $C_1(\epsilon)$ and $C_2(\epsilon)$ in $\epsilon$ expansion.
Since $x_L^{-1-\epsilon}$ and $x_L^{-1-2\epsilon}$ are defined with plus distribution similar to Eq.~\eqref{eq:plusdistdef}, now we obtain the correct $\cO(\epsilon^0)$ formula for $\df\sigma^{[3]}_{\text{E}^2\text{EC}}(x_L,\epsilon)/\df x_L$ in the collinear limit.

The last remaining piece is $\df\sigma^{[3]}_{\text{E}^3\text{C}}(x_L,\epsilon)/\df x_L$. The computation of the self-energy correlator is much easier since its  dependence on $x_L$ is factorized out by $\delta(x_L)$ and the integrals are simply standard cut integrals.  The master integrals can be found in the literature, e.g. \cite{Gehrmann-DeRidder:2003pne,Magerya:2019cvz}. Eventually adding $\df \sigma^{[3]}_{\text{E}^2\text{EC}}(x_L)/\df x_L$ and $\df\sigma^{[3]}_{\text{E}^3\text{C}}(x_L,\epsilon)/\df x_L$ together, we obtain the complete contact terms $\df\sigma^{[3]}_{\text{C}}(x_L,\epsilon)/\df x_L$ for E3C distribution. The results are also summarized in Eq.~\eqref{eq:E3C_contact_q_2loop_1}-\eqref{eq:E3C_contact_g_2loop_2}. Combined with the nonidentical energy weight contributions, we find all $\frac{1}{\epsilon}$ canceled and thus the infrared safety is guaranteed as expected.

\subsubsection{Results of two-loop jet function constants}

With all individual contributions at hand, the full expressions of 2-loop E3Cs in the collinear limit can be written as
\begin{align}
\frac{1}{\sigma_0}\frac{\df\sigma^{[3],\text{2-loop}}_{\text{q}}}{\df x_L}=
&
\,2\,\frac{\df J^{\text{nonid,2-loop}}_q}{\df x_L}+\frac{1}{\sigma_0}\frac{\df\sigma^{[3],\text{2-loop}}_{\text{C,q}}}{\df x_L} \quad(e^+e^- \, \text{annihilation})\,,\label{eq:E3C_q_2loop}\\
\frac{1}{\sigma^\prime_0}\frac{\df\sigma^{[3],\text{2-loop}}_{\text{g}}}{\df x_L}=
&
\,2\,\frac{\df J^{\text{nonid,2-loop}}_g}{\df x_L}+\frac{1}{\sigma^\prime_0}\frac{\df\sigma^{[3],\text{2-loop}}_{\text{C,g}}}{\df x_L} \quad(\text{gluonic Higgs decay})\label{eq:E3C_g_2loop}\,.
\end{align}
Here a factor of $2$ is added because we only consider a single jet in Sec.~\ref{sec:3_1_nonidentical}.
Given the tree-level hard functions, 
$\{H^{(0)}_q,H^{(0)}_g\}=\{2\delta(1-x),0\}$ for $e^+e^-$ annihilation and $\{\tilde{H}^{(0)}_q,\tilde{H}^{(0)}_g\}=\{0,2\delta(1-x)\}$ for the Higgs decay through the effective $Hgg$ coupling, we can extract the two-loop jet constant directly from the $\delta(x_L)$ contribution from Eq.~\eqref{eq:E3C_q_2loop} and Eq.~\eqref{eq:E3C_g_2loop}. We find that the $\mu$ dependence are in full agreement with prediction from RG evolution, providing strong check to our calculation.  The $\mu$ independent part are the new results from this calculation. For the quark jet function, we get
\begin{align}
	\label{eq:j2q_res}
j_2^{q,[3]}=12.3020 \,C_F T_F n_f-26.2764 \,C_A C_F +21.3943 \,C_F^2 \,,
\end{align}
and for gluon jet functions
\begin{align}
	\label{eq:j2g_res}
j_2^{g,[3]}=17.5487 \,C_A T_F n_f -2.05342 \,C_F T_F n_f -5.97991 \,C_A^2+0.904693 \,n_f^2 T_F^2\,. 
\end{align}

\subsection{Perturbative resummation}
\label{sec:pert_res}

We start by defining the logarithmic order for our E3C resummation. The ingredients needed for our E3C resummation are summarized in Table~\ref{tab:ords}. This includes the order of timelike splitting kernel $\hat P(y)$, the boundary information (hard and jet constants), the $\beta$ function for running coupling as well as the fixed-order matching.\footnote{This is the same log counting as $\text{N}^k\text{LL}^{\prime}$ in SCET, except that we omit all ${}^\prime$ for convenience.} Due to the absent of analytic method to solve the RG equation exactly, we also truncate in the number of loops of the RGE solution to the desired logarithmic order~\cite{Dixon:2019uzg}.

\begin{table}
\begin{center}
\begingroup
\renewcommand{\arraystretch}{1.2}
  \begin{tabular}{|c|c|c|c|c|c|}  \hline
   resummation order &     $\hat{P}(y)$ &    $\vec{H}$, $\vec{J}\, \text{ constants}$ &   $\beta[\alpha_s]$ & fixed-order matching\\  
    \hline
    LL  & tree & tree & 1-loop & LO \\
    \hline
    NLL  & 1-loop & 1-loop & 2-loop & NLO \\
    \hline
    NNLL & 2-loop & 2-loop & 3-loop & NNLO \\
    \hline
\end{tabular}
\endgroup
\end{center}
\vspace{-0.2cm}
\caption{Definition of the resummation order and their corresponding fixed-order matching.}
\label{tab:ords}
\end{table}

We first review the LL resummation in $e^+e^-$ annihilation. Based on our resummation setting, it is safe to set $x=1$ in the argument of E3C jet function in Eq.~\eqref{eq:jet_evo}, which only affects the higher-order terms beyond LL. This leads to
\begin{equation}
\frac{d\vec{J}^{[N]}_{\rm{LL}} (\ln\frac{x_LQ^2}{\mu^2})}{d\ln \mu^2}=\vec{J}^{[N]}_{\rm{LL}} (\ln\frac{x_LQ^2}{\mu^2})\cdot\frac{\alpha_s}{4\pi} \int_0^1 dy\, y^N \hat P^{(0)}(y)=-\vec{J}^{[N]}_{\rm{LL}} (\ln\frac{x_LQ^2}{\mu^2})\cdot\frac{\alpha_s}{4\pi}\gamma_T^{(0)}(N+1) \,.
\end{equation}
Here, we introduce the anomalous dimension to be the moment of timelike splitting kernel
\begin{equation}
\gamma_T(N)\equiv -\int_0^1 dy\, y^N \hat P(y)=\left(\frac{\alpha_s}{4\pi}\right)\gamma_T^{(0)}+\left(\frac{\alpha_s}{4\pi}\right)^2 \gamma_T^{(1)}+\cdots \,.
\end{equation}
Then given the boundary condition $\vec{J}^{(0)}=\{2^{-N},2^{-N}\}$, we can directly write down the solution to LL jet function:
\begin{equation}
	\label{eq:LLres}
\vec{J}_{\text{LL}}^{[N]}=2^{-N}(1,1)\cdot \exp\left[-\frac{\gamma_T^{(0)}}{\beta_0} \ln \frac{\alpha_s\left(\sqrt{x_L}Q\right)}{\alpha_s(\mu)}\right] \,.
\end{equation}
Plugging both jet and hard functions into the factorization for the cumulant $\Sigma^{[N]}$ and differentiating it with respect to $x_L$, we obtain the LL resummed physical spectrum for E3C.

Beyond LL, the $x=1$ approximation is no longer valid, and instead we have to solve the jet RGE directly. While it is difficult to obtain a close-form solution for this modified DGLAP equation, we find that a truncated solution in $\alpha_s$ is already in good convergence. Explicitly, we assume the jet function takes the form 
\begin{equation}
\vec{J}^{[N]}=\underbrace{\sum_{i=1}^{\infty} \alpha_s^i L^i \vec{c}_{i,i}}_{\text{LL}}+\underbrace{\sum_{i=1}^{\infty} \alpha_s^i L^{i-1} \vec{c}_{i,i-1}}_{\text{NLL}}+\underbrace{\sum_{i=1}^{\infty} \alpha_s^i L^{i-2} \vec{c}_{i,i-2}}_{\text{NNLL}}+\cdots \, ,
\end{equation}
with $L\equiv\ln\frac{x_L Q^2}{\mu^2}$ and $c_{i,j}$ unknown constants, and solve both the jet RGE and $\beta$ RGE order by order in $\alpha_s$ (which is referred as expanded solution). In practice, we evaluate it numerically up to $\mathcal{O}(\alpha_s^{50})$. Another advantage of using expanded solution is that we only need certain moments of the hard functions. For example, consider one term from the jet function, $\vec{J}^{[N]}\supset \alpha_s^2  \vec{c}_{2,2} L^2$, and plug into Eq.~\eqref{eq:fac_nu}, we find
\begin{align}
&\Sigma^{[N]}\supset \alpha_s^2 \vec{c}_{2,2}\cdot \int_0^1 dx\, x^N \ln^2 \left(\frac{x_L x^2 Q^2}{\mu^2}\right)\cdot \vec{H}_{ee}\left(x, \ln\frac{Q^2}{\mu^2}\right)\notag\\
& =\alpha_s^2 \vec{c}_{2,2}\cdot \bigg[\ln^2 \left(\frac{x_L Q^2}{\mu^2}\right)^2 \int_0^1 dx\, x^N \vec{H}_{ee}\left(x, \ln\frac{Q^2}{\mu^2}\right)\notag\\
&+2 \ln \left(\frac{x_L Q^2}{\mu^2}\right) \int_0^1dx\,  \ln x^2 x^N \vec{H}_{ee}\left(x, \ln\frac{Q^2}{\mu^2}\right)+ \int_0^1dx\,  \ln^2 x^2 x^N \vec{H}_{ee}\left(x, \ln\frac{Q^2}{\mu^2}\right)\bigg]\notag\\
&=\alpha_s^2 \vec{c}_{2,2}\cdot\left[ \ln^2 \left(\frac{x_L Q^2}{\mu^2}\right)+2  \ln \left(\frac{x_L Q^2}{\mu^2}\right) \partial_{N}+4\partial_{N}^2\right] \int_0^1 x^N \vec{H}_{ee}\left(x, \ln\frac{Q^2}{\mu^2}\right) \,, 
\end{align}
where the three terms correspond to the standard moment, the single logarithmic moment and the double logarithmic moment of the E3C hard function. To derive the last line, we also use the following relation
\begin{equation}
 \int_0^1 \ln^k x^2 x^N \vec{H}_{ee}\left(x, \ln\frac{Q^2}{\mu^2}\right)=2^k \partial_{N}^k   \int_0^1 x^N \vec{H}_{ee}\left(x, \ln\frac{Q^2}{\mu^2}\right) \,.
\end{equation}
In the Appendix~\ref{sec:hard_and_jet}, we provide all the hard moments with $N=2,3$ that are required for NNLL resummation.

In this paper, we present results for the NNLL resummation of E3C for $e^+e^-$ annihilation,  and approximate NNLL resummation for jets from the hadronic collision process $pp\rightarrow jj$. For $e^+e^-$ annihilation, we have all ingredients needed for NNLL resummation. And since there is no accurate fixed-order data for E3C at NNLO, we will instead match the NNLL result to NLO. Regarding the dijet production, due to the absence of the two-loop hard constant, we will present the approximate NNLL resummation (which we refer as NNLL$_{\mathrm{approx}}$), with an additional uncertainty coming from the missing two-loop hard constant. Resummation with the accurate two-loop hard function as well as the matching with fixed-order result are left as future improvements.

\section{NNLL resummation in $e^+e^-$ annihilation}
\label{sec:resummation_ee}

With all the ingredients at hand, now we can present the NNLL resummation prediction. In this section, we first consider $e^+e^-$ collision at two different energies: $250$ GeV and $1$ TeV.  In the resummation calculation, we will use $\alpha(m_Z)=0.118$.

\subsection{Resummation results}
\label{sec:resummation_ee_sub1}

Following the discussion in Sec.~\ref{sec:pert_res}, our resummation is performed by perturbatively solving the jet function RG equation to order $\mathcal{O}(\alpha_s^{50})$, plugging back to the cumulant factorization and finally truncating the logarithms $\ln\frac{x_L Q^2}{\mu^2}$ to the desired order.  In the resummation formula, we set canonical jet scale $\mu_j=\mu_h \sqrt{x_L}$ in the factorization, leaving a single hard scale $\mu_h=\mu$ in the resummed expression. We vary the scale $\mu$ to estimate the uncertainty from higher order corrections. Regarding the observables, below we consider three cases: $ N=2$, $ N=3$ and their ratio. 

\begin{figure}[ht]
	\centering
	\setlength{\abovecaptionskip}{-0.01cm}
	\setlength{\belowcaptionskip}{-0.0cm} 
	\includegraphics[width=7.2cm]{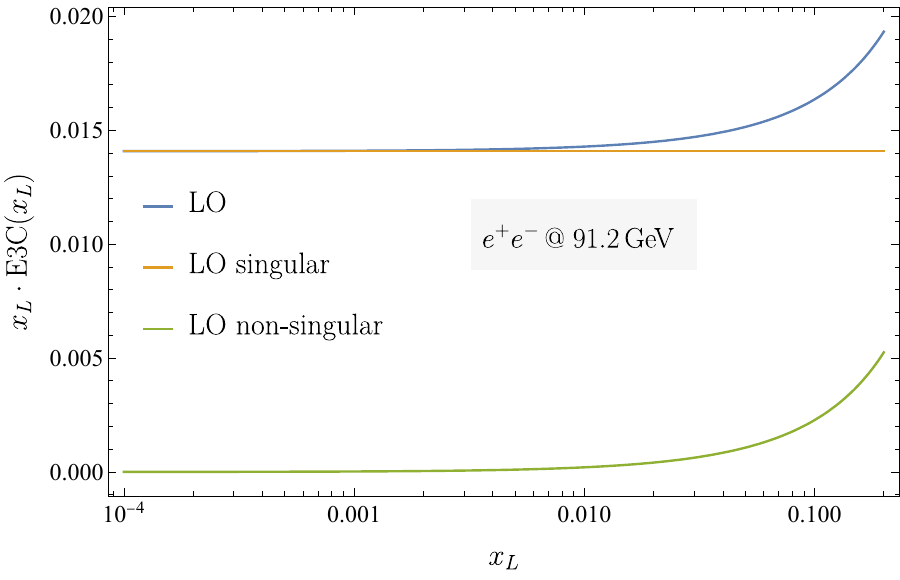}\hfill
	\includegraphics[width=7.2cm]{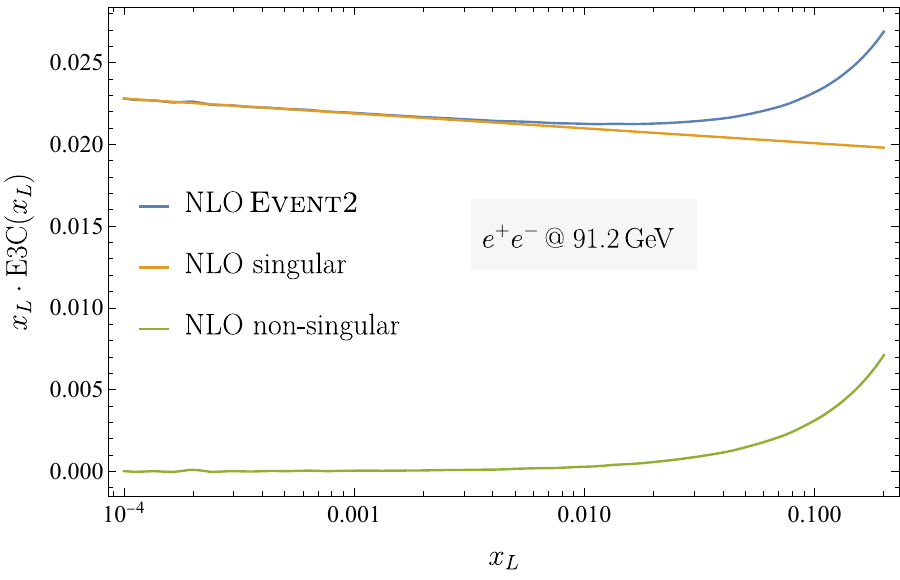}
	\caption{The comparison of fixed-order result and the singular expansion from resummation. The left panel shows good agreement between LO expression and the $\mathcal{O}(\alpha_s)$ singular expansion of E3C resummed result. The difference between them is the non-singular result. The right panel gives the corresponding contributions at NLO, with the numerical full NLO prediction from \textsc{Event2}.}
	\label{fig:e3c_event2_check}
\end{figure}

The $ N=2$ case is precisely the EEC observable, where we directly use the result from Ref.~\cite{Dixon:2019uzg}, and the singular expansion has been verified against the NLO EEC fixed-order calculation. For $ N=3$ case, this is the main result of this paper. In Fig.~\ref{fig:e3c_event2_check}, we first check our $\mathcal{O}(\alpha_s^2)$ expansion with the Monte Carlo program \textsc{Event2}. In the collinear limit, we find excellent agreement between theory and numeric result, while in the meantime, this also suggests the non-singular contribution from fixed-order calculation is negligible in this limit. 

\begin{figure}[ht]
	\centering
	\setlength{\abovecaptionskip}{-0.0cm}
	\setlength{\belowcaptionskip}{-0.0cm} 
	\includegraphics[width=7.3cm]{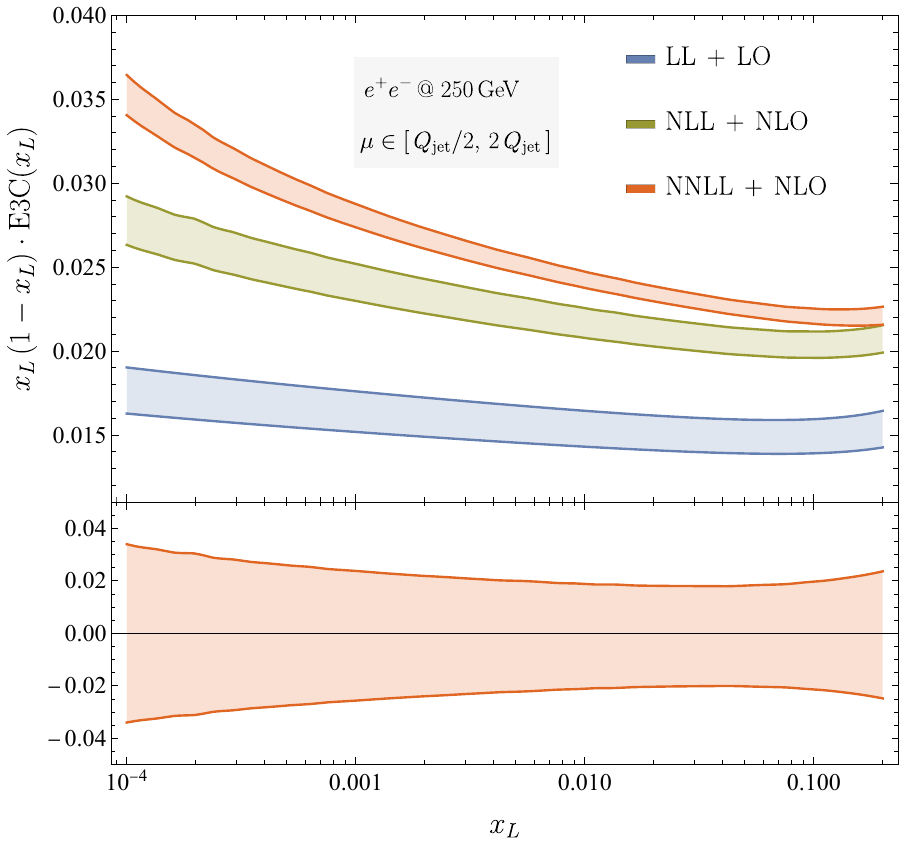}\;
	\includegraphics[width=7.3cm]{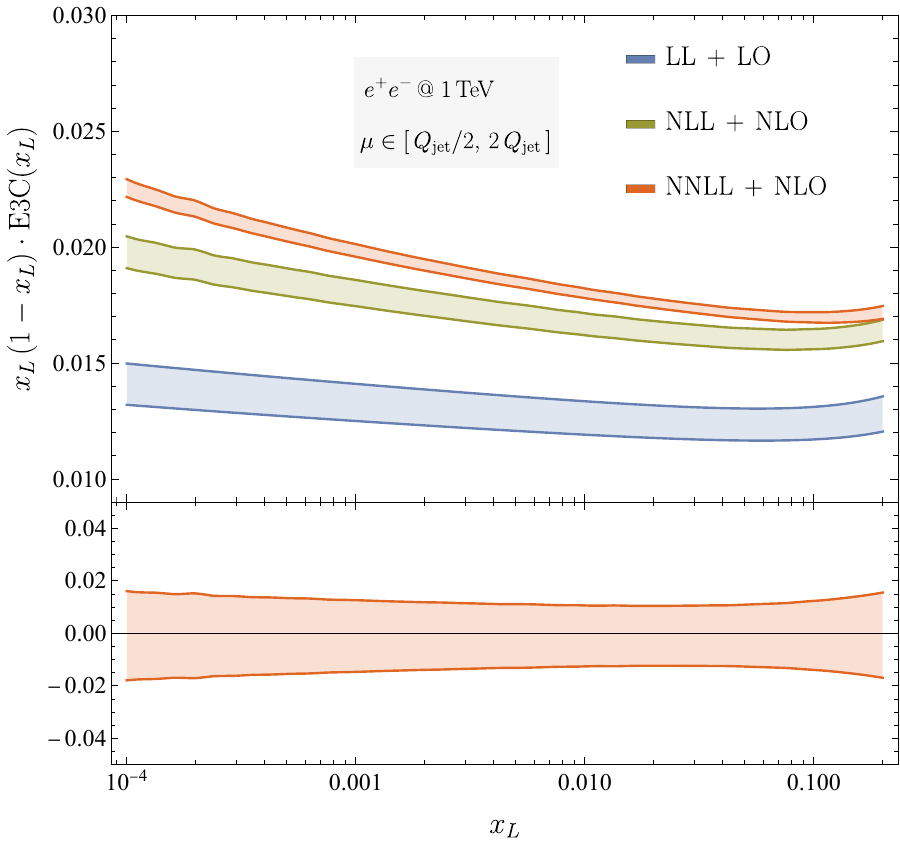}
	\caption{The resummed E3C distribution up to NNLL+NLO, multiplied by a factor $x_L(1-x_L)$, for $e^+e^-$ collision at 250 GeV (left top panel) and 1 TeV (right top panel). Uncertainty bands are obtained by varying the hard scale $\mu$ around the nominal value $Q_{\rm jet} = Q/2$ by a factor of $2$. The lower panels show the relative scale uncertainty of the NNLL+NLO distribution around the central value.}
	\label{fig:e3c_nnll}
\end{figure}

Nevertheless, the matching formula can be written as
\begin{equation}
	\frac{d\sigma^{\text{match}}}{dx_L}=\frac{d\sigma^{\text{resum}}}{dx_L}-\frac{d\sigma^{\text{sing}}}{dx_L}+\frac{d\sigma^{\text{FO}}}{dx_L} \, .
\end{equation}
Here each term is a function of $\alpha_s(\mu)$ evaluated at the hard scale $\mu_h=\mu$.
In Fig.~\ref{fig:e3c_nnll}, we present the E3C resummation up to NNLL, matched to fixed-order. As explained above, due to the absence of NNLO data, we only match NNLL to NLO. The hard scale is chosen to be half of the center-of-mass energy $\mu=Q_{\text{jet}}\equiv Q/2$, the typical energy for each quark jet, and the scale uncertainty is obtained by varying the hard scale by a factor of 2. In both energies, the uncertainty band width goes down as we increase the resummation order, while at $1$ TeV, we have a tighter band because the coupling $\alpha_s$ runs slower at high energy. At NNLL, we find a relative $4 \%$ hard uncertainty for $Q=250$ GeV and $2 \%$ for $Q=1$ TeV. We find large corrections as we go from LL to NNLL, as was also observed previously in \cite{Dixon:2019uzg}, which emphasize the importance of higher order corrections. For higher center-of-mass energy, the convergence between different orders is improved. 

\begin{figure}[ht]
	\centering
	\setlength{\abovecaptionskip}{-0.0cm}
	\setlength{\belowcaptionskip}{-0.0cm} 
	\includegraphics[width=7.3cm]{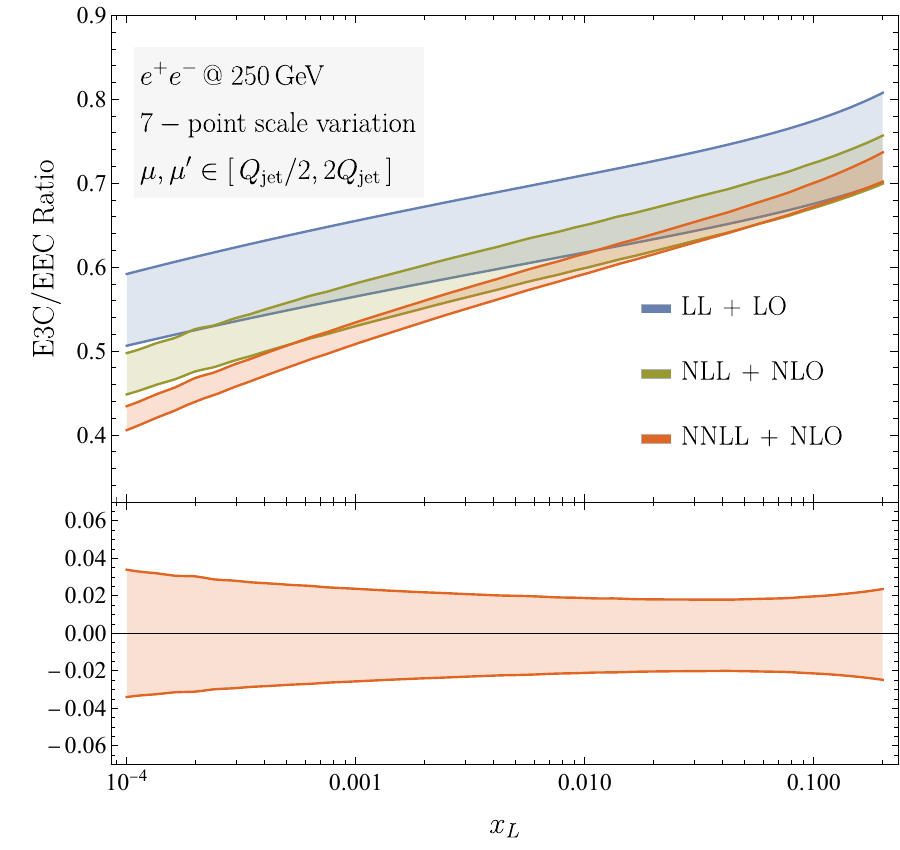}\;
	\includegraphics[width=7.3cm]{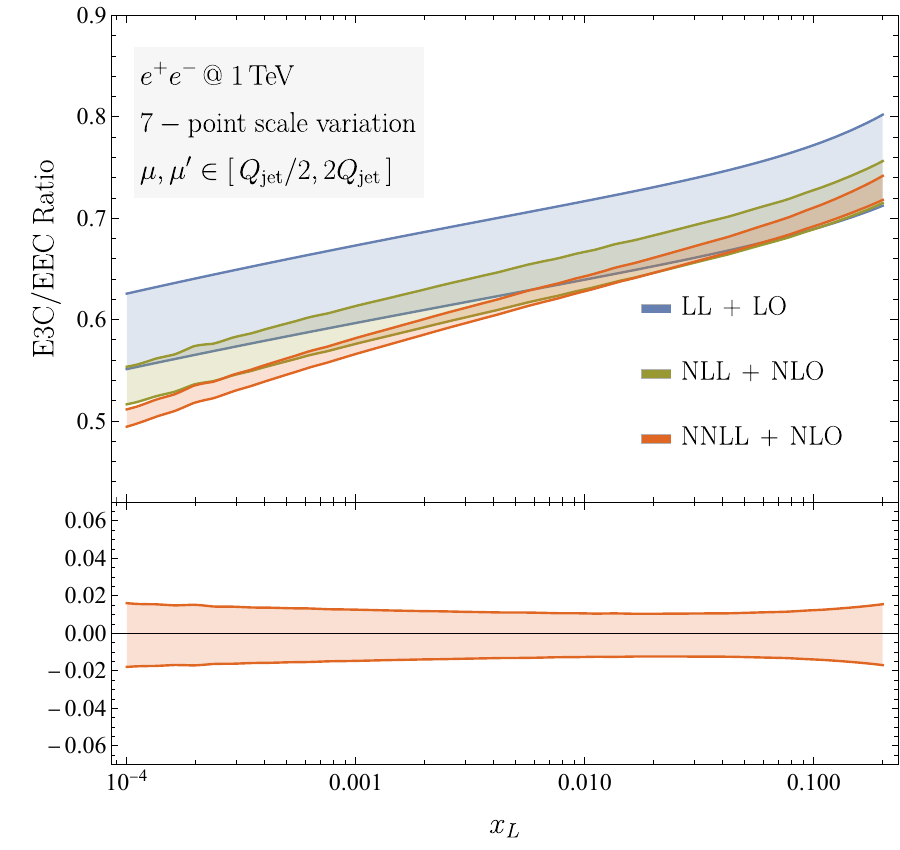}
	\caption{The ratio between resummed E3C and EEC distribution $\Delta_{3,2}(x_L)$ up to NNLL+NLO for $e^+e^-$ collision at 250 GeV (left top panel) and 1 TeV (right top panel). Uncertainty bands are obtained in the seven-point scale variation scheme, namely varying the scales $\mu$ and $\mu^{\prime}$ in the two factors independently by a factor of $2$ around the central value. The lower panels show the relative scale uncertainty of the NNLL+NLO distribution around the central value.}
	\label{fig:ratio_nnll}
\end{figure}

To improve the convergence, we also introduce the ratio of different point energy correlators, namely~\cite{Chen:2020vvp}
\begin{equation}
	\Delta_{m,n}(x_L,\mu, \mu^\prime)\equiv\frac{d\sigma^{[m]}/d x_L}{d\sigma^{[n]}/d x_L},\quad m,n\geq 2 \, ,
\end{equation}
where $\mu$ and $\mu^\prime$ are the hard scale in $\frac{d\sigma^{[m]}}{d x_L}$ and $\frac{d\sigma^{[n]}}{d x_L}$ respectively. In particular, we focus on the ratio between fully matched E3C and EEC, i.e. $\Delta_{3,2}(x_L)$. In Fig.~\ref{fig:ratio_nnll}, we show the NNLL resummed $\Delta_{3,2}(x_L)$ at again $Q=250$ GeV and $Q=1$ TeV, and find good convergence. This implies that the ratio can be used as precision observable. For hard scale uncertainty, we use the seven-point scale variation, which amounts to varying the scales in both numerator and denominator independently by a factor of $2$, to a combination of 
\begin{align}
\bigg(\frac{\mu}{Q_{\text{jet}}},\frac{\mu^\prime}{Q_{\text{jet}}} \bigg)\in \bigg\{
	\bigg(\frac{1}{2},\frac{1}{2}\bigg),\,\bigg(2,2 \bigg),\,\bigg(1, 2\bigg),\,\bigg(1, 1\bigg),\, \bigg(2,1\bigg),\,\bigg(1,\frac{1}{2}\bigg),\,\bigg(\frac{1}{2},1\bigg) \bigg\} \, ,
\end{align}
and take the envelope as the uncertainty estimation. The convergence also indicates that ENC shares similar non-perturbative behavior in the collinear limit and taking the ratio strongly suppresses the power corrections. 

\subsection{Hadronization corrections}

In this subsection, we consider the power-suppressed hadronization corrections in the collinear limit. At present hadronization corrections cannot be computed from first principle. 
For simplicity, we use a phenomenological form for the leading non-perturbative power correction as suggested in \cite{Korchemsky:1999kt}, and fit the unknown parameters from a Monte Carlo program. 
This provides some insights on how to model the hadronization effect for a global fit in the future.

In general, the non-perturbative corrections in infrared-collinear safe observables are (at least) suppressed as $\Lambda_{\rm QCD}/Q$ to some power, where $Q$ is the hard scale of the process. Following from the LL result in Eq.~\eqref{eq:LLres}, we observe that in the collinear limit, there exists a lower scale $\sqrt{x_L}Q$ in the coupling, and the most important non-perturbative correction that could potentially appear is linear in $\Lambda_{\rm QCD}$ and takes the form $\Lambda_{\rm QCD}/(\sqrt{x_L}Q)$, multiplied with an extra kinematic factor $1/x_L$. The sub-leading non-perturbative corrections with additional powers of $\Lambda_{\rm QCD}/(\sqrt{x_L}Q)$ will become necessary down to small $x_L \sim \Lambda_{\rm QCD}^2/Q^2$, where the perturbation theory also breaks down.
For the leading non-perturbative correction we are considering, such structure is in fact recovered for the EEC in the fragmentation modeling of non-perturbative radiations \cite{Basham:1978zq} and and analysis using renormalon or dispersive techniques \cite{Korchemsky:1999kt,Dokshitzer:1999sh,Schindler:2023cww}.

As a qualitative analysis, we use the following parametrization of the leading non-perturbative correction,
\begin{align}
	\label{eq:np_functional}
	\frac{d\sigma^{{\rm NP}-soft}}{d x_L}& = \frac{1}{x_L} \cdot  \bigg(\,\frac{\tilde{\Lambda}}{\sqrt{x_L} \, Q}\,\bigg)^{1+\gamma}  \qquad {\rm (\,soft \; fragmentation\,),} 
\end{align}
we verify the scaling behaviour of the non-perturbative correction in the collinear limit for both EEC and E3C distributions with \textsc{Pythia8}~\cite{Sjostrand:2014zea},
and extract the non-perturbative parameters by fitting from the difference of the hadron level and parton level predictions. Note that the issues of extracting the non-perturbative power corrections from Monte Carlo generators have been pointed out in Ref.~\cite{Becher:2008cf}. In particular, the corrections from the hadronization modeling in the Monte Carlo programs in fact unfaithfully absorb partial subleading-log contributions, as the hadronization modeling has been tuned to reproduce some collider data with limited perturbative accuracy. Therefore, in this paper we only use Monte Carlo to illustrate the impact of power correction for individual EEC and E3C distribution as well as their ratio.

\begin{figure}[ht]
	\centering
	\setlength{\abovecaptionskip}{-0.0cm}
	\setlength{\belowcaptionskip}{-0.0cm} 
	\includegraphics[width=7.2cm]{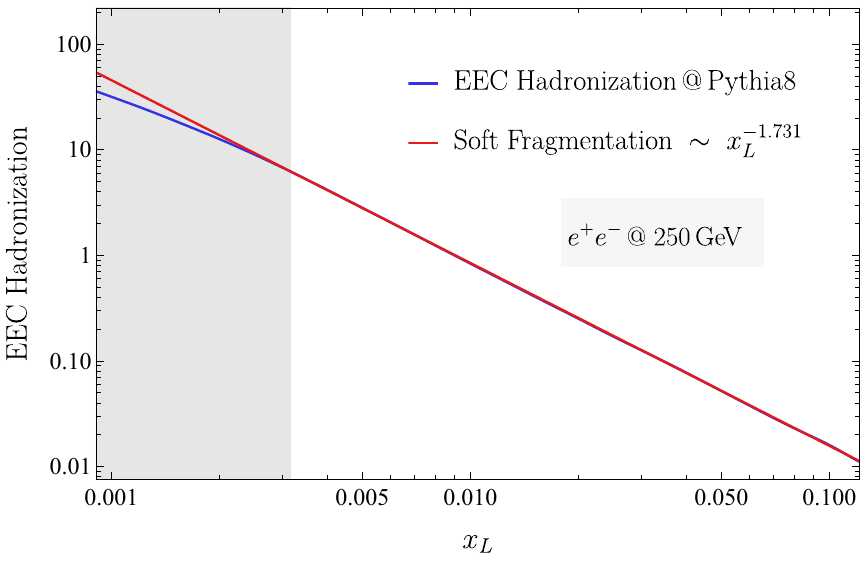} \;
	\includegraphics[width=7.3cm]{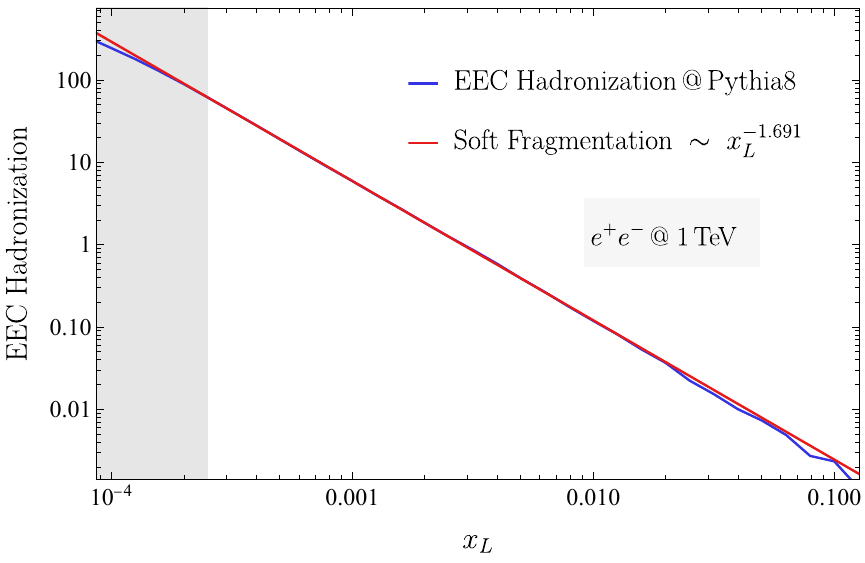} \;
	\includegraphics[width=7.2cm]{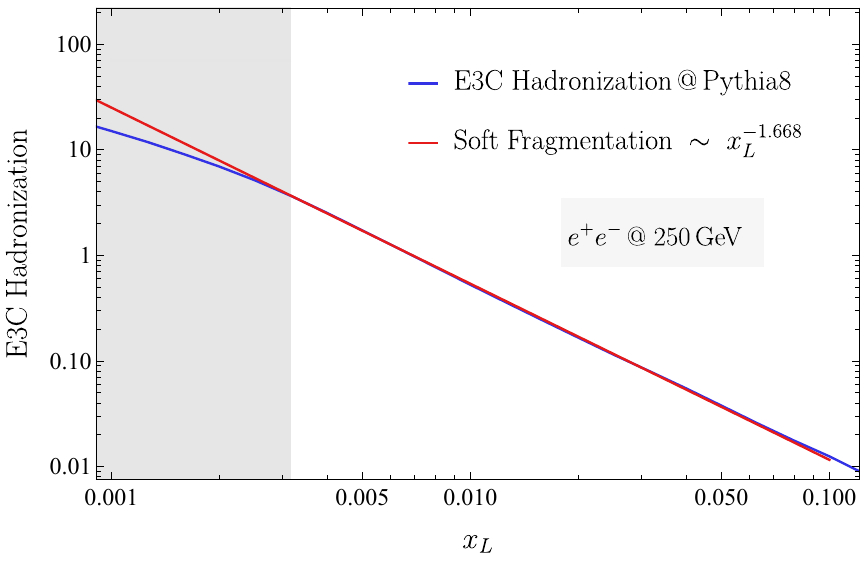} \;\;
	\includegraphics[width=7.2cm]{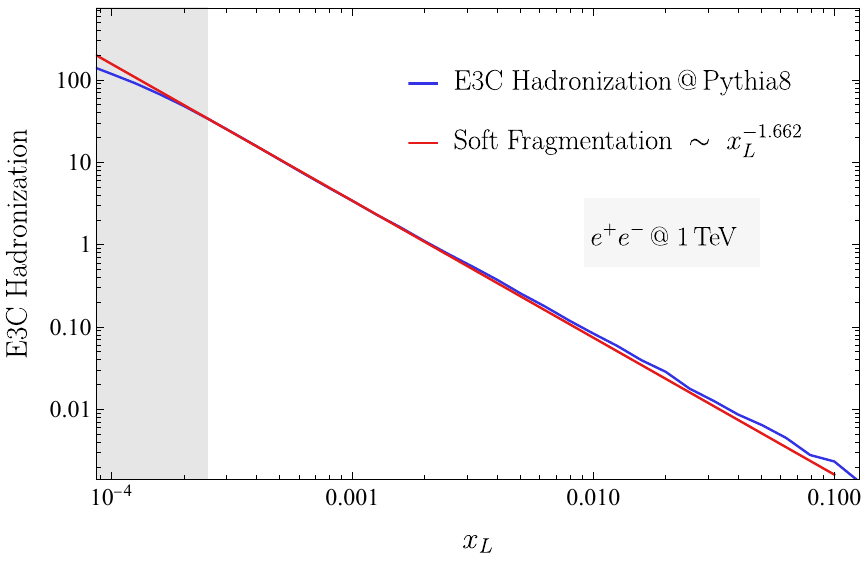} \;
	\caption{Comparison of \textsc{Pythia8} result and the fitting using Eq.~\eqref{eq:np_functional}. The blue curves are the difference of hadron-level and parton-level distribution for EEC and E3C, at both $250$ GeV and $1$ TeV. The red curves are our fitting result with parameters from Eq.~\eqref{eq:EEC_np_para} and~\eqref{eq:E3C_np_para}. The shaded region stands for the range where the parametrization of the leading non-perturbative correction is no longer valid and should be excluded from the fit range.}
	\label{fig:pythia8}
\end{figure}

For our case, we stay in the default settings of \textsc{Pythia8} and obtain the following fit at the 95$\%$ confidence level. At $Q=250$ GeV, we find for EEC and E3C:
\begin{align}
	\label{eq:EEC_np_para}
	\tilde{\Lambda}_{2} &= (0.956 \pm 0.031){\rm\, GeV}\,, \quad\; \gamma_{2} = 0.462 \pm 0.017 \, , \nonumber\\
	\tilde{\Lambda}_{3} &= (0.500 \pm 0.040) {\rm\, GeV} \,, \quad\; \gamma_{3} = 0.335 \pm 0.031  \, .
\end{align}
And in the case with $Q=1$ TeV, we have 
\begin{align}
	\label{eq:E3C_np_para}
	\tilde{\Lambda}_{2} &= (0.775 \pm 0.013){\rm\, GeV}\,, \quad\; \gamma_{2} = 0.383 \pm 0.008 \, , \nonumber\\
	\tilde{\Lambda}_{3} &= (0.435 \pm 0.015) {\rm\, GeV} \,, \quad\; \gamma_{3} = 0.325 \pm 0.012  \, .
\end{align}

We emphasis that for too small $x_L$ value, the leading order non-perturbative approximation itself becomes invalidated. The enhancement of the non-perturbative corrections in the collinear limit must be turned off before entering the fully non-perturbative phase, where the degrees of freedom become freely interacting hadrons and a nice scaling behavior follows~\cite{Komiske:2022enw}. 
In this qualitative analysis, we choose the lower bound of the fit range by finding the extreme point of the distributions from hadron level prediction in \textsc{Pythia8}. Multiplying the extreme point by a factor of 2 gives a good estimate of the lower bound for the range where the non-perturbative correction follows the described scaling behavior. 
In Fig.~\ref{fig:pythia8}, we show the relative hadronization correction from both \textsc{Pythia8}  and  our two-parameter fit. Except the shaded region, our parameterization agrees with the Monte Carlo result and it is sufficient for understanding their structure.

\begin{figure}[ht]
	\centering
	\setlength{\abovecaptionskip}{-0.0cm}
	\setlength{\belowcaptionskip}{-0.0cm} 
	\includegraphics[width=7.3cm]{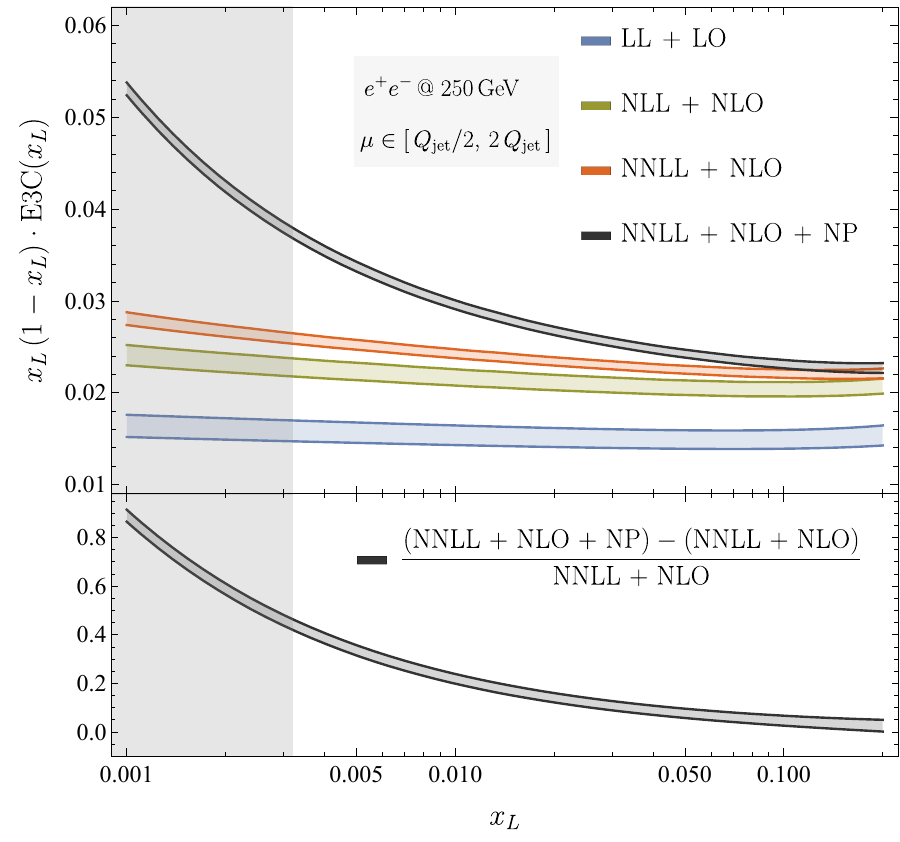} \;
	\includegraphics[width=7.3cm]{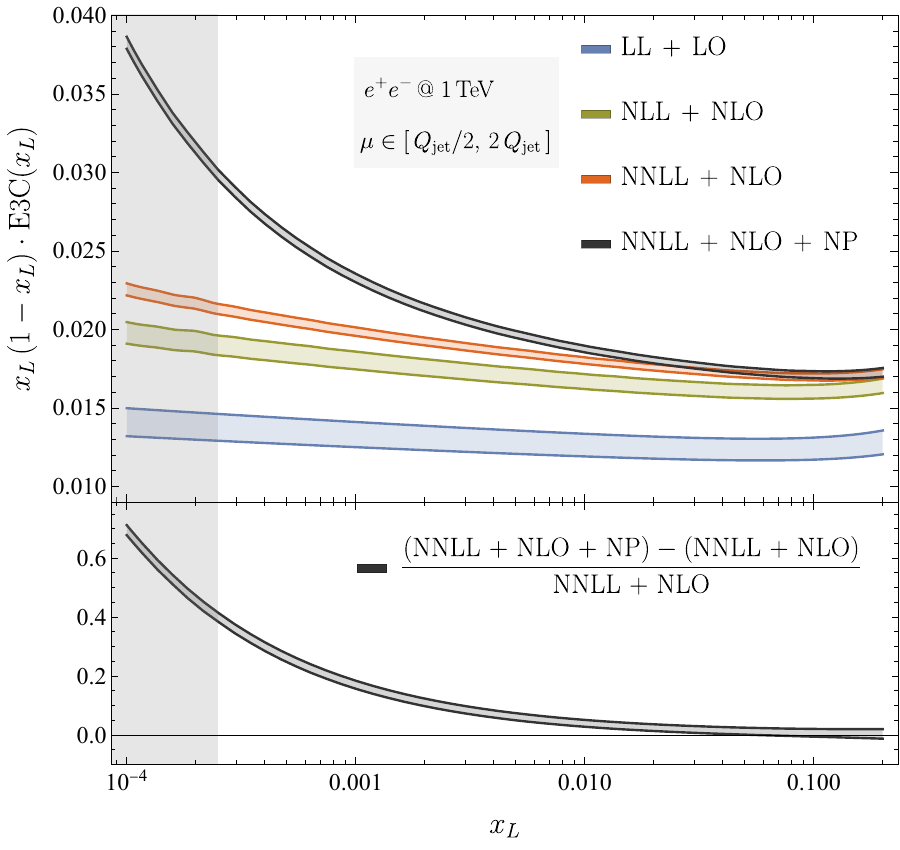} \;
	\caption{The E3C distribution to NNLL+NLO including the non-perturbative (NP) hadronization corrections estimated with \textsc{Pythia8} data, with different plot ranges for collision energies at 250 GeV (left panel) and 1 TeV (right panel). }
	\label{fig:e3c_resum_addnp}
\end{figure}
\begin{figure}[ht]
	\centering
	\setlength{\abovecaptionskip}{-0.0cm}
	\setlength{\belowcaptionskip}{-0.0cm} 
	\includegraphics[width=7.3cm]{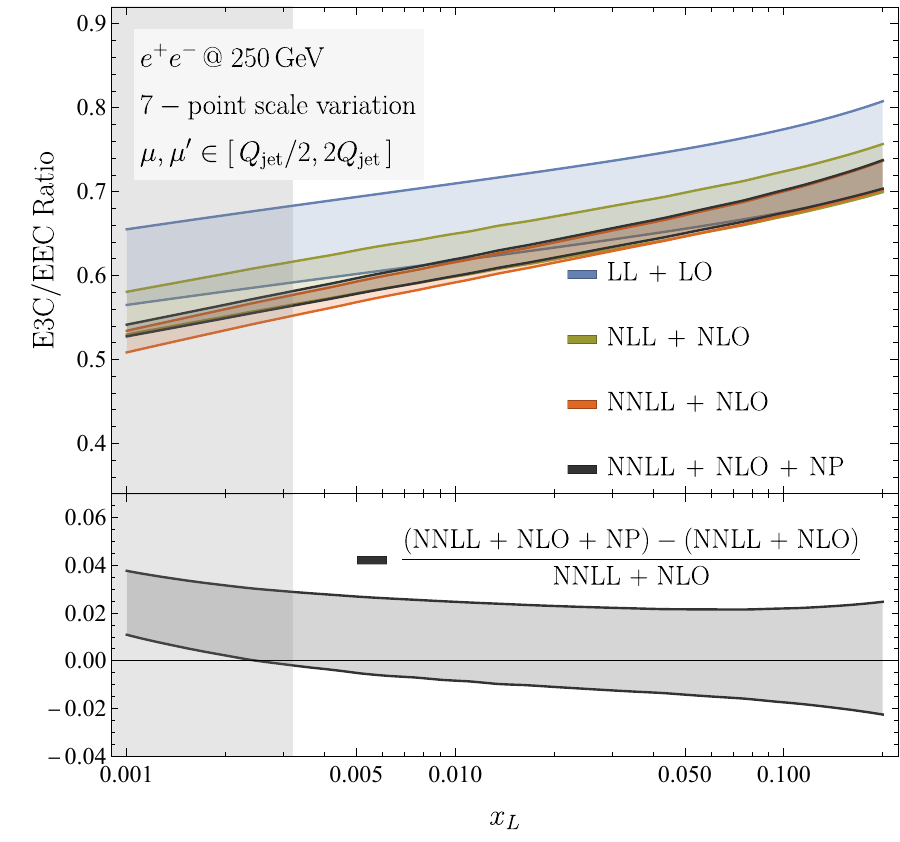} \;
	\includegraphics[width=7.3cm]{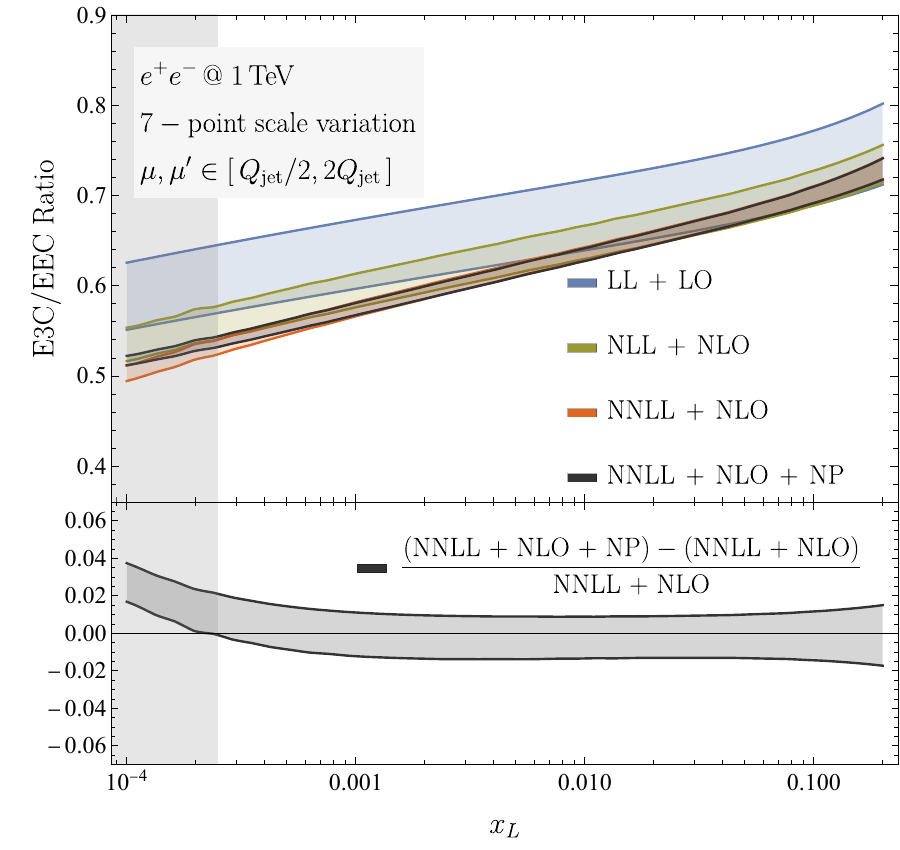} \;
	\caption{The E3C/EEC ratio to NNLL+NLO with non-perturbative (NP) hadronization corrections. The non-perturbative hadronization corrections are estimated as above by $\tilde{\Lambda}/({x_L^{3/2}}Q)$ for both the EEC and E3C distributions with the coefficients fitted from \textsc{Pythia8}. }
	\label{fig:ratio_resum_addnp}
\end{figure}

In Fig.~\ref{fig:e3c_resum_addnp}, we include the non-perturbative correction in the matched E3C resummation, which strongly enhances the extreme collinear limit. At $Q=1$ TeV, the non-perturbative correction changes our NNLL+NLO prediction by only a few percent at $x_L\sim 0.1$, while this modification reaches $50\%$ at $x_L\sim 10^{-4}$. 
This shows that the non-perturbative corrections for energy correlators, though being power suppressed at high energies, can become sizable even at the energy level of future $e^{+}e^{-}$ colliders. However, since EEC and E3C share a close power law in the leading power correction, the enhancement is significantly canceled when considering their ratio $\Delta_{3,2}(x_L)$. As shown in Fig.~\ref{fig:ratio_resum_addnp}, the leading non-perturbative correction only gives rise to roughly $4\%$ effect at $Q=250$ GeV and $2\%$ at $Q=1$ TeV for matched NNLL. This confirms that $\Delta_{3,2}(x_L)$ is insensitive to the hadronization and indeed a good candidate for precise $\alpha_s$ measurement. 

\begin{figure}[ht]
	\centering
	\setlength{\abovecaptionskip}{-0.0cm}
	\setlength{\belowcaptionskip}{-0.0cm} 
	\includegraphics[width=7.3cm]{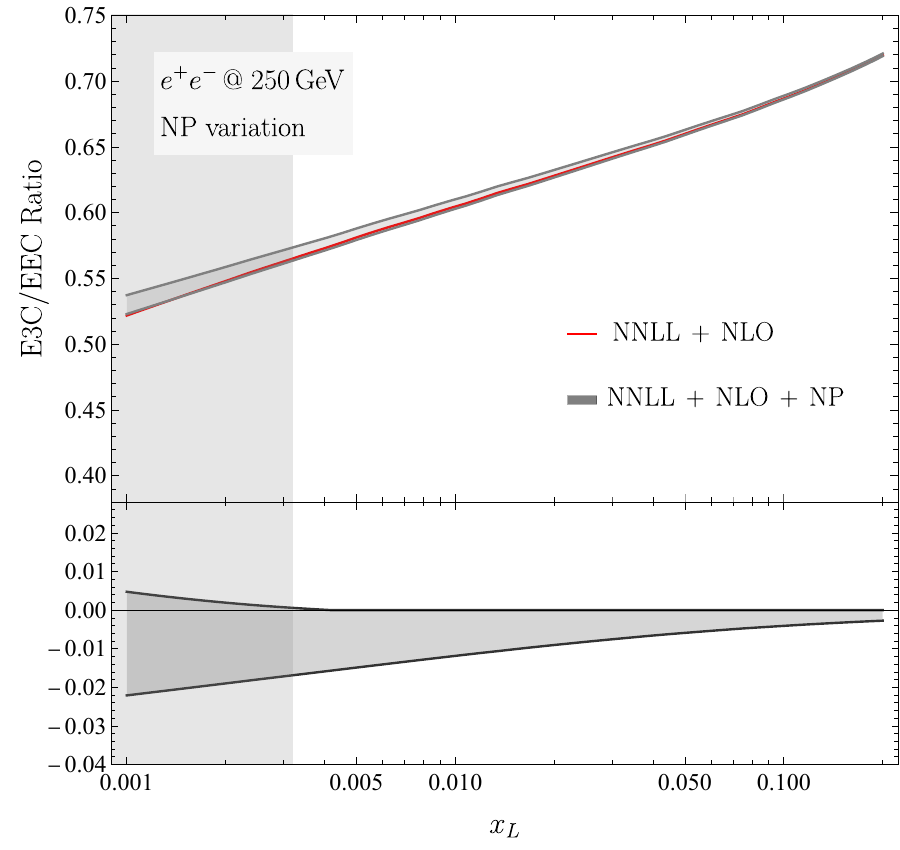} \;
	\includegraphics[width=7.3cm]{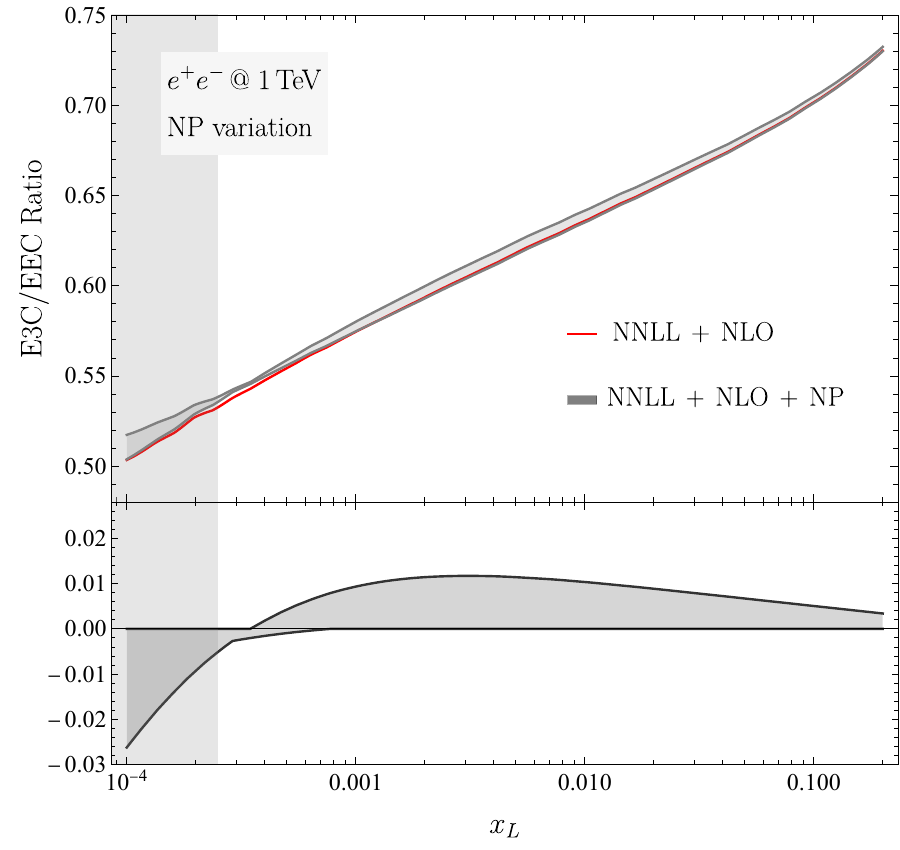} \;
	\caption{The uncertainty from varying the non-perturbative parameters $\tilde\Lambda$ and $\gamma$ in the resummed ratio distribution $\Delta_{3,2}(x_L)$. The bottom panels stand for the relative difference of NNLL+NLO+NP with respect to its central value.}
	\label{fig:ratio_npvar}
\end{figure}

We also investigate the impact on the final resummation results caused by the uncertainties from the two-parameter fit. The statistical error for both $\tilde\Lambda$ and $\gamma$ are given in Eq.~\eqref{eq:EEC_np_para} and~\eqref{eq:E3C_np_para}. Fig.~\ref{fig:ratio_npvar} shows the final uncertainty in the matched NNLL distribution from varying these two NP parameters. In both $Q=250$ GeV and $Q=1$ TeV, excluding the shaded region, the NP uncertainty is much smaller than the hard uncertainty estimated by seven-point variation. In particular, at $Q=1$ TeV, the NP uncertainty is reduced to $1\%$ in the potential fit region. Despite that, we admit that the effect of non-perturbative corrections turns to increase for such small $x_L$ region, and more accurate understanding of the non-perturbative corrections will be required to further improve the precision. 

\subsection{Anticipation of $\alpha_s$ determination}

In this subsection, we discuss the potential of extracting the strong coupling constant $\alpha_s$ from measuring the resumed E3C/EEC ratio $\Delta_{3,2}(x_L)$. 
In literature~\cite{Tulipant:2017ybb, Kardos:2018kqj}, the back-to-back limit of EEC is resummed to NNLL+NLO and has been use for $\alpha_s$ measurement from $e^+e^-$ data. Similar to other event shapes, the non-perturbative correction is significantly large in this region and require careful modeling. And how we profile the resummation and power correction has a sizable effect on the final theory uncertainty. 

Alternatively, we can also do the $\alpha_s$ measurement \textit{only} in the collinear limit.  First of all, as we discussed in Sec.~\ref{sec:resummation_ee_sub1}, the non-singular contribution is almost zero in this limit, and thus it is safe to ignore the higher fixed-order contribution. Secondly, by considering the ratio distribution, $\Delta_{m,n}(x_L)$, the suppressed power corrections will lead to a smaller theory uncertainty and thus more precise $\alpha_s$ determination.  As illustration, we first investigate the sensitivity of $\Delta_{3,2}(x_L)$ when slightly changing the value of $\alpha_s$.
In particular, we vary the value of strong coupling at $Z$-pole $\alpha_s(m_Z)$ by a factor of $5\%$, namely $\alpha_s(m_Z)=\{0.112, 0.118, 0.124\}$ and compare the effect on matched resummation result.

\begin{figure}[ht]
	\centering
	\setlength{\abovecaptionskip}{-0.0cm}
	\setlength{\belowcaptionskip}{-0.0cm} 
	\includegraphics[width=7.3cm]{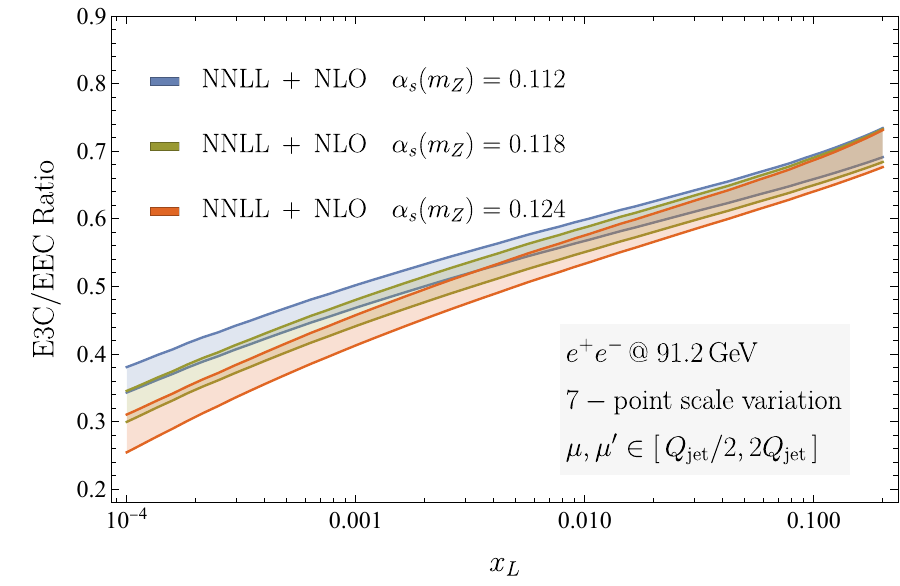} \;
	\includegraphics[width=7.3cm]{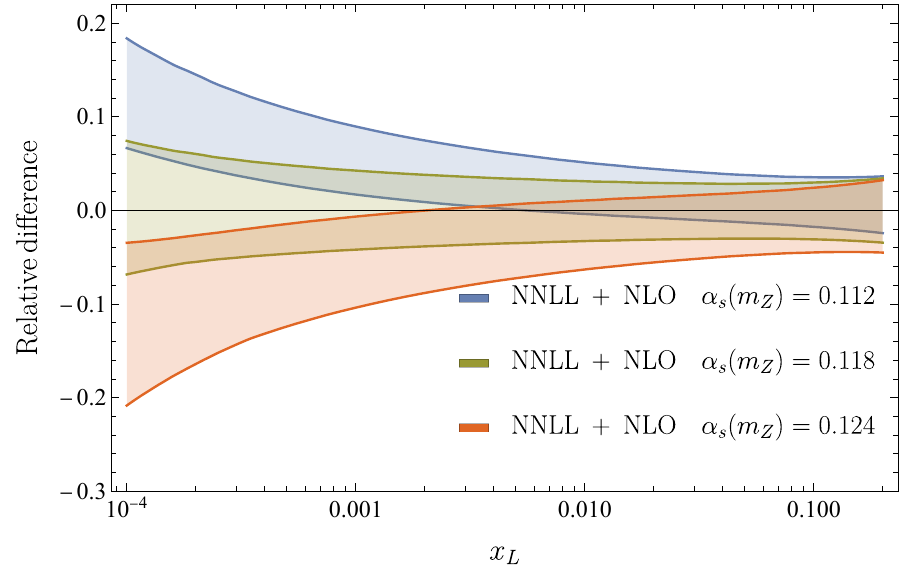} \;
	\caption{Left panel is the matched NNLL ratio distribution $\Delta_{3,2}(x_L)$ with different strong coupling constants: $\alpha_s(m_Z)=\{0.112, 0.118, 0.124\}$ at $Q=91.2$ GeV. The uncertainty is the hard scale variation. Right panel shows the relative deviation of all three bands with respect to the central prediction of $\alpha_s(m_Z)=0.118$.}
	\label{fig:alpha_912}
\end{figure}

\begin{figure}[ht]
	\centering
	\setlength{\abovecaptionskip}{-0.0cm}
	\setlength{\belowcaptionskip}{-0.0cm} 
	\includegraphics[width=7.3cm]{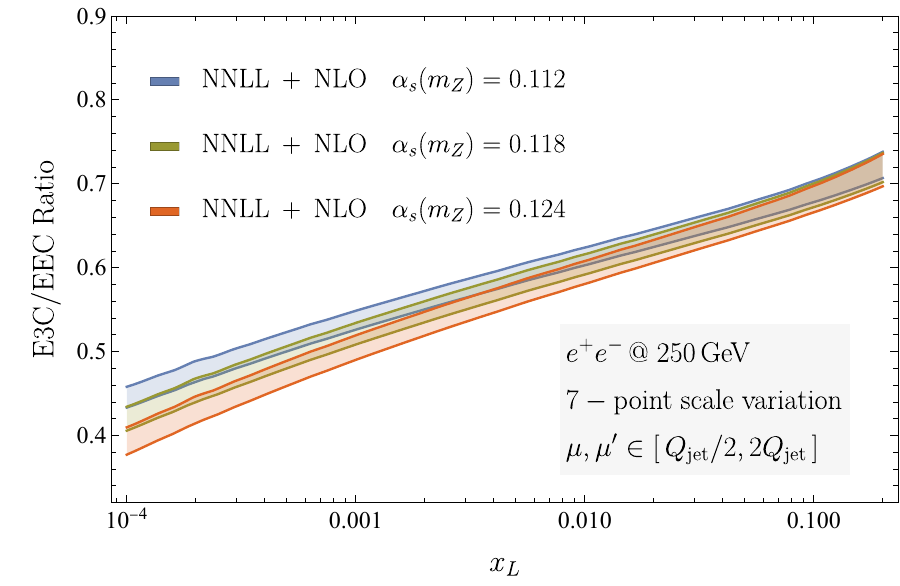} \;
	\includegraphics[width=7.3cm]{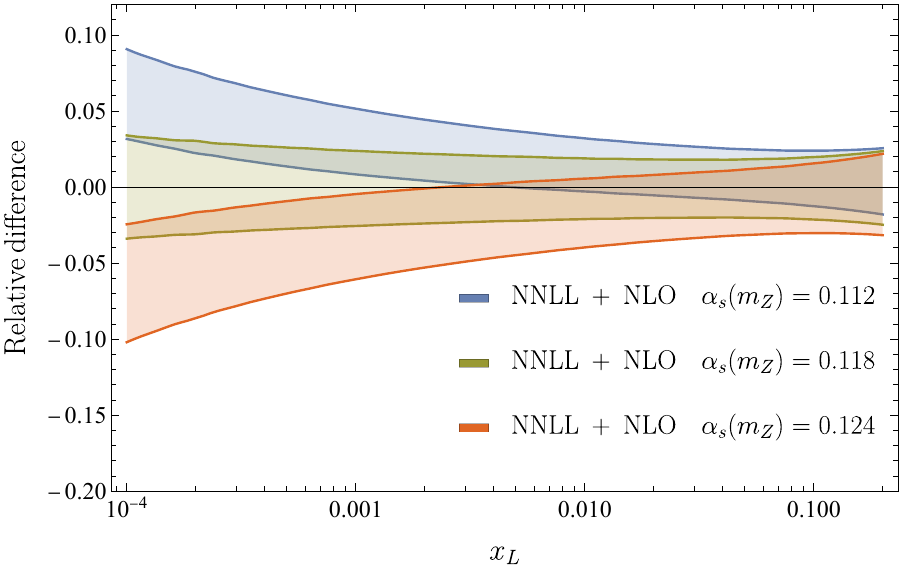} \;\\
	\includegraphics[width=7.3cm]{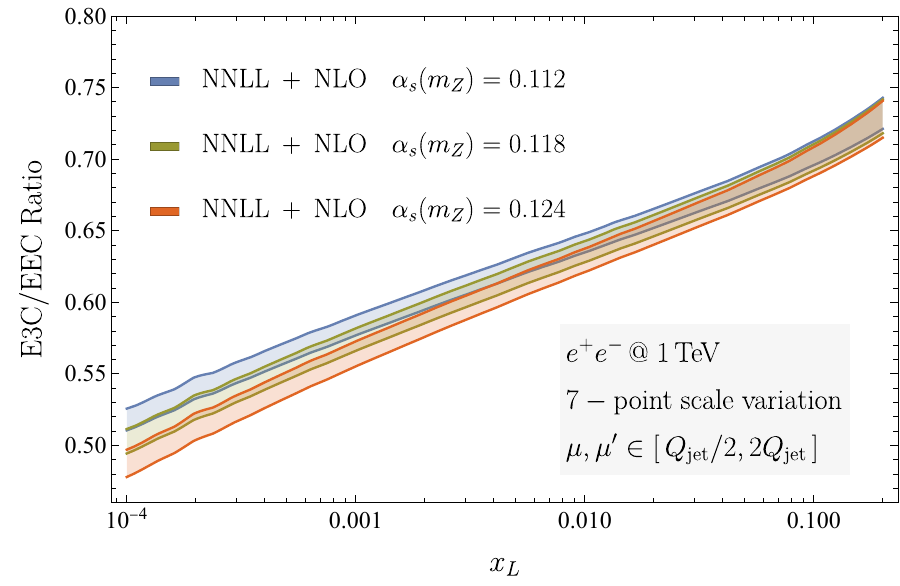} \;
	\includegraphics[width=7.3cm]{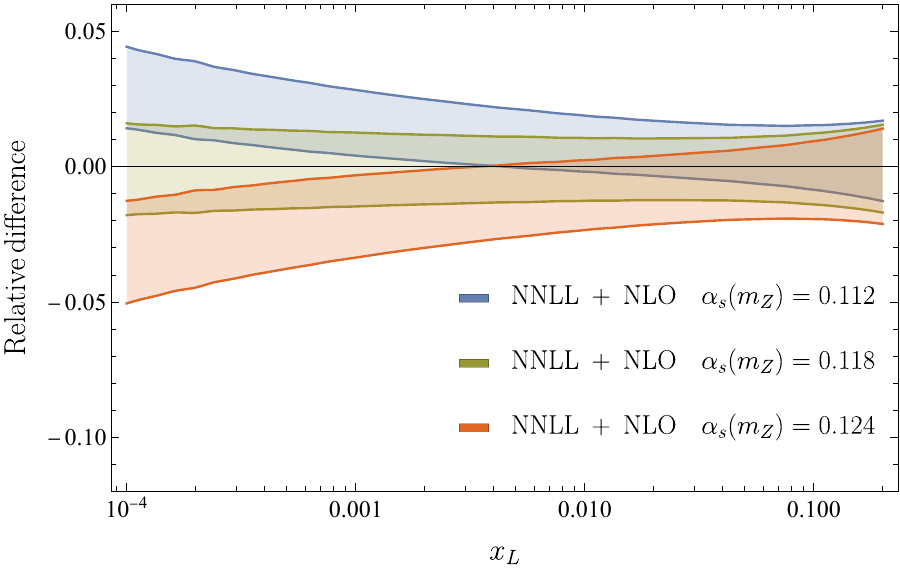} \;
	\caption{Left panels are the matched NNLL ratio distribution $\Delta_{3,2}(x_L)$ with different strong coupling constants: $\alpha_s(m_Z)=\{0.112, 0.118, 0.124\}$ at $Q=250$ GeV and $Q=1$ TeV. Right panels are the relative deviation of all three bands with respect to the central prediction of $\alpha_s(m_Z)=0.118$.}
	\label{fig:alpha_250_1000}
\end{figure}

We first consider the NNLL+NLO $\Delta_{3,2}(x_L)$ at $Q=91.2$ GeV with all three values of $\alpha_s(m_Z)$. As observed in Fig.~\ref{fig:alpha_912}, the slope become sensitive to the $\alpha_s$ in the collinear region $x_L=10^{-3}\sim 10^{-4}$, while the relative difference with respect to $\alpha_s(m_Z)=0.118$ ranges from $10\%$ to $20\%$. The slope sensitivity and the cancellation of hadronization correction have made the ratio of E3C and EEC $\Delta_{3,2}(x_L)$ an advantageous observable for extracting the $\alpha_s$ from $e^+e^-$ annihilation. Similar behaviors also exist at other energies and for completeness, we present the comparison at $Q=250$ GeV and $Q=1$ TeV in Fig.~\ref{fig:alpha_250_1000}.

The fact that the resummed E3C/EEC ratio has larger sensitivity to $\alpha_s$ and reduced non-perturbative corrections in the collinear limit makes it a promising candidate for the $\alpha_s$ determination. To further improve the $\alpha_s$ determination requires improving the resummation accuracy, matching with NNLO fixed-order correction, as well as the non-perturbative modeling.

\section{Approximate NNLL resummation in $pp$ collisions}
\label{sec:resummation_pp}

In this section, we consider the dijet production $pp\rightarrow jj$ at the LHC.
There are several motivations to study energy correlators in $pp$ collisions. First of all, LHC provides unique opportunities to study energy flows correlation in QCD at extremely high energy. While the LEP or future CEPC provides a very clean environment for precise measurements, $pp$ collisions at the LHC can produce multiple jets with very high energies ($p_T \gtrsim 500$ GeV), and high angular resolution can be achieved to probe the underlying dynamics for their formation and evolution. Secondly, as we have observed in the $e^+e^-$ collisions, the non-perturbative corrections for ENC have a relatively simple form compared to other event shape observables (at least in leading power), which might be easier to study non-perturbative QCD.
At the same time, with multiple scales involved, $pp$ collision can provide robust data from high energy to low energy, which is beneficial for understanding non-perturbative effects. 

In this section, we still focus on improving the perturbative predictions for ENC. As in Sec.~\ref{sec:formalism}, the jet functions are universal across different hard processes and the new ingredients are the moments of $pp$ hard function, both regular and logarithmic. The main complication for $pp$ collision is that the hard function now involves convolution with PDFs and algorithmic definition of jet, allowing only numeric calculation of the hard function. 

For the numerical calculation of the hard function, we adopt the anti-$k_t$ jet algorithm and choose the jet radius to be $R_0=0.4$. The complete kinematic cuts are summarized in Eqs.~(\ref{eq:jj_kin1})-(\ref{eq:jj_kin2}).
The $\mu$ independent part of the NLO hard function are presented in Appendix.~\ref{sec:hard_and_jet}. We observes large corrections going from LO to NLO. The $\mu$ dependent part of the NNLO hard function can be derived using the RG equation in \eqref{eq:hard_evo}.  The $\mu$ independent part requires a genuine two-loop corrections and are beyond the scope of this work. Instead we make a simple estimate of the two-loop constant terms, and dubbed the resulting prediction approximate NNLL resummation~(NNLL$_{\text{approx}}$).
Specifically, we use a modified Pad{\' e} approximation to estimate the two-loop hard function constants in both quark channel and gluon channel:
\begin{equation}
	a_s^2 h_0^{(2)} \approx   \kappa  \frac{(a_s h_0^{(1)})^2}{h_0^{(0)}}   \, ,
\end{equation}
where we vary $\kappa$ in the range $[0, 1/2]$ as a naive way to estimate our theory uncertainties on the missing two-loop constants. For the splitting function, $\beta$ function, as well as the jet functions, we used the ones required by NNLL accuracy as shown in Table~\ref{tab:ords}.

\begin{figure}[ht]
	\centering
	\setlength{\abovecaptionskip}{-0.0cm}
	\setlength{\belowcaptionskip}{-0.0cm} 
	\includegraphics[width=7.2cm]{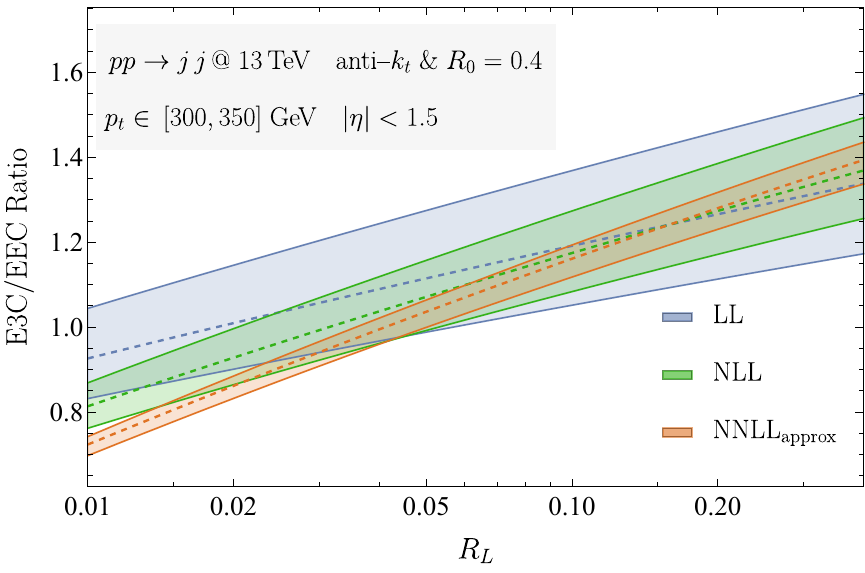} \;
	\includegraphics[width=7.2cm]{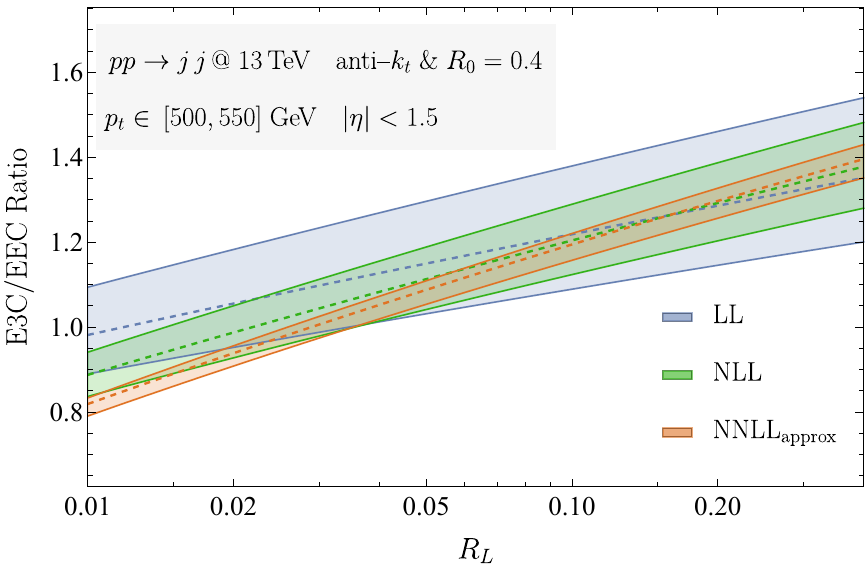} \;
	\caption{Normalized E3C/EEC ratio $\Delta_{3,2}(R_L)$ for $pp \to jj$ with $\alpha_s(m_Z)=0.118$, and jet $p_t$ ranges $[ 300,350]$ GeV (left) and $[ 500,550]$ GeV (right). Uncertainty
		bands are obtained in the 7-point scale variation scheme, with additional uncertainty from varying the estimation of NNLO hard function constant for $\rm NNLL_{approx}$. }
	\label{fig:LHC-E3CRatio-dijet}
\end{figure}

In Fig.~\ref{fig:LHC-E3CRatio-dijet}, we show the E3C/EEC ratio $\Delta_{3,2}(R_L)$ up to NNLL$_{\text{approx}}$, with the hard uncertainty estimated by seven-point variation.
Due to the lack of knowledge of the genuine two-loop hard function moment, we have chosen to normalize the E3C/EEC distribution in the range of $R_L \in [0.01,0.4]$ to reduce the impact from not knowing the full two-loop hard function.
We find good convergence for both $p_t$ ranges: $[300,350]$ GeV and $[500,550]$ GeV. 
In the future, it would be interesting the compute the two-loop hard function, as well as match the resummed results to fixed order to improve the prediction around $R_L \sim R_0$.

\subsection{Anticipation of $\alpha_s$ determination}

Similar to $e^+e^-$ annihilation, in this subsection we discuss the potential of extracting the strong coupling constant $\alpha_s$ from the resummed $\Delta_{3,2}(R_L)$ distribution in $pp\rightarrow jj$. In particular, we also investigate the slope sensitivity of the distribution with respect to different values of $\alpha_s$. For hadron colliders, we need to change the PDFs as we vary the strong coupling among $\alpha_s(m_Z)=0.118\pm 0.06$. For this purpose, we use three PDF sets: {\texttt{NNPDF31\_nnlo\_as\_0112}}, {\texttt{NNPDF31\_nnlo\_as\_0118}} and {\texttt{NNPDF31\_nnlo\_as\_0124}} when calculating the hard function using the method in \cite{Liu:2023fsq}.

\begin{figure}[ht]
\centering
\setlength{\abovecaptionskip}{-0.0cm}
\setlength{\belowcaptionskip}{-0.0cm} 
\includegraphics[width=0.49\linewidth]{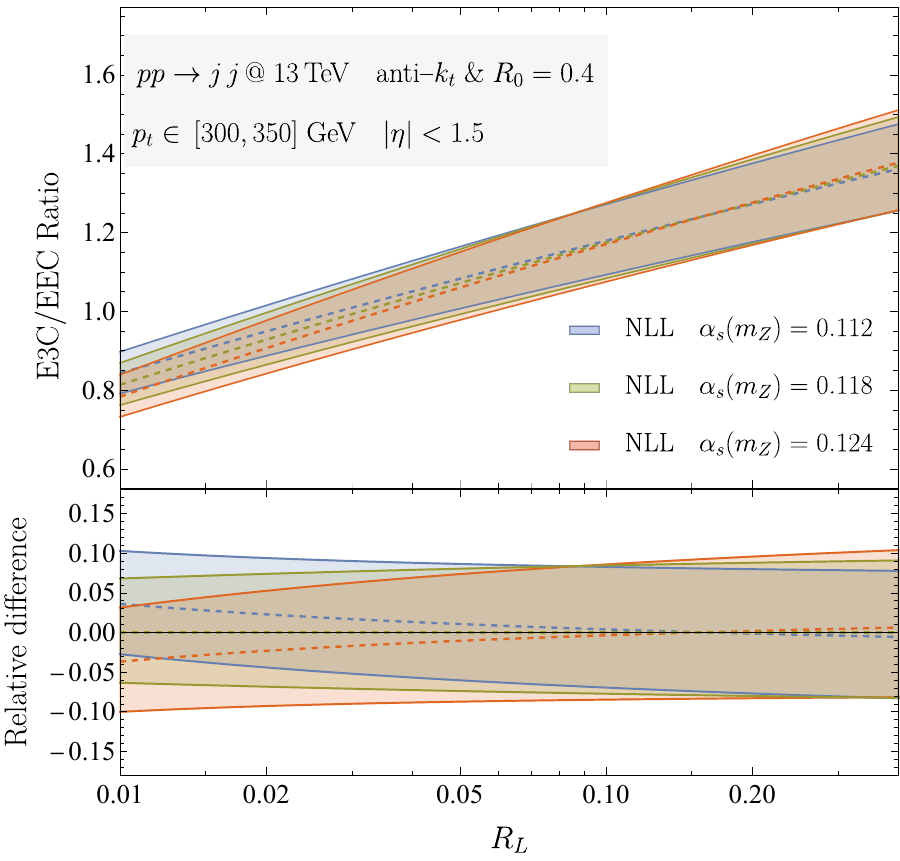}\hfill
\includegraphics[width=0.49\linewidth]{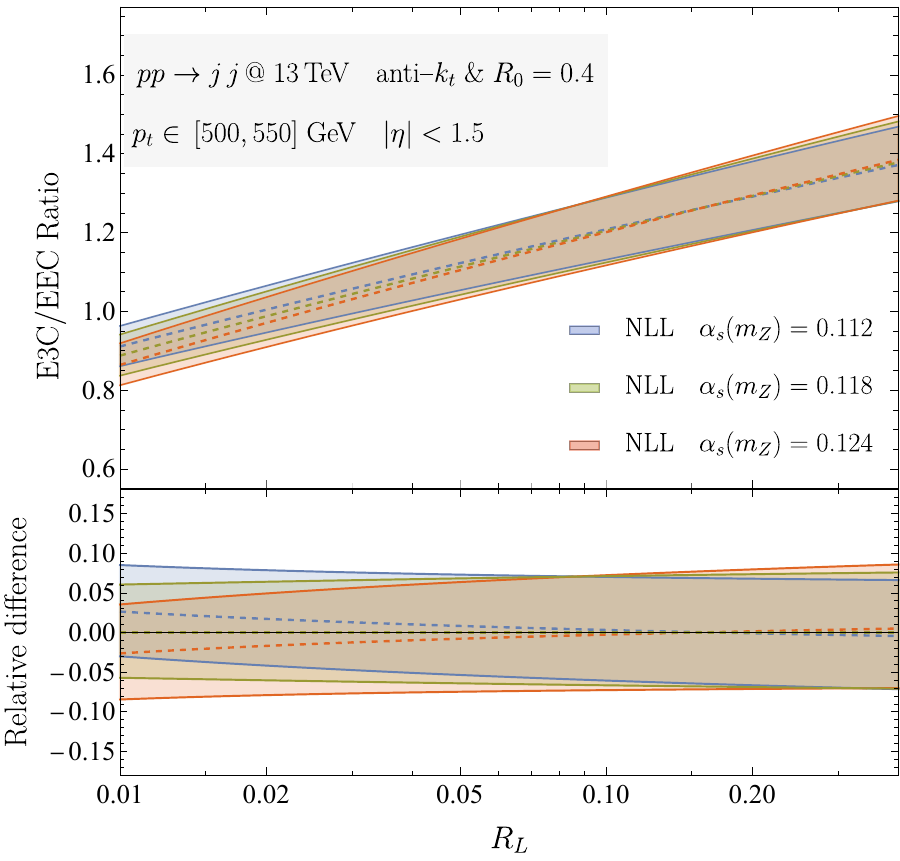}
\includegraphics[width=0.49\linewidth]{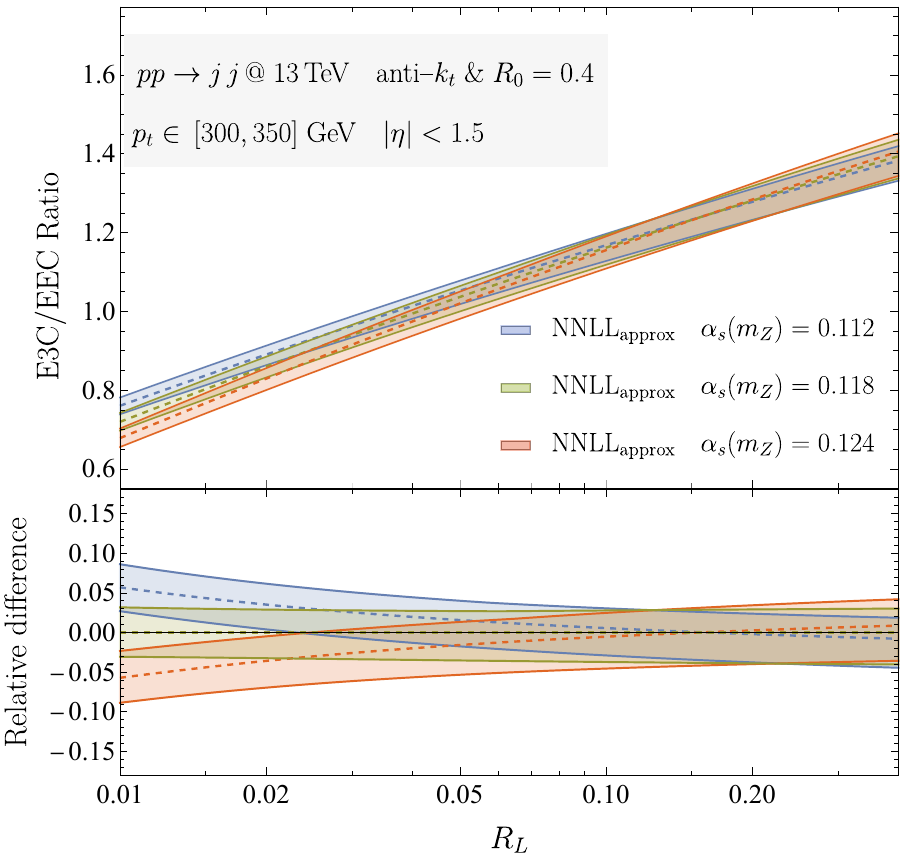}\hfill
\includegraphics[width=0.49\linewidth]{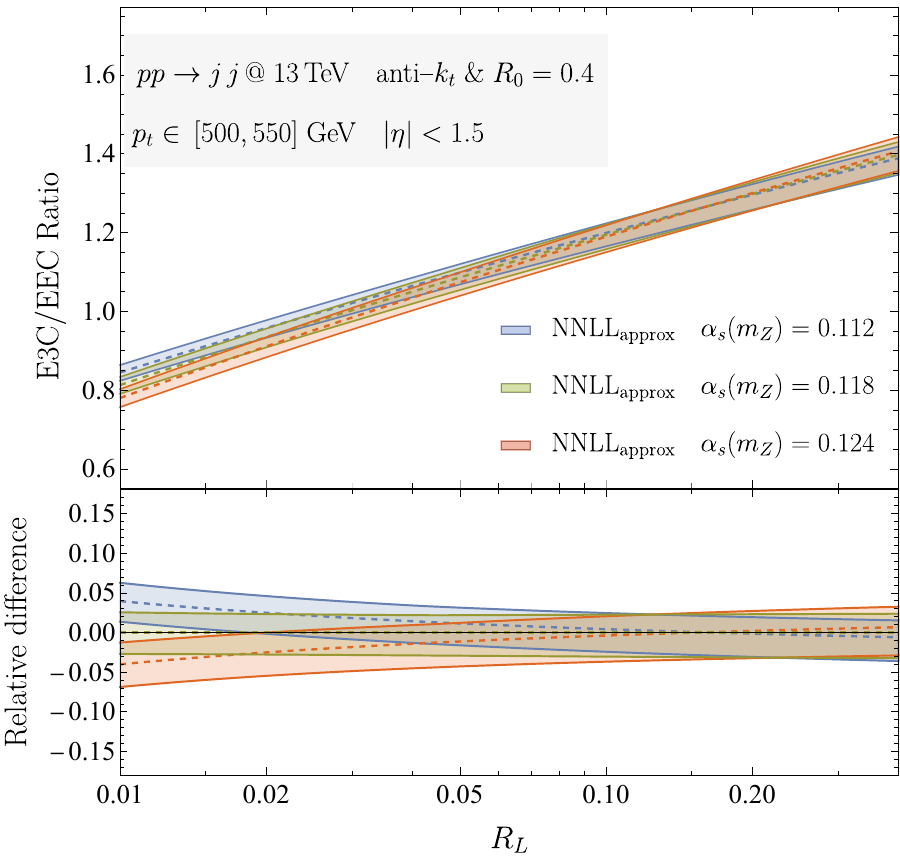}
\caption{Normalized NLL (upper) and $\rm NNLL_{approx}$ (lower) resummation result for E3C/EEC ratio at $pp \to jj$ with $\alpha_s(m_Z)=0.118 \pm 0.06$, i.e., varied by about $5\%$, for two different jet $p_t$ ranges [300,350] GeV and [500,550] GeV. Lower panels show the relative difference from the result at the central scale with $\alpha_s(m_Z)=0.118$.} 
\label{fig:LHC-VaryAlphaS}
\end{figure}

As shown in Fig.~\ref{fig:LHC-VaryAlphaS}, for each $p_t$ range, the uncertainty is significantly reduced from NLL to NNLL$_{\text{approx}}$, leading to distinguishable slopes with respect to different $\alpha_s$. This suggests that ratios of energy correlators are good candidate for extracting $\alpha_s$. We note that there is larger slope variation for lower $p_t$ of the jet, in agreement with the expectation that the measurement at lower energy is more sensitive to $\alpha_s$ due to asymptotic free nature of QCD.

\section{Conclusion}
\label{sec:conclusion}

In this paper we have performed a systematic study of resummation of projected three-point energy correlator E3C~\cite{Chen:2020vvp}, and its ratio to EEC, in both $e^+e^-$ collider and $pp$ collider. We have achieved the first NNLL accuracy for the $e^+e^-$ case, and $\rm NNLL_{approx}$ accuracy for the $pp$ case.
Our results show that good perturbative convergence can be achieved for the ratios of projected energy correlators. The current theoretical uncertainties are at a level of a few percent, and can be further improved in the future when the higher order ingredients become available. We have also shown that the ratio observable is sensitive to variation of $\alpha_s$, therefore provides a good candidate for precision $\alpha_s$ determination using jet substructure. 

To achieve the above theory accuracy, one of the main new ingredients is the two-loop E3C jet function computed in this work. The calculation includes three pieces: double-real, real-virtual and double-virtual. The last two contributions only involve a single $\delta$ measurement function in the phase space integral and share a similar form as the analytic EEC calculation at NLO~\cite{Dixon:2018qgp}. Regarding the double-real emissions, which amounts to integrating the fully-differential EEEC distribution within the collinear kinematic space, we used two different approaches and find the same results. The first method is to subtract the infrared divergence in the collinear EEEC jet function, integrate it separately with $d$-dimension kinematic space, and expand the finite terms in $\epsilon$. The second approach benefits from the recently developed parametric IBP, where we can also simplify the integrand with IBP reduction and calculate the integrals via differential equations.

Regarding the ENC resummation, for $e^+e^-$ annihilation, we solve the E3C jet RGE (which is a modified DGLAP equation) order by order in $\alpha_s$ with the two-loop boundary, and push the resummation up to NNLL.  For $pp$ collisions, we calculate the combined hard function moments using the method in \cite{Liu:2023fsq} for dijet production. We present the complete NLL and the approximate NNLL resummation result, where the approximation is due to the missing of genuine two-loop hard function constant. The uncertainty is reduced compared with the previous results~\cite{Chen:2020vvp,Komiske:2022enw,Lee:2022ige}. For the fixed-order matching, we notice that the singular contribution dominates the collinear limit and the non-singular contribution from matching has only small effects in the $e^+e^-$ case.  Nevertheless, we perform the matching for $e^+e^-$ given the fixed-order result is already available, but leave the matching with fixed-order in the $pp$ case for the future study.

For a complete phenomenological analysis and precise $\alpha_s$ extraction at hadron collider, there are still several ingredients needed in the future. Perturbatively, we need to compute both two-loop hard function and the NLO non-singular distribution for $pp\rightarrow jj$, in order to achieve a full NNLL story. 
More over, it would be interesting to solve the RG equation exactly following \cite{Cao:2023oef}, and compare the results with the truncation method. 
 At the same time, for both $e^+e^-$ and $pp$, it would be interesting to better understand the hadronization power corrections to help further reduce theoretical uncertainties. We hope that all these efforts can lead to a precision determination of $\alpha_s$ from jet substructure in the future.
 
\acknowledgments

The authors thank Hao Chen, Kyle Lee, Meng Xiao, Tong-Zhi Yang, Yulei Ye for useful discussions. XYZ also thanks the MIT CTP for its hospitality while part of this work was performed. The work of WC, YL, ZX, and HXZ was supported by the National Natural Science Foundation
of China under the Grant No. 11975200.
The work of JG was sponsored by the National Natural Science Foundation
of China under the Grant No.12275173 and No.11835005.

\bigskip
\appendix

\section{Hard and jet functions}
\label{sec:hard_and_jet}

\subsection*{$e^+e^-$ Hard function}

The ENC hard function for $e^+e^-$ can be obtained from the semi-inclusive hadron fragmentation function. At NNLL, following our resummation procedure, we need the regular up to two-loop, single logarithmic up to one-loop and the double logarithmic moments at tree level with respect to the energy fraction $x$:
\begin{align}
	\int_0^1 dx \, x^N \, H_{q,g}(x,\mu=Q)\ &=\ \sum\limits_{L=0}^\infty
	\left( \frac{\alpha_s}{4\pi} \right)^{L} h_L^{q,g}(N) \,, \nn\\
	\int_0^1 dx \, x^N \, \ln x \, H_{q,g}(x,\mu=Q)\ &=\ \sum\limits_{L=1}^\infty
	\left( \frac{\alpha_s}{4\pi} \right)^{L} \dot{h}_L^{q,g}(N) \,,  \nn\\
	\int_0^1 dx \, x^N \, \ln^2 x \, H_{q,g}(x,\mu=Q)\ &=\ \sum\limits_{L=1}^\infty
	\left( \frac{\alpha_s}{4\pi} \right)^{L} \ddot{h}_L^{q,g}(N) \,.
\end{align}

For EEC ($N=2$), we have 
\begin{align}
	h_0^q &= 2 \,, \qquad
	h_0^g = 0 \,, \qquad\qquad
	h_1^q = \frac{131}{4} \, C_F \,, \qquad
	h_1^g = - \frac{71}{12} \, C_F \,, \nn\\
	h_2^q &= \left( 64 \zeta_4 - \frac{1172}{3} \zeta_3 - 166 \zeta_2
	+ \frac{2386397}{2592} \right) C_A C_F\nn\\
	&+ \left( - 128 \zeta_4 + \frac{1016}{3} \zeta_3 + \frac{1751}{18} \zeta_2
	- \frac{1105289}{5184} \right) C_F^2 
	+ \left( 32 \zeta_3 + \frac{118}{15} \zeta_2 - \frac{8530817}{54000} \right)
	C_F T_F n_f \,, \nn\\
	h_2^g &= \left( - \frac{76}{3} \zeta_3 + \frac{188}{45} \zeta_2
	- \frac{29802739}{324000} \right) C_A C_F
	+ \left( \frac{124}{3} \zeta_3 + \frac{523}{18} \zeta_2
	- \frac{674045}{5184} \right) C_F^2 \,, \nn\\
	\dot{h}_0^q&=0\,,\qquad \dot{h}_1^q = \left( 40 \zeta_3 + \frac{61}{3}  \zeta_2
	- \frac{5303}{72} \right) C_F  \,,\qquad	\dot{h}_0^g=0\,, \qquad
	\dot{h}_1^g = \left( - \frac{7}{3} \zeta_2 + \frac{31}{4} \right) C_F \,,\nn\\
	\ddot{h}_0^q&=0\,, \qquad   \ddot{h}_0^g=0\,.
\end{align}
Note that the EEC hard moments are also summarized in the appendix of Ref.~\cite{Dixon:2019uzg}). However, the normalization condition in~\cite{Dixon:2019uzg} is different from ours, due to the scaled energy $E_i/(Q/2)$ there in contrast with $E_i/Q$ here in the definition of the jet function. 
For E3C ($N=3$), we find
\begin{align}
	h_0^q&=\, 2,\qquad h_0^g=0,\qquad h_1^q=\frac{11909}{300}C_F,\qquad h_1^g=-\frac{547}{150}C_F\,,\nn\\
	h_2^q&=\, \left(-\frac{942}{5}\zeta_3-\frac{17}{45}\zeta_2+\frac{17147309}{32400}\right) C_A C_F
	+ \left(32\zeta_3+\frac{322}{25}\zeta_2-\frac{6169957}{30000}\right)C_F n_f T_F\nn\\
	&
	+ \left(-\frac{2012}{15}\zeta_3-\frac{8987}{30}\zeta _2+\frac{3256506739}{3240000}\right)C_F^2\,,\nn\\
	h_2^g&= \left(\frac{52}{5}\zeta_3+\frac{4396}{225}\zeta _2-\frac{101763773}{810000}\right)C_A C_F+ \left(\frac{392}{15}\zeta_3+\frac{397}{15}\zeta_2-\frac{163115357}{1620000}\right)C_F^2
	\,,\nn\\
	\dot{h}_0^q&=0\,,\qquad \dot{h}_1^q= \left(40\zeta_3+\frac{337}{15}\zeta_2-\frac{709693}{9000}\right)C_F \,,\nn\\
	\dot{h}_0^g&=0\,, \qquad
	\dot{h}_1^g=\left(-\frac{22}{15}\zeta_2+\frac{16739}{4500}\right) C_F \,,\nn\\
	\ddot{h}_0^q&=0\,, \qquad   \ddot{h}_0^g=0\,.
\end{align}

For completeness, we also provide the E3C ($N=3$) hard moments for the gluonic Higgs decay, which is needed for extracting the two-loop gluon jet constants. Here we use $\tilde{h}$ to distinguish from the $e^+e^-$ case.
\begin{align}
	\tilde{h}_0^q&=\, 0,\qquad \tilde{h}_0^g=2,
	\qquad
	\tilde{h}_1^q=-\frac{2461}{450} n_f T_F,
	\qquad 
	\tilde{h}_1^g=\frac{11491}{150}C_A-\frac{494}{45}n_f T_F\,,\nn\\
	\tilde{h}_2^q&=\,n_f T_F \left[C_A \left(\frac{88}{3}\zeta_3+\frac{3428}{75}\zeta_2-\frac{219509243}{810000}\right)+\left(\frac{1727}{225}\zeta_2-\frac{187858397}{1620000}\right) C_F\right]\nn\\
	&
	+\left(-\frac{352}{45}\zeta_2+\frac{7224}{125}\right) n_f^2 T_F^2
	\,,\nn\\
	\tilde{h}_2^g&=
	n_f T_F \left[C_A \left(-\frac{208}{3}\zeta_3+\frac{1264}{15}\zeta_2-\frac{38190113}{40500}\right)
	+C_F \left(96 \zeta_3-\frac{242}{225}\zeta_2-\frac{113165189}{810000}\right)\right]\nn\\
	&
	+C_A^2 \left(-388\zeta_3-\frac{31684}{75}\zeta _2+\frac{837482633}{270000}\right)
	+n_f^2 T_F^2\left(-\frac{64}{9}\zeta_2+\frac{44252}{675}\right) 
	\,,\nn\\
	\dot{\tilde h}_0^q&=0\,,\qquad \dot{\tilde{h}}_1^q= n_f T_F\left(-\frac{22}{15}\zeta_2+\frac{404}{125}\right)
	\,,\nn\\
	\dot{\tilde h}_0^g&=0\,,\qquad \dot{\tilde{h}}_1^g=
	C_A \left(40 \zeta_3+\frac{346}{15}\zeta_2-\frac{2134817}{27000}\right)+\left(-\frac{8}{3}\zeta _2+\frac{5369}{1350}\right) n_f T_F \,,\nn\\
	\ddot{\tilde h}_0^q&=0\,,\qquad \ddot{\tilde h}_0^g=0\,.
\end{align}

\subsection*{$pp \to jj $ Hard function}

The following table gives the hard function moments for $pp\rightarrow jj$ calculated in \textsc{Madgraph5} in two different $p_t$ ranges: $[300,350]$ GeV and $[500,550]$ GeV, needed for the resummation of both EEC ($N=2$) and E3C ($N=3$).

\begin{center}
	\begingroup
\begin{tabular}{||c c c c c c c||} 
 \hline
 & & & & & & \\ [-2.5ex]
 \multicolumn{7}{||c||}{$pp \to jj$ at 13 TeV, with  {\texttt{NNPDF31\_nnlo\_as\_0118}}} \\  [1.2ex]
 \hline 
 \hline
  & & & & & & \\ [-2.5ex]
 (300,350) GeV & $h_0^q$ & $h_0^g$ & $a_s\, h_1^q$ & $a_s\,h_1^g$ & $a_s\,\dot{h}_1^q$ & $a_s\,\dot{h}_1^g$ \\   [0.8ex]
 \hline
 $N=2$ & 0.3571 & 0.6429 & 0.1003 & 0.3304 & 0.0546 & 0.2149 \\ 
 \hline
 $N=3$ & 0.3571 & 0.6429 & 0.1463 & 0.4996 & 0.0393 & 0.1379 \\
 \hline 
 \hline
  & & & & & & \\ [-2.5ex]
 (500,550) GeV & $h_0^q$ & $h_0^g$ & $a_s\, h_1^q$ & $a_s\,h_1^g$ & $a_s\,\dot{h}_1^q$ & $a_s\,\dot{h}_1^g$ \\   [0.8ex]
 \hline
 $N=2$ & 0.4417 & 0.5583 & 0.1337 & 0.2473 & 0.0568 & 0.1816 \\ 
 \hline
 $N=3$ & 0.4417 & 0.5583 & 0.1820 & 0.3894 & 0.0417 & 0.1150 \\
 \hline
\end{tabular}
\vspace{-0.1cm}
\captionof{table}{Values for hard function moments in $pp$ collision for different $p_t$ ranges. The NLO corrections turn out to be significant.}
\vspace{-0.1cm}
\endgroup
\end{center}

\begin{figure}[H]
\centering
\includegraphics[width=0.45\textwidth]{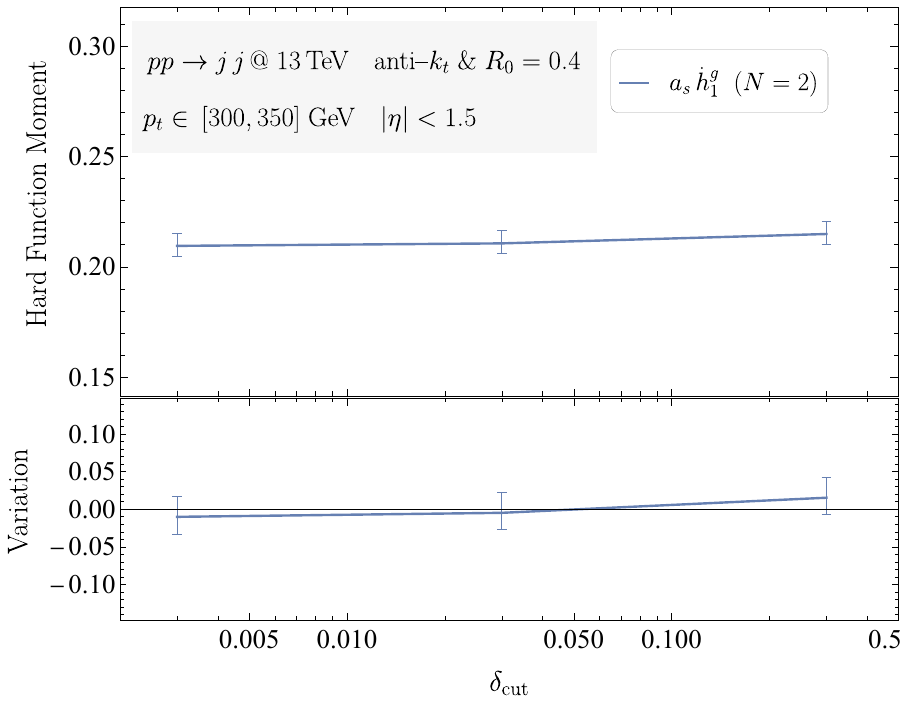}
\hfill
\includegraphics[width=0.45\textwidth]{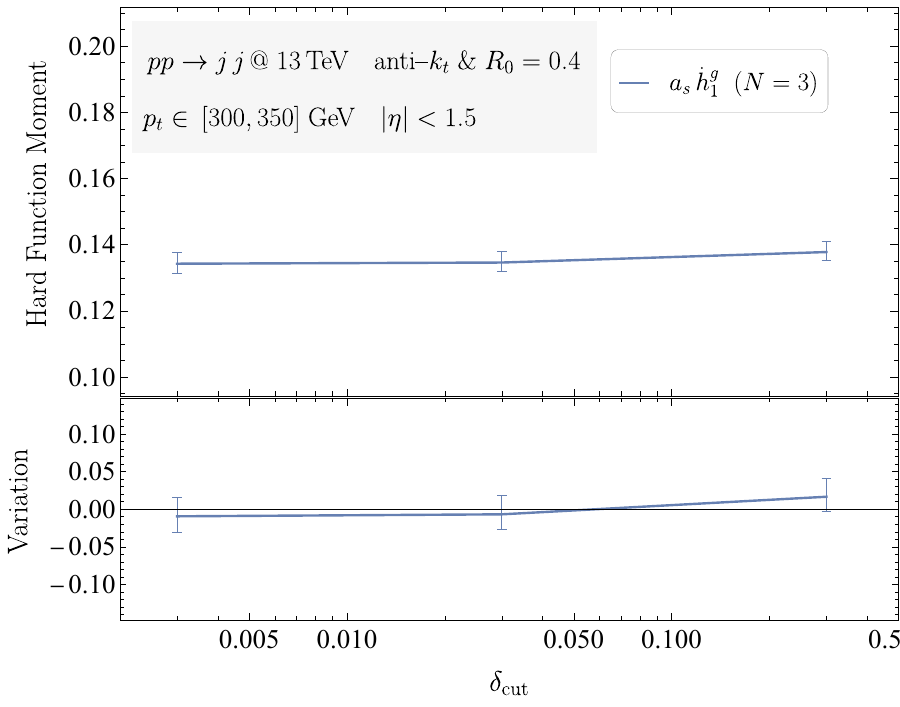}\\
\includegraphics[width=0.45\textwidth]{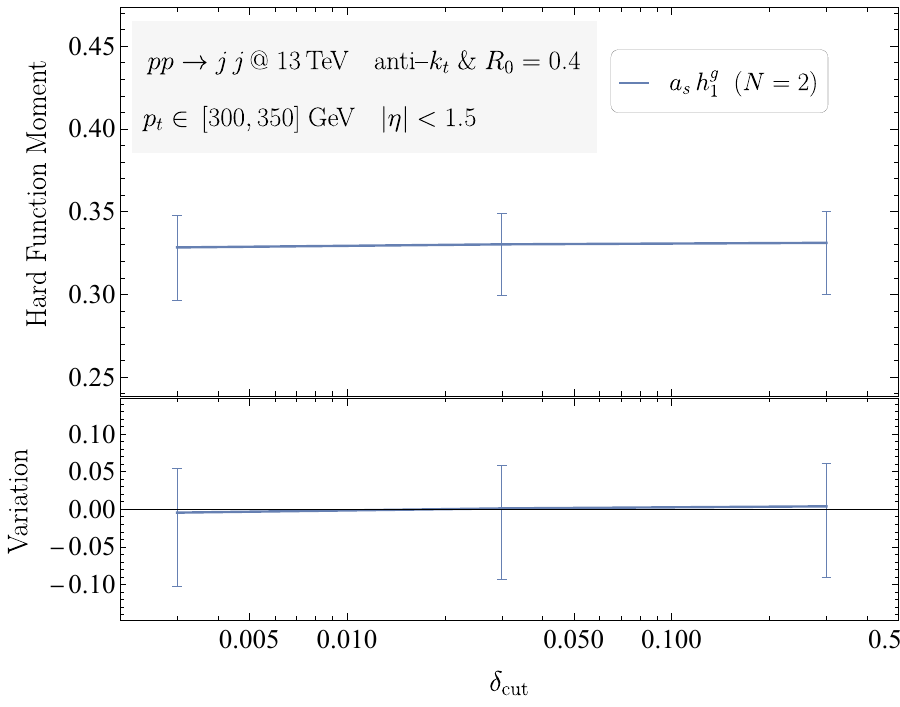}
\hfill
\includegraphics[width=0.45\textwidth]{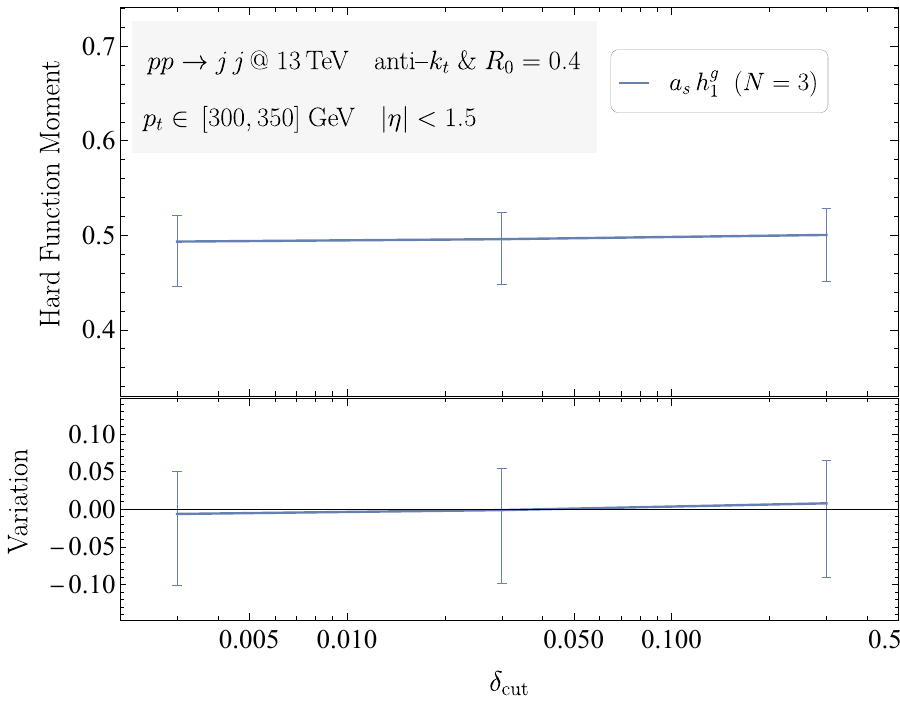}\\
\includegraphics[width=0.45\textwidth]{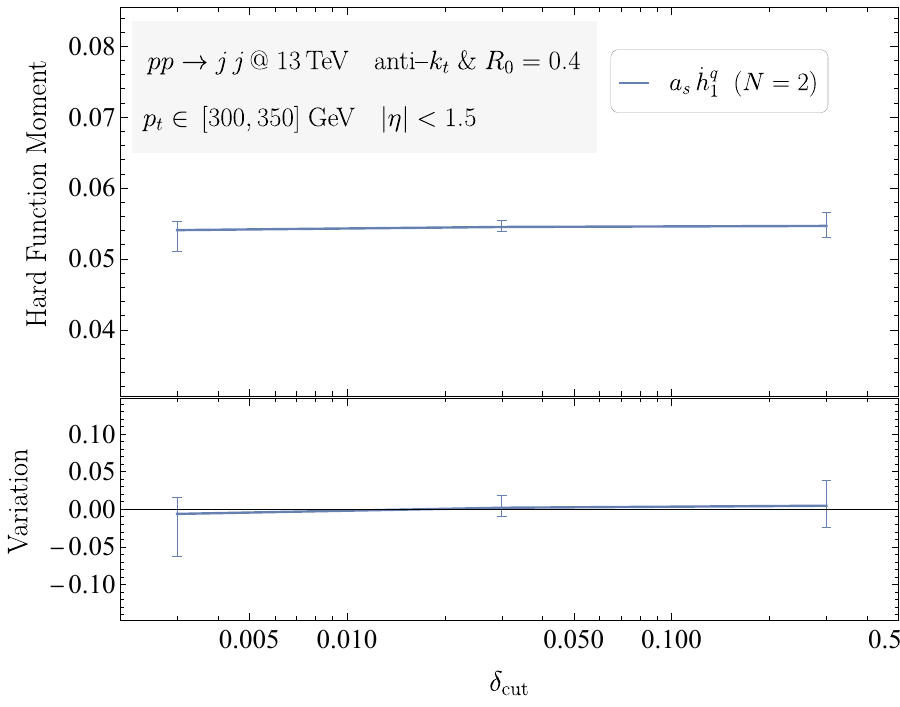}
\hfill
\includegraphics[width=0.45\textwidth]{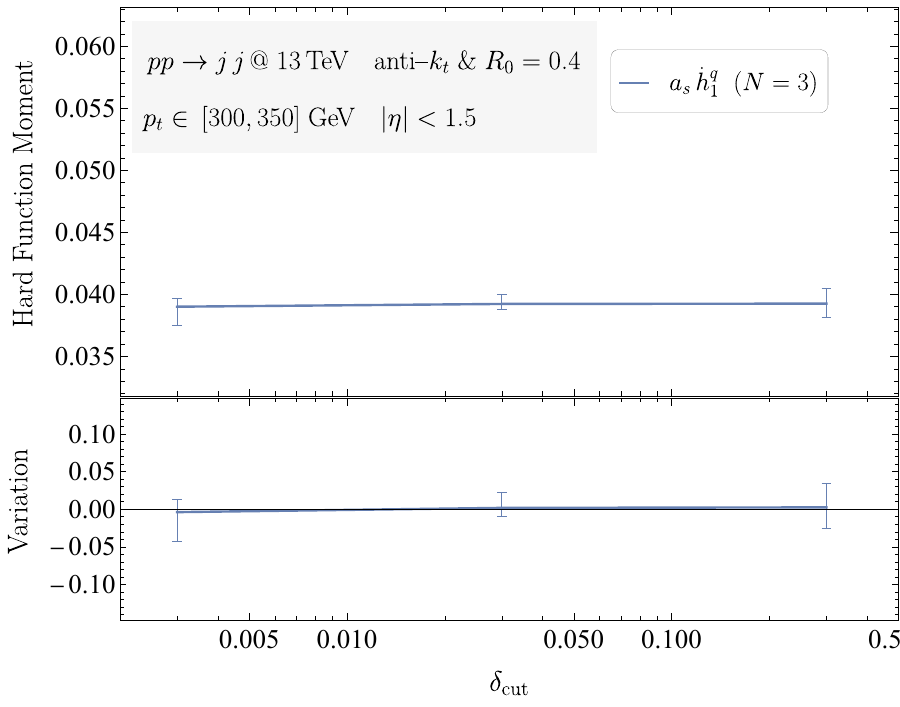}\\
\includegraphics[width=0.45\textwidth]{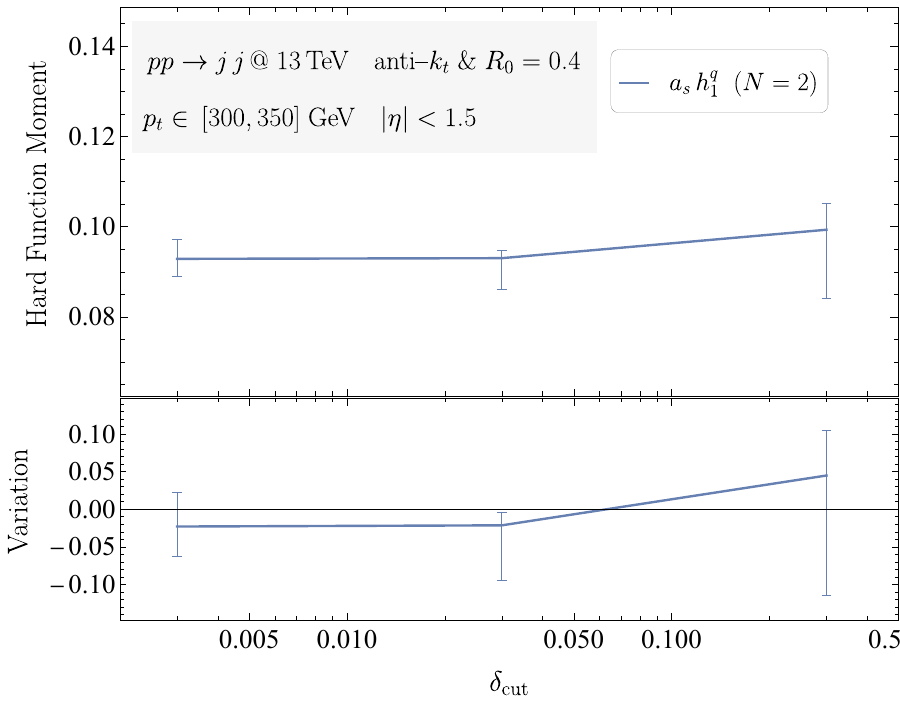}
\hfill
\includegraphics[width=0.45\textwidth]{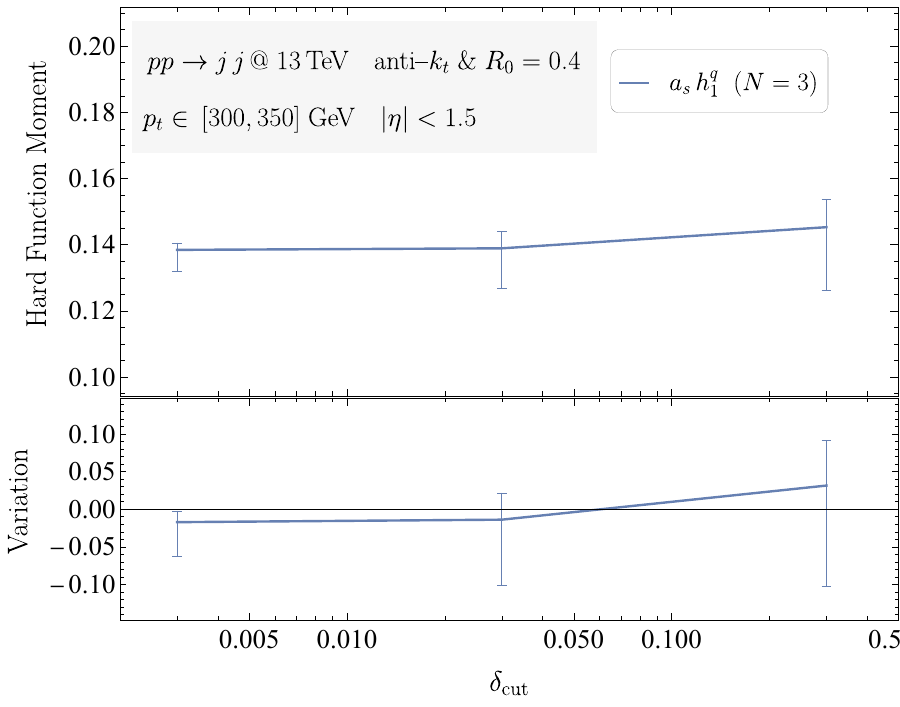}
\caption{NLO hard function moments for $N=2$ (left), and $N=3$ (right), with  $p_t \in [300,350]$ GeV and $\delta_{\rm cut} \in \{ 0.003,\, 0.03,\, 0.3 \}$. The lower panels show the relative variation of the moments compared with the average value for three $\delta_{\rm cut}$. The error bars represent the Monte-Carlo numeric uncertainty given by \textsc{Madgraph5}. }
\label{fig:Moments-Cuts}
\end{figure}

As one of the important checks of our calculation, we show in Fig.~\ref{fig:Moments-Cuts} the independence of the slicing parameter $\delta_{\rm cut}$  when evaluating the hard function moments using the method in \cite{Liu:2023fsq}. 
The values of the moments are in agreement within the numeric uncertainty for three values of $\delta_{\rm cut}$ across two orders of magnitude, namely $\delta_{\rm cut} \in \{0.003, 0.03, 0.3\}$. 

\subsection*{Jet function}

For ENC, solving the jet function RGE requires the regular anomalous dimensions and their derivatives, and at NNLL, similar to hard function, we need the regular terms up to two-loop, the first derivative up to one-loop as well as the second derivative at tree-level.

The QCD timelike splitting function is expanded in $\frac{\alpha_s}{4\pi}$
\begin{equation}
	P_{ij}(x)=\sum_{L=0}^{\infty}\left(\frac{\alpha_s}{4\pi}\right)^{L+1} P_{ij}^{(L)}(x) \,,
\end{equation}
and the anomalous dimension for ENC is defined to be the (N+1) Mellin moment of the splitting function. Explicitly,
\begin{align}
	\gamma_{T,ij}^{(L)}&\equiv -\int_0^1 \df x \,x^N \,P_{ij}^{(L)}(x)\,,\nn\\
	\dot \gamma_{T,ij}^{(L)}&\equiv -\int_0^1 \df x\, \ln x \, x^N P_{ij}^{(L)}(x)\,,\nn\\
	\ddot \gamma_{T,ij}^{(L)}&\equiv -\int_0^1 \df x\, \ln^2 x \, x^N P_{ij}^{(L)}(x)\,.
\end{align}
Here the dot also represents the derivative with respect to $N$. Note that $\{i,j\}=\{q,g\}$ and the anomalous dimension is a $2\times 2$ matrix. 

The results for EEC ($N=2$) are derived and summarized in the appendix of Ref.~\cite{Dixon:2019uzg}, so here we provide the expressions for E3C ($N=3$).

At LO, we find
\begin{align}
	\gamma_{T,qq}^{(0)} &= \frac{157}{30} \, C_F\,, \qquad
	\gamma_{T,gq}^{(0)}= -\frac{11}{15} \, C_F \,, \qquad
	\gamma_{T,qg}^{(0)} = \frac{11}{30} \, n_f \,, \qquad
	\gamma_{T,gg}^{(0)} = \frac{21}{5}\, C_A + \frac{2}{3} \, n_f \,,\nn\\
	\dot{\gamma}_{T,qq}^{(0)}&= \left( 4\zeta_2 - \frac{10169}{1800} \right) C_F \,, \quad
	\dot{\gamma}_{T,gq}^{(0)}= \frac{247}{900} \, C_F \,, \quad
	\dot{\gamma}_{T,qg}^{(0)} = \frac{137}{1800} \, n_f \,, \quad
	\dot{\gamma}_{T,gg}^{(0)} = \left( 4\zeta_2 - \frac{2453}{450} \right) C_A \,,   \nn\\
	\ddot{\gamma}_{T,qq}^{(0)}&= \left( - 8\zeta_3 + \frac{507103}{54000} \right) C_F \,, \quad
	\ddot{\gamma}_{T,gq}^{(0)}= - \frac{5489}{27000} \, C_F \,, \quad
	\ddot{\gamma}_{T,qg}^{(0)}= - \frac{1919}{54000} \, n_f  \,, \nn\\
	\ddot{\gamma}_{T,gg}^{(0)}&= \left( - 8\zeta_3 + \frac{124511}{13500} \right) C_A \,, 
\end{align}
and at NLO, we have
\begin{align}
	\gamma_{T,qq}^{(1)}&=
	\left( -\frac{628}{15}+\frac{2905763}{54000} \right) C_F^2 
	+ \frac{16157}{675} C_A C_F
	- \frac{13427}{3000} \, C_F n_f \,, \nn \\
	\gamma_{T,gq}^{(1)} &= \left(\frac{88}{15}\zeta_2-\frac{104389}{27000}  \right) C_F^2
	-\frac{142591}{13500} C_A C_F  \,, \nn \\
	\gamma_{T,qg}^{(1)} &= \left(\frac{44}{15}\zeta_2 -\frac{60391}{27000} \right) C_A n_f
	- \frac{166729}{54000} \, C_F n_f - \frac{6}{25} \, n_f^2 \,, \nn \\
	\gamma_{T,gg}^{(1)} &=
	\left( -\frac{168}{5}\zeta_2+\frac{90047}{1500}\right) C_A^2
	+ \left(- \frac{16}{3}\zeta_2 +\frac{2273}{1350} \right) C_A n_f 
	+ \frac{57287}{27000} \, C_F n_f \,, \nn\\
	\dot{\gamma}_{T,qq}^{(1)}&= \left(-120\zeta_4+ \frac{422}{3}\zeta_3+\frac{10169}{150}\zeta_2-\frac{162656941}{1080000}\right) C_F^2\nn\\
	&+ \left(20\zeta_4-\frac{1181}{15}\zeta_3+\frac{268}{9}\zeta_2+\frac{992579}{36000} \right) C_F C_A  + \left(\frac{16}{3}\zeta_3-\frac{40}{9}\zeta_2-\frac{433757}{1620000} \right)  C_F n_f \,, \nn \\
	\dot{\gamma}_{T,gq}^{(1)}&= \left(-\frac{286}{15}\zeta_3+\frac{1034}{225}\zeta_2+\frac{15207541}{810000} \right) C_F C_A
	+ \left(\frac{44}{5}\zeta_3-\frac{71}{9}\zeta_2+\frac{235643}{540000} \right) C_F^2 \,, \nn \\
	\dot{\gamma}_{T,qg}^{(1)} &= \left(\frac{11}{5}\zeta_3-\frac{25}{18}\zeta_2-\frac{1490669}{1620000} \right) C_A n_f
	+ \left(-\frac{22}{3}\zeta_3+\frac{217}{225}\zeta_2+\frac{8521133}{1080000} \right) C_F n_f \nn\\ 
	&+ \left(-\frac{22}{45}\zeta_2+\frac{10121}{13500} \right) n_f^2 \,, \nn \\
	\dot{\gamma}_{T,gg}^{(1)} &= \left(-100\zeta_4+\frac{772}{15}\zeta_3+\frac{21418}{225}\zeta_2-\frac{42705619}{405000} \right) C_A^2 
	+ \left(\frac{32}{3}\zeta_3-\frac{40}{9}\zeta_2-\frac{21958}{3375} \right) C_A n_f\nn\\
	&- \frac{59659}{540000} \, C_F n_f \,,
\end{align}
as well as NNLO:
\begin{align}
	\gamma_{T,qq}^{(2)}&=
	\left( \frac{1439}{75} \zeta_3+ \frac{136066373}{972000} \right) C_F C_A^2  \nn \\
	&+ \left( \frac{628}{3}\zeta_4+ \frac{172466}{225} \zeta_3- \frac{113212}{225}\zeta_2-\frac{443247883}{9720000} \right) C_F^2 C_A \nn \\
	&+\left( 1256 \zeta_4-\frac{14936}{15}\zeta_3-\frac{2251148}{3375}\zeta_2+ \frac{47976425617}{48600000} \right) C_F^3 \nn \\
	&+\left(-\frac{2126}{45}\zeta_3 + \frac{8492}{3375}\zeta_2-\frac{57923471}{4050000} \right) C_A C_F n_f\nn\\
	&+\left(-\frac{2656}{225}\zeta_3+\frac{88163}{1125}\zeta_2-\frac{638186993}{8100000} \right) C_F^2 n_f
	-\frac{19711}{18000} \, C_F n_f^2 \,, \nn \\
	\gamma_{T,gq}^{(2)} &=
	\left(\frac{6448}{75}\zeta_3-\frac{10898}{375}\zeta_2-\frac{2010250477}{12150000} \right) C_F C_A^2 \nn\\ &+\left(\frac{88}{3}\zeta_4-\frac{31346}{225}\zeta_3+\frac{234407}{1125}\zeta_2-\frac{1694499413}{24300000} \right) C_F^2 C_A \nn \\
	&+ \left(-176 \zeta_4 +\frac{1796}{15}\zeta_3+\frac{79268}{3375}\zeta_2-\frac{1061823161}{24300000} \right) C_F^3  \nn \\
	&+ \left(\frac{704}{45}\zeta_3-\frac{3736}{675}\zeta_2+\frac{2334509}{405000} \right)C_A C_F n_f+ \left(-\frac{88}{45}\zeta_3+\frac{152}{225}\zeta_2-\frac{14837573}{4050000} \right) C_F^2 n_f\,, \nn \\
	\gamma_{T,qg}^{(2)} &=
	\left(-\frac{220}{3}\zeta_4+\frac{1004}{225}\zeta_3+\frac{323629}{6750}\zeta_2-\frac{140682763}{6075000} \right) C_A^2 n_f \nn\\
	&+\left(\frac{6503}{225}\zeta_3+\frac{19387}{1125}\zeta_2-\frac{509985949}{24300000} \right) C_A C_F n_f \nn \\
	&+\left(\frac{622}{225}\zeta_3+\frac{79361}{6750}\zeta_2-\frac{2412861131}{48600000}\right) C_F^2 n_f
	+\left(\frac{176}{45}\zeta_3+\frac{389}{675}\zeta_2-\frac{51449}{9000}\right) C_A n_f^2 \nn \\
	&+ \left(\frac{497}{135}\zeta_2-\frac{915539}{300000}\right) C_F n_f^2
	- \frac{86}{375} \, n_f^3\,, \nn \\
	\gamma_{T,gg}^{(2)} &=
	\left(840 \zeta_4-\frac{3752}{25}\zeta_3-\frac{342578}{375}\zeta_2+\frac{1069405919}{1350000} \right)C_A^3 \nn \\
	&+\left(\frac{400}{3} \zeta_4-\frac{29534}{225}\zeta_3-\frac{30316}{675}\zeta_2+\frac{129284923}{2430000} \right) C_A^2 n_f \nn \\
	&+\left(\frac{2744}{45}\zeta_3-\frac{2158}{125}\zeta_2-\frac{188283293}{6075000} \right) C_A C_F n_f
	+\left(-\frac{352}{225}\zeta_3+\frac{4037}{3375}\zeta_2+\frac{27742123}{24300000} \right)C_F^2 n_f \nn \\
	&+\left(-\frac{64}{9}\zeta_3+\frac{160}{27}\zeta_2-\frac{71341}{27000} \right) C_A n_f^2+\left(-\frac{484}{675}\zeta_2-\frac{165553}{270000} \right) C_F n_f^2\,.
\end{align}

\section{$\beta$-function RGE and running coupling }

The well-known QCD $\beta$-function is written as
\begin{equation}
	\label{eq:beta}
	\frac{\mathrm{d}\alpha_s(\mu)}{\mathrm{d}\ln \mu}= \beta (\alpha_s(\mu)),\quad 
	\beta (\alpha)=- 2 \alpha\, \left[ \left( \frac{\alpha}{4 \pi} \right) \beta_0 + \left(
	\frac{\alpha}{4 \pi} \right)^2 \beta_1 + \left( \frac{\alpha}{4 \pi}
	\right)^3 \beta_2 + \cdots \right]\,,
\end{equation}
where the coefficient up to three loops are given by~\cite{Tarasov:1980au,Larin:1993tp,vanRitbergen:1997va,Czakon:2004bu}
\begin{align}
	\beta_0 &=\frac{11}{3} C_A - \frac{4}{3} T_F n_f 
	\,, \\
	\beta_1 &= \frac{34}{3} C_A^2 - \frac{20}{3} C_A T_F n_f - 4 C_F T_F n_f 
	\nn \,,\\
	\beta_2 &= n_f^2 T_F^2 \left(\frac{158 }{27}C_A+\frac{44
	}{9}C_F\right)+n_f T_F \left(2
	C_F^2-\frac{205 }{9}C_FC_A-\frac{1415 }{27}C_A^2\right)+\frac{2857 }{54}C_A^3
	\,,\nn\\
	\beta_3 &= \frac{1093}{729} n_f^3
	+\left(\frac{50065}{162} + \frac{6472}{81}\zeta_3\right) n_f^2
	+\left(-\frac{1078361}{162} - \frac{6508}{27} \zeta_3 \right) n_f + 3564 \zeta_3 + \frac{149753}{6} \nn\,.
\end{align}

At one-loop, the $\beta$-RGE can be solved exactly. At two-loop and beyond, there are different solutions. In terms of $L\equiv \ln\frac{\mu^2}{\Lambda_{\text{QCD}}^2}$, a expanded solution can be written as:
\begin{multline}
	   \alpha_s (\mu) = \frac{4 \pi}{\beta_0} \left[ \frac{1}{L} -
	   \frac{\beta_1}{\beta_0^2 L^2} \ln L + \frac{\beta_1^2}{\beta_0^4 L^3}
	   (\ln^2 L - \ln L - 1) + \frac{\beta_2}{\beta_0^3 L^3} \right. \\
	   \left. + \frac{\beta_1^3}{\beta_0^6 L^4} \left( - \ln^3 L + \frac{5}{2}
	   \ln^2 L + 2 \ln L - \frac{1}{2} \right) - 3 \frac{\beta_1
		   \beta_2}{\beta_0^5 L^4} \ln L + \frac{\beta_3}{2 \beta_0^4 L^4} \frac{}{}
	   \right]\,.
	 \end{multline}
Here we can obtain the two-loop running coupling for NLL resumation by setting $\beta_{2}=\beta_{3}=0$ and three-loop running coupling for NNLL by only $\beta_{3}=0$.

Alternatively, one can iteratively solve the RGE order by order in a formal expansion parameter $\epsilon\sim\frac{\beta_n}{\beta_0}$, with $n\geq 1$. For NLL, the two-loop running coupling is written as 
\begin{equation}
	\alpha_s(\mu) = \alpha_s(Q)\left[X+\alpha_s(Q)\frac{\beta_1}{4\pi \beta_0}\ln X\right]^{-1}, \quad X\equiv 1+\frac{\alpha_s(Q)}{2\pi}\beta_0\ln\frac{\mu}{Q}
	\,,\end{equation}
and at three loops for NNLL
\begin{equation}
	\alpha_s(\mu) = \alpha_s(Q)\left\{X+\alpha_s(Q)\frac{\beta_1}{4\pi \beta_0}\ln X+\frac{\alpha_s^2(Q)}{16\pi^2}\left[\frac{\beta_2}{\beta_0}\left(1-\frac{1}{X}\right)+\frac{\beta_1^2}{\beta_0^2}\left(\frac{1}{X}-1+\frac{\ln X}{X}\right)\right]\right\}^{-1}
	\,.\end{equation}

For the resummation in this paper, we use the iterative solution (the latter one) and set the coupling at $Q=91.2$ GeV to be the world average value $\alpha_s(m_Z)=0.118$.

\section{Squeeze limit of EEEC jet functions}
\label{sec:eeec_squeeze}

In this section, we provide the perturbative data for the squeeze limit of the EEEC jet function in Eq.~\eqref{eq:eeec_jet_expanded}, which is needed for E3C jet function calculation. Given the conformal parameterization,
\begin{equation}
	x_1=x_L z \bar z, \quad x_2=x_L(1-z)(1-\bar z),\quad x_3=x_L \,,
\end{equation}
the squeeze limits correspond to $z\rightarrow 0, 1, \infty$, related by a $\mathbb{S}_3$ symmetry. Without loss of generality, we provide the $z\rightarrow 1$ limit for the shapes function up to $\mathcal{O}(\epsilon^2)$. In the quark jet, we find for $G(z)$

\begin{align}
	G_q(z)&\overset{z\to1}{\approx} C_FT_Fn_f\bigg( {\color{darkred}\frac{13}{4800 (1-z) (1-\bar z)} } + \frac{z-2}{1440 (1-\bar z)^2}+\frac{\bar z}{1440 (1-z)^2} -\frac{39 z+1}{28800 (1-z)^2}\nn\\
	&+\frac{13}{9600 (1-\bar z)} \bigg)+C_F C_A\bigg( {\color{darkred} \frac{91}{4800 (1-z) (1-\bar z)}} +\frac{2-z}{2880 (1-\bar z)^2}-\frac{\bar z}{2880 (1-z)^2}\nn\\
	&-\frac{273 z-293}{28800
		(1-z)^2}+\frac{91}{9600 (1-\bar z)} \bigg)+C_F^2\bigg({\color{darkred} \frac{1}{20 (1-z) (\bar z-1)} }  -\frac{z+\bar z-2}{40 (z-1) (1-\bar z)} \bigg) \, , 
\end{align}
and for $F(z)$:
\begin{align}
	F_q(z)&\overset{z\to1}{\approx} C_FT_Fn_f\bigg( {\color{darkred} \frac{649}{28800 (1-z) (1-\bar z)}} -\frac{259}{43200 (1-z)^2}-\frac{259}{43200 (1-\bar z)^2}  \bigg)\nn\\
	&+C_F C_A\bigg( {\color{darkred} \frac{561}{3200 (1-z)(1-\bar z)}}+ \frac{229}{86400 (1-z)^2}+\frac{229}{86400 (1-\bar z)^2}\bigg)\nn\\
	&+C_F^2\bigg( {\color{darkred} \frac{3307}{7200 (1-z)(1-\bar z)}}\bigg) \, ,
\end{align}
as well as the $H(z)$:
\begin{align}
	H_q(z)&\overset{z\to1}{\approx} C_FT_Fn_f\bigg( {\color{darkred}  \frac{664193-23400\pi^2}{4320000(1-z) (1-\bar z)}  } + \frac{1800 \pi ^2-53191}{1296000 (1-z)^2}+\frac{1800 \pi ^2-53191}{1296000
		(1-\bar z)^2}\bigg)\nn\\
		&+C_F C_A\bigg( {\color{darkred}    \frac{1805867 - 54600 \pi^2}{1440000(1-z)(1-\bar z)}  } +\frac{45421-1800 \pi ^2}{2592000 (1-\bar z)^2}-\frac{1800 \pi ^2-45421}{2592000
			(1-z)^2} \bigg)\nn\\
			&+C_F^2 \bigg( {\color{darkred}  \frac{352451 - 10800 \pi^2}{108000 (1-z)(1-\bar z)}  } \bigg) \, .
\end{align}
Here the red stands for the most singular term, which contributes to $\frac{1}{\epsilon}$ divergence in the E3C jet function calculation. For the gluon jet, we also find
\begin{align}
	G_g(z)&\overset{z\to1}{\approx} C_FT_Fn_f\bigg( {\color{darkred}\frac{3}{320 (1-z) (1-\bar z)} }+\frac{3}{640(1-z)}+\frac{3}{640(1-\bar z)}\bigg)\nn\\
	&+C_A T_F n_f\bigg({ \color{darkred}\frac{7}{800 (1-z) (1-\bar z)} } +\frac{z-2}{1440 (1-\bar z)^2}+\frac{\bar z}{1440 (1-z)^2}-\frac{63 z-43}{14400
	(1-z)^2}\nn\\
	&+\frac{7}{1600 (1-\bar z)}  \bigg) +C_A^2\bigg(   {\color{darkred} \frac{49}{800(1-z)(1-\bar z)}   }    +\frac{2-z}{2880 (1-\bar z)^2}-\frac{\bar z}{2880 (1-z)^2}\nn\\
	&-\frac{441 z-451}{14400
		(1-z)^2}+\frac{49}{1600 (1-\bar z)}   \bigg) \, ,
\end{align}
\begin{align}
	F_g(z)&\overset{z\to1}{\approx} C_FT_Fn_f\bigg( {\color{darkred}\frac{241}{3200 (1-z) (1-\bar z)} }\bigg)+C_A T_F n_f\bigg(  {\color{darkred}   \frac{343}{4800(1-z)(1-\bar z)} }  -\frac{259}{43200 (1-z)^2}\nn\\
	&-\frac{259}{43200 (1-\bar z)^2} \bigg)+C_A^2\bigg( {\color{darkred}    \frac{557}{960(1-z)(1-\bar z)}} +\frac{229}{86400 (1-z)^2}  +\frac{229}{86400 (1-\bar z)^2}\bigg)\nn \, ,
\end{align}
\begin{align}
	H_g(z)&\overset{z\to1}{\approx} C_FT_Fn_f\bigg( {\color{darkred}\frac{434309 - 16200 \pi^2}{864000 (1-z) (1-\bar z)} }\bigg)+C_A T_F n_f\bigg(  {\color{darkred}   \frac{1033981-37800\pi^2}{2160000(1-z)(1-\bar z)} } \nn\\
	& +\frac{1800 \pi ^2-53191}{1296000 (1-z)^2}+\frac{1800 \pi ^2-53191}{1296000 (1-\bar z)^2}\bigg)+C_A^2\bigg({\color{darkred}  \frac{2999389 - 88200 \pi^2}{720000 (1-z)(1-\bar z)} }\nn\\
	&-\frac{1800 \pi ^2-45421}{2592000	(1-z)^2} -\frac{1800 \pi ^2-45421}{2592000 (1-\bar z)^2}\bigg) \, .
\end{align}

\section{Result of two-loop E3C jet function calculation}

We list the individual results for the two-loop jet function calculation in Sec.~\ref{sec:calculate_jet_fn}. As we discussed above, the calculation is reorganized as nonidentical energy weight contribution and contact terms. For the nonidentical energy weight in Sec.~\ref{sec:3_1_nonidentical}, we find for the quark jet
\begin{align}
\label{eq:E3C_quark_jet_1}
    \frac{d J^{\text{nonid}}_q}{dx_L}&=\left(\frac{\alpha_s}{4\pi}\right)^2\bigg\{\delta(x_L)f_q(\mu,Q,\epsilon)+\frac{1}{x_L}\bigg[C_F T_F n_f\bigg(-\frac{13}{200\epsilon}+\frac{13}{100}\ln\left(\frac{Q^2x_L}{\mu^2}\right)\notag\\
    &-0.44158(3)\bigg)+C_F^2\left(-\frac{6}{5\epsilon}+\frac{12}{5}\ln \left(\frac{Q^2x_L}{\mu^2}\right)-10.963(1) \right)\notag\\
    &+C_F C_A\left(-\frac{91}{200\epsilon}+\frac{91}{100} \ln \left(\frac{Q^2x_L}{\mu^2}\right) -4.3743(7)\right)\bigg]\bigg\} \,,
\end{align}
with the coefficient of the $\delta(x_L)$ being
\begin{align}
\label{eq:E3C_quark_jet_2}
    f_q(\mu,Q,\epsilon)&=C_F T_F n_f \bigg[\frac{13}{400\epsilon^2}+\frac{1}{\epsilon}\left(\frac{13}{200}\ln\left(\frac{\mu^2}{Q^2}\right)+0.22079(2)\right)+0.44158(3)\ln\left(\frac{\mu^2}{Q^2}\right)\notag\\
    &+\frac{13}{200}\ln^2\left(\frac{\mu^2}{Q^2}\right)+0.5441(8)\bigg]+C_F C_A\bigg[\frac{91}{400\epsilon^2}+\frac{1}{\epsilon}\left(\frac{91}{200}\ln\left(\frac{\mu^2}{Q^2}\right)+2.1871(8)\right)\notag\\
    &+4.3743(7)\ln\left(\frac{\mu^2}{Q^2}\right)+\frac{91}{200}\ln^2\left(\frac{\mu^2}{Q^2}\right)+10.483(2)\bigg]+C_F^2\bigg[24.60(4)+\frac{3}{5\epsilon^2}\notag\\
    &+\frac{1}{\epsilon}\left(\frac{6}{5}\ln\left(\frac{\mu^2}{Q^2}\right)+5.4815(3)\right)+10.963(1)\ln\left(\frac{\mu^2}{Q^2}\right)+\frac{6}{5}\ln^2\left(\frac{\mu^2}{Q^2}\right)\bigg]  \,.
\end{align}
The $\ln \left(\frac{Q^2x_L}{\mu^2}\right)$ term is verified by the jet RGE.
For a gluon jet, the $\cO(\alpha_s^2)$ contribution is 
\begin{align}
\label{eq:E3C_gluon_jet_1}
    \frac{d J^{\text{nonid}}_g}{dx_L}&=\left(\frac{\alpha_s}{4\pi}\right)^2\bigg\{\delta(x_L)f_g(\mu,Q,\epsilon)+\frac{1}{x_L}\bigg[C_F T_F n_f \left(-\frac{9}{40\epsilon}+\frac{9}{20}\ln\left(\frac{Q^2x_L}{\mu^2}\right)\right.\notag\\
    &\left.-1.8862(6)\right)+C_A T_F n_f \left(-\frac{21}{100\epsilon}+\frac{21}{50}\ln\left(\frac{Q^2x_L}{\mu^2}\right)-1.5376(9)\right)\notag\\
    &+C_A^2\left(-\frac{147}{100\epsilon}+\frac{147}{50}\ln\left(\frac{Q^2x_L}{\mu^2}\right)-14.031(3)\right)\bigg]\bigg\}  \,,
\end{align}
with the corresponding coefficient 
\begin{align}
\label{eq:E3C_gluon_jet_2}
    f_g(\mu,Q,\epsilon)&=C_A T_F n_f\bigg[\frac{21}{200\epsilon^2}+\frac{1}{\epsilon}\left(\frac{21}{100}\ln\left(\frac{\mu^2}{Q^2}\right)+0.7688(5)\right)+1.5376(9) \ln\left(\frac{\mu^2}{Q^2}\right)\notag\\
    &+\frac{21}{100}\ln^2\left(\frac{\mu^2}{Q^2}\right) +2.350(8) \bigg]+C_F T_F n_f \bigg[ \frac{9}{80\epsilon^2}+\frac{1}{\epsilon}\left(\frac{9}{40}\ln\left(\frac{\mu^2}{Q^2}\right)+0.9431(3)\right)\notag\\
    &+1.886(3) \ln\left(\frac{\mu^2}{Q^2}\right)+\frac{9}{40}\ln^2\left(\frac{\mu^2}{Q^2}\right)+3.757(1)\bigg]+C_A^2\bigg[33.188(4)+ \frac{147}{200\epsilon^2}\notag\\
    &+\frac{1}{\epsilon}\left(\frac{147}{100}\ln\left(\frac{\mu^2}{Q^2}\right)+7.01569(5)\right)+14.031(3) \ln\left(\frac{\mu^2}{Q^2}\right) +\frac{147}{100}\ln^2\left(\frac{\mu^2}{Q^2}\right) \bigg]  \,.
\end{align}

Regarding the contact term in Sec.~\ref{sec:3_2_contact}, for $e^+e^-$ annihilation, we have the sum of E$^2$EC and E$^3$C
\begin{align}
\label{eq:E3C_contact_q_2loop_1}
\frac{1}{\sigma_0}\frac{\df\sigma^{[3],\text{2-loop}}_{\text{C,q}}(x_L,\epsilon)}{\df x_L}
=&\left(\frac{\alpha_s}{4\pi}\right)^2 \Biggl\{ \delta(x_L) r_q(\mu,Q,\epsilon)
+
\left[\frac{1}{x_L}\right]_+\biggl[C_A C_F \biggl(\frac{91}{100 \epsilon }+\frac{1189}{200} \ln \left(\frac{\mu ^2}{Q^2}\right)\nn\\
&
-6 \zeta_3+\frac{25 \pi ^2}{6}-\frac{52307}{18000}\biggr)
   +C_F n_fT_F \left(\frac{13}{100 \epsilon}-\frac{31}{25} \ln \left(\frac{\mu ^2}{Q^2}\right)-\frac{14809}{2000}\right)\nn\\
   &
   +C_F^2 \left(\frac{12}{5 \epsilon }+\frac{24}{5} \ln \left(\frac{\mu
   ^2}{Q^2}\right)+12 \zeta_3-\frac{43 \pi
   ^2}{6}+\frac{274081}{3600}\right)\biggr]\nn\\
   &
   +\left[\frac{\ln (x_L)}{x_L}\right]_+ 
   \left(-\frac{1343}{200}C_A
   C_F+\frac{113}{100} C_F n_f T_F+\frac{87}{80}C_F^2\right)
   \Biggr\}  \,,
\end{align}
with the singular part $r_q(\mu,Q,\epsilon)$
\begin{align}
\label{eq:E3C_contact_q_2loop_2}
    r_q(\mu,Q,\epsilon) &=C_A C_F \Biggl[-\frac{91}{200 \epsilon ^2}+\frac{1}{\epsilon}\Biggl(-\frac{91}{100} \ln \left(\frac{\mu^2}{Q^2}\right)+3 \zeta_3-\frac{25 \pi^2}{12}+\frac{452921}{36000}\Biggr)-\frac{91}{100} \ln ^2\left(\frac{\mu^2}{Q^2}\right)\notag\\
    &+\left(6 \zeta_3+\frac{890167}{36000}-\frac{25 \pi ^2}{6}\right) \ln
   \left(\frac{\mu ^2}{Q^2}\right)-\frac{347 \zeta_3}{2}+\frac{7 \pi ^4}{20}-\frac{6697 \pi
   ^2}{1800}+\frac{47220317}{270000}\Biggr]
   \nn\\
   &+C_F n_f T_F \Biggl[-\frac{13}{200 \epsilon^2}
   +\frac{1}{\epsilon}\Biggl(-\frac{13}{100} \ln \left(\frac{\mu
   ^2}{Q^2}\right)-\frac{5299}{12000}\Biggr)-\frac{13}{100} \ln^2\left(\frac{\mu ^2}{Q^2}\right)\notag\\
   &-\frac{4349}{6000}\ln \left(\frac{\mu
   ^2}{Q^2}\right)+4 \zeta_3+\frac{137 \pi ^2}{400}-\frac{1413979}{720000}\Biggr]+C_F^2 \Biggl[-\frac{6}{5 \epsilon ^2}\notag\\
   &+\frac{1}{\epsilon }\Biggl(-\frac{12}{5} \ln\left(\frac{\mu ^2}{Q^2}\right)-6 \zeta_3+\frac{43 \pi
   ^2}{12}-\frac{281641}{7200}\Biggr)-\frac{12}{5} \ln^2\left(\frac{\mu ^2}{Q^2}\right)\notag\\
   &+\left(-12 \zeta_3-\frac{281641}{3600}+\frac{43 \pi^2}{6}\right) \ln \left(\frac{\mu ^2}{Q^2}\right)+293 \zeta_3-\frac{7\pi ^4}{10}+\frac{15371\pi^2}{1440}-\frac{380074411}{864000}\Biggr]  \, .
\end{align}
Similarly, in the gluonic Higgs decay, we get
\begin{align}
\label{eq:E3C_contact_g_2loop_1}
\frac{1}{\sigma^\prime_0}\frac{\df\sigma^{[3],\text{2-loop}}_{\text{C,g}}(x_L,\epsilon)}{\df x_L}
=&\lambda(\mu)\left(\frac{\alpha_s}{4\pi}\right)^2\Bigg\{\delta(x_L) r_g(\mu,Q,\epsilon)
+
\left[\frac{1}{x_L}\right]_+
\bigg\{
n_f^2 T_F^2
   \bigg(-\frac{3}{5} \ln \left(\frac{\mu
   ^2}{Q^2}\right)-\frac{131}{60}\bigg)\nn\\
   &
+n_f T_F \bigg[C_A \left(\frac{21}{50 \epsilon }-\frac{171}{100} \ln \left(\frac{\mu
   ^2}{Q^2}\right)+\frac{7 \pi
   ^2}{15}-\frac{140917}{9000}\right)\nn\\
   &
   +C_F \left(\frac{9}{10} \ln \left(\frac{\mu
   ^2}{Q^2}\right)+\frac{9}{20 \epsilon }+\frac{1579}{400}\right)\bigg]\nn\\
   &
   +C_A^2
   \bigg(\frac{147}{50
   \epsilon }+\frac{1743}{100} \ln \left(\frac{\mu ^2}{Q^2}\right)+6 \zeta_3-\frac{97 \pi ^2}{30}+\frac{211829}{2250}\bigg)
   \bigg\}\nn\\
   &
   +\left[\frac{\ln (x_L)}{x_L}\right]_+
   \bigg[n_f T_F \left(\frac{51
   }{25}C_A-\frac{69 }{40}C_F\right)-\frac{133 }{25}C_A^2+\frac{2}{5} n_f^2
   T_F^2\bigg]
\Bigg\}  \,,
\end{align}
with the gluonic singular term $r_g(\mu,Q,\epsilon)$
\begin{align}
\label{eq:E3C_contact_g_2loop_2}
    r_g(\mu,Q,\epsilon)&=
 C_A  T_F n_f \biggl[-\frac{21}{100 \epsilon
   ^2}+\frac{1}{\epsilon }\biggl(-\frac{21}{50} \ln \left(\frac{\mu ^2}{Q^2}\right)-\frac{7 \pi^2}{30}+\frac{6887}{9000}\biggr)-\frac{1163}{150} \ln^2\left(\frac{\mu^2}{Q^2}\right)
\notag\\
&
+\left(-\frac{948847}{18000}-\frac{7 \pi ^2}{15}\right) \ln\left(\frac{\mu^2}{Q^2}\right)
   -\frac{211 \zeta_3}{10}+\frac{3037 \pi ^2}{1800}-\frac{5585159}{67500}\biggr]
   +   C_F T_F n_f \notag\\
   &\biggl[-\frac{9}{40 \epsilon ^2}+\frac{1}{\epsilon }\biggl(-\frac{9}{20} \ln \left(\frac{\mu
   ^2}{Q^2}\right)-\frac{1509}{800}\biggr)-\frac{9}{20} \ln
   ^2\left(\frac{\mu ^2}{Q^2}\right)-\frac{3109}{400} \ln \left(\frac{\mu
   ^2}{Q^2}\right)+15 \zeta_3\notag\\
   &+\frac{5 \pi
   ^2}{8}-\frac{230393}{6000}\biggr]+C_A^2 \biggl\{-\frac{147}{100 \epsilon ^2}+\frac{1}{\epsilon }\biggl[-\frac{147}{50} \ln \left(\frac{\mu ^2}{Q^2}\right)-3
   \zeta_3+\frac{97 \pi ^2}{60}-\frac{474857}{18000}\biggr]\nn\\
   &+\frac{2143}{300} \ln ^2\left(\frac{\mu
   ^2}{Q^2}\right)+\left(-6 \zeta_3+\frac{261281}{18000}+\frac{97 \pi ^2}{30}\right) \ln
   \left(\frac{\mu ^2}{Q^2}\right)+\frac{1133 \zeta_3}{10}-\frac{7 \pi ^4}{20}\notag\\
   &+\frac{373 \pi
   ^2}{100}-\frac{12512789}{90000}
   \biggr\}+n_f^2 T_F^2 \biggl[\frac{4}{3} \ln ^2\left(\frac{\mu
   ^2}{Q^2}\right)+\frac{2971}{300} \ln \left(\frac{\mu ^2}{Q^2}\right)-\frac{23 \pi
   ^2}{45}+\frac{579043}{27000}\biggr]  \,,
\end{align}
where $\lambda$ is the effective $Hgg$ coupling\footnote{For the case of gluonic Higgs decays, we normalize the E3C into the form where the LO E3C is $\frac{1}{\sigma^\prime_0}\frac{\df\sigma^{[3]}_0 }{\df x_L}=\lambda(\mu) \left(\frac{1}{4}\delta(x_L) + \frac{3}{4}\delta(1-x_L)\right)$ in $d=4-2\epsilon$ dimensions. } \cite{Gehrmann:2010ue}. These results are then used to extract the two-loop jet constants.

\section{Fixed-order expansion}

In this section, we provide the singular expansion of projected energy correlator up to NNLO $\mathcal{O}(\alpha_s^3)$ in $e^+e^-$ annihilation. This can be achieved by expanding our resummed distribution with canonical scale $\mu=Q$. For EEC, we find
\begin{align}
	\frac{1}{\sigma_0}\frac{d\sigma^{[2]}}{dx_L}&=\left(\frac{\alpha_s}{4\pi}\right)C_F\frac{3}{2 x_L}+\left(\frac{\alpha_s}{4\pi}\right)^2 C_F\bigg\{\bigg[\frac{53}{30}n_f T_F+\frac{25}{4}C_F-\frac{107}{15}C_A\bigg]\frac{\ln x_L}{x_L}\notag\\
	&+\bigg[-\frac{4913}{450}n_f T_F+\bigg(-\frac{8263}{216}+\frac{43}{9}\pi^2-8\zeta_3\bigg)C_F+\bigg(\frac{35336}{675}-\frac{25}{9}\pi^2+4\zeta_3\bigg)C_A\bigg]\frac{1}{x_L}\bigg\}\notag\\
	&+\left(\frac{\alpha_s}{4\pi}\right)^3C_F\bigg\{ \bigg[ \frac{8059}{300}C_A^2-\frac{340}{9}C_F C_A+\frac{625}{48}C_F^2-\frac{16259}{900}C_A T_F n_f+\frac{4619}{360}C_F T_F n_f\notag\\
	&+\frac{92}{45}n_f^2 T_F^2 \bigg]\frac{\ln^2 x_L}{x_L} +\bigg[ -\frac{17734}{675}n_f^2 T_F^2 +\bigg(-\frac{64 \zeta_3}{3}-\frac{6760183}{32400}+\frac{416 \pi ^2}{27} \bigg) C_F T_F n_F \notag\\
	&+\bigg( \frac{32 \zeta_3}{3}+\frac{6644267}{27000}-\frac{36 \pi ^2}{5}\bigg) C_A T_F n_f+\bigg(-\frac{172 \zeta_3}{3}-\frac{723533}{2592}+\frac{1849 \pi ^2}{54} \bigg)C_F^2\notag\\
	&+\bigg(-\frac{74 \zeta_3}{3}-\frac{2916859}{6750}+\frac{503 \pi ^2}{30} \bigg) C_A^2 +\bigg( \frac{262 \zeta_3}{3}+\frac{105425}{144}-\frac{550 \pi ^2}{9} \bigg) C_F C_A \bigg]\frac{\ln x_L}{x_L}\notag\\
	&+\bigg[\bigg(\frac{88031}{1125}+\frac{4\pi^2}{5}\bigg)n_f^2 T_F^2 +\bigg(-\frac{15988 \zeta _3}{45}+\frac{236 \pi ^4}{135}-\frac{15161 \pi ^2}{360}+\frac{164829499}{243000} \bigg) C_F T_F n_F \notag\\
	&+\bigg( \frac{3679 \zeta _3}{15}-\frac{118 \pi ^4}{135}+\frac{379579 \pi ^2}{16200}-\frac{1025118113}{1080000} \bigg) C_A T_F n_F \notag\\
	&+\bigg(8 \pi ^2 \zeta _3+52 \zeta _3+208 \zeta _5-\frac{167 \pi ^4}{27}-\frac{18805 \pi ^2}{1296}+\frac{742433}{1944} \bigg) C_F^2 \notag\\
	&+\bigg(4 \pi ^2 \zeta _3-\frac{47483 \zeta _3}{90}+56 \zeta _5-\frac{481 \pi ^4}{540}-\frac{906257 \pi ^2}{16200}+\frac{964892417}{540000} \bigg) C_A^2 \notag\\
	&+\bigg(-12 \pi ^2 \zeta _3+\frac{10604 \zeta _3}{15}-216 \zeta _5+\frac{847 \pi ^4}{180}+\frac{137305 \pi ^2}{1296}-\frac{105395741}{51840}\bigg) C_F C_A\bigg]\frac{1}{x_L} \bigg\} \, .
\end{align}
Similarly, for E3C, we have
\begin{align}
	\frac{1}{\sigma_0}\frac{d\sigma^{[3]}}{dx_L}&=\left(\frac{\alpha_s}{4\pi}\right)C_F\frac{9}{8x_L}+\left(\frac{\alpha_s}{4\pi}\right)^2 C_F\bigg\{\bigg[\frac{139}{100}n_f T_F+\frac{471}{80}C_F-\frac{979}{200}C_A\bigg]\frac{\ln x_L}{x_L}\notag\\
	&+\bigg[-\frac{24863}{3000}n_f T_F-\frac{21}{10}C_F+\frac{66769}{3000} C_A\bigg]\frac{1}{x_L}\bigg\}\notag\\
	&+\left(\frac{\alpha_s}{4\pi}\right)^3C_F\bigg\{ \bigg[ \frac{17743}{1000}C_A^2-\frac{412753}{12000}C_F C_A+\frac{24649}{1600}C_F^2-\frac{19019}{1500}C_A T_F n_f\notag\\
	&+\frac{35369}{3000}C_F T_F n_f+\frac{128}{75}n_f^2 T_F^2 \bigg]\frac{\ln^2 x_L}{x_L}+\bigg[-\frac{4559891}{22500}C_A^2-\frac{814823}{48000}C_F^2\notag\\
	&+\bigg(\frac{34399441}{120000}-\frac{11 \pi ^2}{2}\bigg) C_F C_A + \bigg(2 \pi ^2-\frac{1026851}{10000} \bigg)C_F T_F n_f+ \frac{3055907}{22500}C_A T_F n_f\notag\\
	&-\frac{23494}{1125}n_f^2 T_F^2 \bigg]\frac{\ln x_L}{x_L}+\bigg[j_2^{q,[3]}\bigg(\frac{157}{15}-\frac{44 C_A}{3 C_F}+\frac{16 n_f T_F}{3C_F}\bigg)-\frac{22}{15}j_2^{g,[3]} \notag\\
	&+ \bigg(\frac{106027}{54000}-\frac{22 \pi ^2}{225}\bigg) n_f^2 T_F^2 +\bigg( \frac{1827 \zeta _3}{25}-\frac{3877 \pi ^2}{3000}-\frac{3239027203}{10800000} \bigg) C_F T_F n_f \notag\\
	&+\bigg(-\frac{1037 \zeta _3}{50}-\frac{2167 \pi ^2}{4500}-\frac{24958553}{3600000} \bigg) C_A T_F n_f \notag\\
	&+\bigg( \frac{3267 \zeta _3}{20}-\frac{111313 \pi ^2}{14400}-\frac{6031520921}{17280000} \bigg) C_F^2 + \bigg( -\frac{829 \zeta _3}{100}+\frac{4433 \pi ^2}{2250}+\frac{363491521}{5400000} \bigg) C_A^2 \notag\\
	&+\bigg(-\frac{42321 \zeta _3}{200}+\frac{284797 \pi ^2}{36000}+\frac{4941457181}{7200000}\bigg) C_F C_A \bigg]\frac{1}{x_L} \bigg\} \, ,
\end{align}
with the two-loop jet constant $j_2^{q/g,[3]}$ from  Eq.~\eqref{eq:j2q_res}-\eqref{eq:j2g_res}.

\bibliography{E3C_Ref}{}

\providecommand{\href}[2]{#2}\begingroup\raggedright\begin{thebibliography}{100}

\bibitem{Basham:1978zq}
C.~L. Basham, L.~S. Brown, S.~D. Ellis, and S.~T. Love, {\it {Energy
  Correlations in electron-Positron Annihilation in Quantum Chromodynamics:
  Asymptotically Free Perturbation Theory}},  {\em Phys. Rev.} {\bf D19} (1979)
  2018.

\bibitem{Basham:1978bw}
C.~L. Basham, L.~S. Brown, S.~D. Ellis, and S.~T. Love, {\it {Energy
  Correlations in electron - Positron Annihilation: Testing QCD}},  {\em Phys.
  Rev. Lett.} {\bf 41} (1978) 1585.

\bibitem{Belitsky:2013ofa}
A.~V. Belitsky, S.~Hohenegger, G.~P. Korchemsky, E.~Sokatchev, and
  A.~Zhiboedov, {\it {Energy-Energy Correlations in N=4 Supersymmetric
  Yang-Mills Theory}},  {\em Phys. Rev. Lett.} {\bf 112} (2014), no.~7 071601,
  [\href{http://arxiv.org/abs/1311.6800}{{\tt arXiv:1311.6800}}].

\bibitem{Henn:2019gkr}
J.~M. Henn, E.~Sokatchev, K.~Yan, and A.~Zhiboedov, {\it {Energy-energy
  correlation in $N$=4 super Yang-Mills theory at next-to-next-to-leading
  order}},  {\em Phys. Rev. D} {\bf 100} (2019), no.~3 036010,
  [\href{http://arxiv.org/abs/1903.05314}{{\tt arXiv:1903.05314}}].

\bibitem{Dixon:2018qgp}
L.~J. Dixon, M.-X. Luo, V.~Shtabovenko, T.-Z. Yang, and H.~X. Zhu, {\it
  {Analytical Computation of Energy-Energy Correlation at Next-to-Leading Order
  in QCD}},  {\em Phys. Rev. Lett.} {\bf 120} (2018), no.~10 102001,
  [\href{http://arxiv.org/abs/1801.03219}{{\tt arXiv:1801.03219}}].

\bibitem{Luo:2019nig}
M.-X. Luo, V.~Shtabovenko, T.-Z. Yang, and H.~X. Zhu, {\it {Analytic
  Next-To-Leading Order Calculation of Energy-Energy Correlation in
  Gluon-Initiated Higgs Decays}},  {\em JHEP} {\bf 06} (2019) 037,
  [\href{http://arxiv.org/abs/1903.07277}{{\tt arXiv:1903.07277}}].

\bibitem{Gao:2020vyx}
J.~Gao, V.~Shtabovenko, and T.-Z. Yang, {\it {Energy-energy correlation in
  hadronic Higgs decays: analytic results and phenomenology at NLO}},  {\em
  JHEP} {\bf 02} (2021) 210, [\href{http://arxiv.org/abs/2012.14188}{{\tt
  arXiv:2012.14188}}].

\bibitem{Hofman:2008ar}
D.~M. Hofman and J.~Maldacena, {\it {Conformal collider physics: Energy and
  charge correlations}},  {\em JHEP} {\bf 05} (2008) 012,
  [\href{http://arxiv.org/abs/0803.1467}{{\tt arXiv:0803.1467}}].

\bibitem{Chen:2019bpb}
H.~Chen, M.-X. Luo, I.~Moult, T.-Z. Yang, X.~Zhang, and H.~X. Zhu, {\it {Three
  point energy correlators in the collinear limit: symmetries, dualities and
  analytic results}},  {\em JHEP} {\bf 08} (2020), no.~08 028,
  [\href{http://arxiv.org/abs/1912.11050}{{\tt arXiv:1912.11050}}].

\bibitem{Chen:2020adz}
H.~Chen, I.~Moult, and H.~X. Zhu, {\it {Quantum Interference in Jet
  Substructure from Spinning Gluons}},  {\em Phys. Rev. Lett.} {\bf 126}
  (2021), no.~11 112003, [\href{http://arxiv.org/abs/2011.02492}{{\tt
  arXiv:2011.02492}}].

\bibitem{Chen:2022jhb}
H.~Chen, I.~Moult, J.~Sandor, and H.~X. Zhu, {\it {Celestial Blocks and
  Transverse Spin in the Three-Point Energy Correlator}},
  \href{http://arxiv.org/abs/2202.04085}{{\tt arXiv:2202.04085}}.

\bibitem{Chen:2022swd}
H.~Chen, I.~Moult, J.~Thaler, and H.~X. Zhu, {\it {Non-Gaussianities in
  Collider Energy Flux}},  \href{http://arxiv.org/abs/2205.02857}{{\tt
  arXiv:2205.02857}}.

\bibitem{Chang:2022ryc}
C.-H. Chang and D.~Simmons-Duffin, {\it {Three-point energy correlators and the
  celestial block expansion}},  \href{http://arxiv.org/abs/2202.04090}{{\tt
  arXiv:2202.04090}}.

\bibitem{Yang:2022tgm}
T.-Z. Yang and X.~Zhang, {\it {Analytic Computation of Three-point Energy
  Correlator in QCD}},  \href{http://arxiv.org/abs/2208.01051}{{\tt
  arXiv:2208.01051}}.

\bibitem{Yan:2022cye}
K.~Yan and X.~Zhang, {\it {Three-point energy correlator in $\mathcal{N}=4$
  super Yang-Mills Theory}},  \href{http://arxiv.org/abs/2203.04349}{{\tt
  arXiv:2203.04349}}.

\bibitem{Chen:2020vvp}
H.~Chen, I.~Moult, X.~Zhang, and H.~X. Zhu, {\it {Rethinking jets with energy
  correlators: Tracks, resummation, and analytic continuation}},  {\em Phys.
  Rev. D} {\bf 102} (2020), no.~5 054012,
  [\href{http://arxiv.org/abs/2004.11381}{{\tt arXiv:2004.11381}}].

\bibitem{Dixon:2019uzg}
L.~J. Dixon, I.~Moult, and H.~X. Zhu, {\it {Collinear limit of the
  energy-energy correlator}},  {\em Phys. Rev. D} {\bf 100} (2019), no.~1
  014009, [\href{http://arxiv.org/abs/1905.01310}{{\tt arXiv:1905.01310}}].

\bibitem{Kologlu:2019mfz}
M.~Kologlu, P.~Kravchuk, D.~Simmons-Duffin, and A.~Zhiboedov, {\it {The
  light-ray OPE and conformal colliders}},  {\em JHEP} {\bf 01} (2021) 128,
  [\href{http://arxiv.org/abs/1905.01311}{{\tt arXiv:1905.01311}}].

\bibitem{Korchemsky:2019nzm}
G.~P. Korchemsky, {\it {Energy correlations in the end-point region}},  {\em
  JHEP} {\bf 01} (2020) 008, [\href{http://arxiv.org/abs/1905.01444}{{\tt
  arXiv:1905.01444}}].

\bibitem{Komiske:2022enw}
P.~T. Komiske, I.~Moult, J.~Thaler, and H.~X. Zhu, {\it {Analyzing N-point
  Energy Correlators Inside Jets with CMS Open Data}},
  \href{http://arxiv.org/abs/2201.07800}{{\tt arXiv:2201.07800}}.

\bibitem{Lee:2022ige}
K.~Lee, B.~Me\c{c}aj, and I.~Moult, {\it {Conformal Colliders Meet the LHC}},
  \href{http://arxiv.org/abs/2205.03414}{{\tt arXiv:2205.03414}}.

\bibitem{L3:1992nwf}
{\bf L3} Collaboration, B.~Adeva et~al., {\it {Studies of hadronic event
  structure and comparisons with QCD models at the Z0 resonance}},  {\em Z.
  Phys. C} {\bf 55} (1992) 39--62.

\bibitem{SLD:1994idb}
{\bf SLD} Collaboration, K.~Abe et~al., {\it {Measurement of alpha-s (M(Z)**2)
  from hadronic event observables at the Z0 resonance}},  {\em Phys. Rev. D}
  {\bf 51} (1995) 962--984, [\href{http://arxiv.org/abs/hep-ex/9501003}{{\tt
  hep-ex/9501003}}].

\bibitem{DELPHI:1996oqw}
{\bf DELPHI} Collaboration, P.~Abreu et~al., {\it {Measurement of event shape
  and inclusive distributions at S**(1/2) = 130-GeV and 136-GeV}},  {\em Z.
  Phys. C} {\bf 73} (1997) 229--242.

\bibitem{ALEPH:2003obs}
{\bf ALEPH} Collaboration, A.~Heister et~al., {\it {Studies of QCD at e+ e-
  centre-of-mass energies between 91-GeV and 209-GeV}},  {\em Eur. Phys. J. C}
  {\bf 35} (2004) 457--486.

\bibitem{DELPHI:2004omy}
{\bf DELPHI} Collaboration, J.~Abdallah et~al., {\it {The Measurement of
  alpha(s) from event shapes with the DELPHI detector at the highest LEP
  energies}},  {\em Eur. Phys. J. C} {\bf 37} (2004) 1--23,
  [\href{http://arxiv.org/abs/hep-ex/0406011}{{\tt hep-ex/0406011}}].

\bibitem{OPAL:2004wof}
{\bf OPAL} Collaboration, G.~Abbiendi et~al., {\it {Measurement of event shape
  distributions and moments in e+ e- ---\ensuremath{>} hadrons at 91-GeV -
  209-GeV and a determination of alpha(s)}},  {\em Eur. Phys. J. C} {\bf 40}
  (2005) 287--316, [\href{http://arxiv.org/abs/hep-ex/0503051}{{\tt
  hep-ex/0503051}}].

\bibitem{Dissertori:2007xa}
G.~Dissertori, A.~Gehrmann-De~Ridder, T.~Gehrmann, E.~W.~N. Glover,
  G.~Heinrich, and H.~Stenzel, {\it {First determination of the strong coupling
  constant using NNLO predictions for hadronic event shapes in e+ e-
  annihilations}},  {\em JHEP} {\bf 02} (2008) 040,
  [\href{http://arxiv.org/abs/0712.0327}{{\tt arXiv:0712.0327}}].

\bibitem{Davison:2009wzs}
R.~A. Davison and B.~R. Webber, {\it {Non-Perturbative Contribution to the
  Thrust Distribution in e+ e- Annihilation}},  {\em Eur. Phys. J. C} {\bf 59}
  (2009) 13--25, [\href{http://arxiv.org/abs/0809.3326}{{\tt
  arXiv:0809.3326}}].

\bibitem{Bethke:2009ehn}
{\bf JADE} Collaboration, S.~Bethke, S.~Kluth, C.~Pahl, and J.~Schieck, {\it
  {Determination of the Strong Coupling alpha(s) from hadronic Event Shapes
  with O(alpha**3(s)) and resummed QCD predictions using JADE Data}},  {\em
  Eur. Phys. J. C} {\bf 64} (2009) 351--360,
  [\href{http://arxiv.org/abs/0810.1389}{{\tt arXiv:0810.1389}}].

\bibitem{Dissertori:2009ik}
G.~Dissertori, A.~Gehrmann-De~Ridder, T.~Gehrmann, E.~W.~N. Glover,
  G.~Heinrich, G.~Luisoni, and H.~Stenzel, {\it {Determination of the strong
  coupling constant using matched NNLO+NLLA predictions for hadronic event
  shapes in e+e- annihilations}},  {\em JHEP} {\bf 08} (2009) 036,
  [\href{http://arxiv.org/abs/0906.3436}{{\tt arXiv:0906.3436}}].

\bibitem{Abbate:2010xh}
R.~Abbate, M.~Fickinger, A.~H. Hoang, V.~Mateu, and I.~W. Stewart, {\it {Thrust
  at $N^{3}LL$ with Power Corrections and a Precision Global Fit for
  $\alpha_{s}(mZ)$}},  {\em Phys. Rev. D} {\bf 83} (2011) 074021,
  [\href{http://arxiv.org/abs/1006.3080}{{\tt arXiv:1006.3080}}].

\bibitem{Abbate:2012jh}
R.~Abbate, M.~Fickinger, A.~H. Hoang, V.~Mateu, and I.~W. Stewart, {\it
  {Precision Thrust Cumulant Moments at $N^3$LL}},  {\em Phys. Rev. D} {\bf 86}
  (2012) 094002, [\href{http://arxiv.org/abs/1204.5746}{{\tt
  arXiv:1204.5746}}].

\bibitem{Hoang:2014wka}
A.~H. Hoang, D.~W. Kolodrubetz, V.~Mateu, and I.~W. Stewart, {\it
  {$C$-parameter distribution at N$^3$LL' including power corrections}},  {\em
  Phys. Rev. D} {\bf 91} (2015), no.~9 094017,
  [\href{http://arxiv.org/abs/1411.6633}{{\tt arXiv:1411.6633}}].

\bibitem{Hoang:2015hka}
A.~H. Hoang, D.~W. Kolodrubetz, V.~Mateu, and I.~W. Stewart, {\it {Precise
  determination of $\alpha_s$ from the $C$-parameter distribution}},  {\em
  Phys. Rev. D} {\bf 91} (2015), no.~9 094018,
  [\href{http://arxiv.org/abs/1501.04111}{{\tt arXiv:1501.04111}}].

\bibitem{Becher:2008cf}
T.~Becher and M.~D. Schwartz, {\it {A precise determination of $\alpha_s$ from
  LEP thrust data using effective field theory}},  {\em JHEP} {\bf 07} (2008)
  034, [\href{http://arxiv.org/abs/0803.0342}{{\tt arXiv:0803.0342}}].

\bibitem{Chien:2010kc}
Y.-T. Chien and M.~D. Schwartz, {\it {Resummation of heavy jet mass and
  comparison to LEP data}},  {\em JHEP} {\bf 08} (2010) 058,
  [\href{http://arxiv.org/abs/1005.1644}{{\tt arXiv:1005.1644}}].

\bibitem{Luisoni:2020efy}
G.~Luisoni, P.~F. Monni, and G.~P. Salam, {\it {$C$-parameter hadronisation in
  the symmetric 3-jet limit and impact on $\alpha_s$ fits}},  {\em Eur. Phys.
  J. C} {\bf 81} (2021), no.~2 158,
  [\href{http://arxiv.org/abs/2012.00622}{{\tt arXiv:2012.00622}}].

\bibitem{Bhattacharya:2022dtm}
A.~Bhattacharya, M.~D. Schwartz, and X.~Zhang, {\it {Sudakov Shoulder
  Resummation for Thrust and Heavy Jet Mass}},
  \href{http://arxiv.org/abs/2205.05702}{{\tt arXiv:2205.05702}}.

\bibitem{Bhattacharya:2023qet}
A.~Bhattacharya, J.~K.~L. Michel, M.~D. Schwartz, I.~W. Stewart, and X.~Zhang,
  {\it {NNLL Resummation of Sudakov Shoulder Logarithms in the Heavy Jet Mass
  Distribution}},  \href{http://arxiv.org/abs/2306.08033}{{\tt
  arXiv:2306.08033}}.

\bibitem{Marzani:2019evv}
S.~Marzani, D.~Reichelt, S.~Schumann, G.~Soyez, and V.~Theeuwes, {\it {Fitting
  the Strong Coupling Constant with Soft-Drop Thrust}},  {\em JHEP} {\bf 11}
  (2019) 179, [\href{http://arxiv.org/abs/1906.10504}{{\tt arXiv:1906.10504}}].

\bibitem{Hannesdottir:2022rsl}
H.~S. Hannesdottir, A.~Pathak, M.~D. Schwartz, and I.~W. Stewart, {\it
  {Prospects for strong coupling measurement at hadron colliders using
  soft-drop jet mass}},  {\em JHEP} {\bf 04} (2023) 087,
  [\href{http://arxiv.org/abs/2210.04901}{{\tt arXiv:2210.04901}}].

\bibitem{LeBlanc:2022bwd}
M.~LeBlanc, B.~Nachman, and C.~Sauer, {\it {Going off topics to demix quark and
  gluon jets in \ensuremath{\alpha}$_{S}$ extractions}},  {\em JHEP} {\bf 02}
  (2023) 150, [\href{http://arxiv.org/abs/2206.10642}{{\tt arXiv:2206.10642}}].

\bibitem{Li:2021zcf}
Y.~Li, I.~Moult, S.~S. van Velzen, W.~J. Waalewijn, and H.~X. Zhu, {\it
  {Extending Precision Perturbative QCD with Track Functions}},  {\em Phys.
  Rev. Lett.} {\bf 128} (2022), no.~18 182001,
  [\href{http://arxiv.org/abs/2108.01674}{{\tt arXiv:2108.01674}}].

\bibitem{Chang:2013rca}
H.-M. Chang, M.~Procura, J.~Thaler, and W.~J. Waalewijn, {\it {Calculating
  Track-Based Observables for the LHC}},  {\em Phys. Rev. Lett.} {\bf 111}
  (2013) 102002, [\href{http://arxiv.org/abs/1303.6637}{{\tt
  arXiv:1303.6637}}].

\bibitem{Jaarsma:2022kdd}
M.~Jaarsma, Y.~Li, I.~Moult, W.~Waalewijn, and H.~X. Zhu, {\it {Renormalization
  Group Flows for Track Function Moments}},
  \href{http://arxiv.org/abs/2201.05166}{{\tt arXiv:2201.05166}}.

\bibitem{Mitov:2006ic}
A.~Mitov, S.~Moch, and A.~Vogt, {\it {Next-to-Next-to-Leading Order Evolution
  of Non-Singlet Fragmentation Functions}},  {\em Phys. Lett. B} {\bf 638}
  (2006) 61--67, [\href{http://arxiv.org/abs/hep-ph/0604053}{{\tt
  hep-ph/0604053}}].

\bibitem{Chen:2020uvt}
H.~Chen, T.-Z. Yang, H.~X. Zhu, and Y.~J. Zhu, {\it {Analytic Continuation and
  Reciprocity Relation for Collinear Splitting in QCD}},
  \href{http://arxiv.org/abs/2006.10534}{{\tt arXiv:2006.10534}}.

\bibitem{Almasy:2011eq}
A.~A. Almasy, S.~Moch, and A.~Vogt, {\it {On the Next-to-Next-to-Leading Order
  Evolution of Flavour-Singlet Fragmentation Functions}},  {\em Nucl. Phys. B}
  {\bf 854} (2012) 133--152, [\href{http://arxiv.org/abs/1107.2263}{{\tt
  arXiv:1107.2263}}].

\bibitem{Gehrmann:2022cih}
T.~Gehrmann and R.~Sch\"urmann, {\it {Photon fragmentation in the antenna
  subtraction formalism}},  {\em JHEP} {\bf 04} (2022) 031,
  [\href{http://arxiv.org/abs/2201.06982}{{\tt arXiv:2201.06982}}].

\bibitem{Alwall:2011uj}
J.~Alwall, M.~Herquet, F.~Maltoni, O.~Mattelaer, and T.~Stelzer, {\it {MadGraph
  5 : Going Beyond}},  {\em JHEP} {\bf 06} (2011) 128,
  [\href{http://arxiv.org/abs/1106.0522}{{\tt arXiv:1106.0522}}].

\bibitem{Alwall:2014hca}
J.~Alwall, R.~Frederix, S.~Frixione, V.~Hirschi, F.~Maltoni, O.~Mattelaer,
  H.~S. Shao, T.~Stelzer, P.~Torrielli, and M.~Zaro, {\it {The automated
  computation of tree-level and next-to-leading order differential cross
  sections, and their matching to parton shower simulations}},  {\em JHEP} {\bf
  07} (2014) 079, [\href{http://arxiv.org/abs/1405.0301}{{\tt
  arXiv:1405.0301}}].

\bibitem{Buckley:2014ana}
A.~Buckley, J.~Ferrando, S.~Lloyd, K.~Nordstr\"om, B.~Page, M.~R\"ufenacht,
  M.~Sch\"onherr, and G.~Watt, {\it {LHAPDF6: parton density access in the LHC
  precision era}},  {\em Eur. Phys. J. C} {\bf 75} (2015) 132,
  [\href{http://arxiv.org/abs/1412.7420}{{\tt arXiv:1412.7420}}].

\bibitem{Liu:2023fsq}
C.~Liu, X.~Shen, B.~Zhou, and J.~Gao, {\it {Automated calculation of Jet
  fragmentation at NLO in QCD}},  \href{http://arxiv.org/abs/2305.14620}{{\tt
  arXiv:2305.14620}}.

\bibitem{Cacciari:2008gp}
M.~Cacciari, G.~P. Salam, and G.~Soyez, {\it {The anti-$k_t$ jet clustering
  algorithm}},  {\em JHEP} {\bf 04} (2008) 063,
  [\href{http://arxiv.org/abs/0802.1189}{{\tt arXiv:0802.1189}}].

\bibitem{Marcantonini:2008qn}
C.~Marcantonini and I.~W. Stewart, {\it {Reparameterization Invariant Collinear
  Operators}},  {\em Phys. Rev. D} {\bf 79} (2009) 065028,
  [\href{http://arxiv.org/abs/0809.1093}{{\tt arXiv:0809.1093}}].

\bibitem{Bauer:2000yr}
C.~W. Bauer, S.~Fleming, D.~Pirjol, and I.~W. Stewart, {\it {An Effective field
  theory for collinear and soft gluons: Heavy to light decays}},  {\em Phys.
  Rev. D} {\bf 63} (2001) 114020,
  [\href{http://arxiv.org/abs/hep-ph/0011336}{{\tt hep-ph/0011336}}].

\bibitem{Bauer:2000ew}
C.~W. Bauer, S.~Fleming, and M.~E. Luke, {\it {Summing Sudakov logarithms in B
  ---\ensuremath{>} X(s gamma) in effective field theory}},  {\em Phys. Rev. D}
  {\bf 63} (2000) 014006, [\href{http://arxiv.org/abs/hep-ph/0005275}{{\tt
  hep-ph/0005275}}].

\bibitem{Bauer:2001yt}
C.~W. Bauer, D.~Pirjol, and I.~W. Stewart, {\it {Soft collinear factorization
  in effective field theory}},  {\em Phys. Rev. D} {\bf 65} (2002) 054022,
  [\href{http://arxiv.org/abs/hep-ph/0109045}{{\tt hep-ph/0109045}}].

\bibitem{Bauer:2001ct}
C.~W. Bauer and I.~W. Stewart, {\it {Invariant operators in collinear effective
  theory}},  {\em Phys. Lett. B} {\bf 516} (2001) 134--142,
  [\href{http://arxiv.org/abs/hep-ph/0107001}{{\tt hep-ph/0107001}}].

\bibitem{Beneke:2002ph}
M.~Beneke, A.~P. Chapovsky, M.~Diehl, and T.~Feldmann, {\it {Soft collinear
  effective theory and heavy to light currents beyond leading power}},  {\em
  Nucl. Phys. B} {\bf 643} (2002) 431--476,
  [\href{http://arxiv.org/abs/hep-ph/0206152}{{\tt hep-ph/0206152}}].

\bibitem{Chen:2019mqc}
W.~Chen, {\it {Reduction of Feynman Integrals in the Parametric
  Representation}},  {\em JHEP} {\bf 02} (2020) 115,
  [\href{http://arxiv.org/abs/1902.10387}{{\tt arXiv:1902.10387}}].

\bibitem{Chen:2019fzm}
W.~Chen, {\it {Reduction of Feynman Integrals in the Parametric Representation
  II: Reduction of Tensor Integrals}},  {\em Eur. Phys. J. C} {\bf 81} (2021),
  no.~3 244, [\href{http://arxiv.org/abs/1912.08606}{{\tt arXiv:1912.08606}}].

\bibitem{Chen:2020wsh}
W.~Chen, {\it {Reduction of Feynman integrals in the parametric representation
  III: integrals with cuts}},  {\em Eur. Phys. J. C} {\bf 80} (2020), no.~12
  1173, [\href{http://arxiv.org/abs/2007.00507}{{\tt arXiv:2007.00507}}].

\bibitem{PhysRev.133.B1549}
T.~D. Lee and M.~Nauenberg, {\it Degenerate systems and mass singularities},
  {\em Phys. Rev.} {\bf 133} (Mar, 1964) B1549--B1562.

\bibitem{Kinoshita:1962ur}
T.~Kinoshita, {\it {Mass singularities of Feynman amplitudes}},  {\em J. Math.
  Phys.} {\bf 3} (1962) 650--677.

\bibitem{Lee:2012cn}
R.~N. Lee, {\it {Presenting LiteRed: a tool for the Loop InTEgrals REDuction}},
   \href{http://arxiv.org/abs/1212.2685}{{\tt arXiv:1212.2685}}.

\bibitem{Lee:2013mka}
R.~N. Lee, {\it {LiteRed 1.4: a powerful tool for reduction of multiloop
  integrals}},  {\em J. Phys. Conf. Ser.} {\bf 523} (2014) 012059,
  [\href{http://arxiv.org/abs/1310.1145}{{\tt arXiv:1310.1145}}].

\bibitem{Smirnov:2019qkx}
A.~V. Smirnov and F.~S. Chuharev, {\it {FIRE6: Feynman Integral REduction with
  Modular Arithmetic}},  {\em Comput. Phys. Commun.} {\bf 247} (2020) 106877,
  [\href{http://arxiv.org/abs/1901.07808}{{\tt arXiv:1901.07808}}].

\bibitem{Meyer:2017joq}
C.~Meyer, {\it {Algorithmic transformation of multi-loop master integrals to a
  canonical basis with CANONICA}},  {\em Comput. Phys. Commun.} {\bf 222}
  (2018) 295--312, [\href{http://arxiv.org/abs/1705.06252}{{\tt
  arXiv:1705.06252}}].

\bibitem{Lee:2014ioa}
R.~N. Lee, {\it {Reducing differential equations for multiloop master
  integrals}},  {\em JHEP} {\bf 04} (2015) 108,
  [\href{http://arxiv.org/abs/1411.0911}{{\tt arXiv:1411.0911}}].

\bibitem{Lee:2020zfb}
R.~N. Lee, {\it {Libra: A package for transformation of differential systems
  for multiloop integrals}},  {\em Comput. Phys. Commun.} {\bf 267} (2021)
  108058, [\href{http://arxiv.org/abs/2012.00279}{{\tt arXiv:2012.00279}}].

\bibitem{Campbell:1997hg}
J.~M. Campbell and E.~W.~N. Glover, {\it {Double unresolved approximations to
  multiparton scattering amplitudes}},  {\em Nucl. Phys. B} {\bf 527} (1998)
  264--288, [\href{http://arxiv.org/abs/hep-ph/9710255}{{\tt hep-ph/9710255}}].

\bibitem{Catani:1998nv}
S.~Catani and M.~Grazzini, {\it {Collinear factorization and splitting
  functions for next-to-next-to-leading order QCD calculations}},  {\em Phys.
  Lett. B} {\bf 446} (1999) 143--152,
  [\href{http://arxiv.org/abs/hep-ph/9810389}{{\tt hep-ph/9810389}}].

\bibitem{Ritzmann:2014mka}
M.~Ritzmann and W.~J. Waalewijn, {\it {Fragmentation in Jets at NNLO}},  {\em
  Phys. Rev. D} {\bf 90} (2014), no.~5 054029,
  [\href{http://arxiv.org/abs/1407.3272}{{\tt arXiv:1407.3272}}].

\bibitem{Gehrmann-DeRidder:1997fom}
A.~Gehrmann-De~Ridder and E.~W.~N. Glover, {\it {A Complete O (alpha alpha-s)
  calculation of the photon + 1 jet rate in e+ e- annihilation}},  {\em Nucl.
  Phys. B} {\bf 517} (1998) 269--323,
  [\href{http://arxiv.org/abs/hep-ph/9707224}{{\tt hep-ph/9707224}}].

\bibitem{Tkachov:1981wb}
F.~V. Tkachov, {\it {A Theorem on Analytical Calculability of Four Loop
  Renormalization Group Functions}},  {\em Phys. Lett. B} {\bf 100} (1981)
  65--68.

\bibitem{Chetyrkin:1981qh}
K.~G. Chetyrkin and F.~V. Tkachov, {\it {Integration by Parts: The Algorithm to
  Calculate beta Functions in 4 Loops}},  {\em Nucl. Phys. B} {\bf 192} (1981)
  159--204.

\bibitem{Lee:2014tja}
R.~N. Lee, {\it {Modern techniques of multiloop calculations}},  in {\em {49th
  Rencontres de Moriond on QCD and High Energy Interactions}}, pp.~297--300,
  2014.
\newblock \href{http://arxiv.org/abs/1405.5616}{{\tt arXiv:1405.5616}}.

\bibitem{Maierhofer:2017gsa}
P.~Maierh\"ofer, J.~Usovitsch, and P.~Uwer, {\it {Kira\textemdash{}A Feynman
  integral reduction program}},  {\em Comput. Phys. Commun.} {\bf 230} (2018)
  99--112, [\href{http://arxiv.org/abs/1705.05610}{{\tt arXiv:1705.05610}}].

\bibitem{Klappert:2019emp}
J.~Klappert and F.~Lange, {\it {Reconstructing rational functions with
  FireFly}},  {\em Comput. Phys. Commun.} {\bf 247} (2020) 106951,
  [\href{http://arxiv.org/abs/1904.00009}{{\tt arXiv:1904.00009}}].

\bibitem{Klappert:2020aqs}
J.~Klappert, S.~Y. Klein, and F.~Lange, {\it {Interpolation of dense and sparse
  rational functions and other improvements in FireFly}},  {\em Comput. Phys.
  Commun.} {\bf 264} (2021) 107968,
  [\href{http://arxiv.org/abs/2004.01463}{{\tt arXiv:2004.01463}}].

\bibitem{Klappert:2020nbg}
J.~Klappert, F.~Lange, P.~Maierh\"ofer, and J.~Usovitsch, {\it {Integral
  reduction with Kira 2.0 and finite field methods}},  {\em Comput. Phys.
  Commun.} {\bf 266} (2021) 108024,
  [\href{http://arxiv.org/abs/2008.06494}{{\tt arXiv:2008.06494}}].

\bibitem{Kotikov:1990kg}
A.~V. Kotikov, {\it {Differential equations method: New technique for massive
  Feynman diagrams calculation}},  {\em Phys. Lett. B} {\bf 254} (1991)
  158--164.

\bibitem{Remiddi:1997ny}
E.~Remiddi, {\it {Differential equations for Feynman graph amplitudes}},  {\em
  Nuovo Cim. A} {\bf 110} (1997) 1435--1452,
  [\href{http://arxiv.org/abs/hep-th/9711188}{{\tt hep-th/9711188}}].

\bibitem{Henn:2013pwa}
J.~M. Henn, {\it {Multiloop integrals in dimensional regularization made
  simple}},  {\em Phys. Rev. Lett.} {\bf 110} (2013) 251601,
  [\href{http://arxiv.org/abs/1304.1806}{{\tt arXiv:1304.1806}}].

\bibitem{goncharov1mpl}
A.~B. Goncharov, {\it Multiple polylogarithms and mixed tate motives},  2001.

\bibitem{Goncharov:1998kja}
A.~B. Goncharov, {\it {Multiple polylogarithms, cyclotomy and modular
  complexes}},  {\em Math. Res. Lett.} {\bf 5} (1998) 497--516,
  [\href{http://arxiv.org/abs/1105.2076}{{\tt arXiv:1105.2076}}].

\bibitem{Borwein:1999js}
J.~M. Borwein, D.~M. Bradley, D.~J. Broadhurst, and P.~Lisonek, {\it {Special
  values of multiple polylogarithms}},  {\em Trans. Am. Math. Soc.} {\bf 353}
  (2001) 907--941, [\href{http://arxiv.org/abs/math/9910045}{{\tt
  math/9910045}}].

\bibitem{Naterop:2019xaf}
L.~Naterop, A.~Signer, and Y.~Ulrich, {\it {handyG \textemdash{}Rapid numerical
  evaluation of generalised polylogarithms in Fortran}},  {\em Comput. Phys.
  Commun.} {\bf 253} (2020) 107165,
  [\href{http://arxiv.org/abs/1909.01656}{{\tt arXiv:1909.01656}}].

\bibitem{Cutkosky:1960sp}
R.~E. Cutkosky, {\it {Singularities and discontinuities of Feynman
  amplitudes}},  {\em J. Math. Phys.} {\bf 1} (1960) 429--433.

\bibitem{Anastasiou:2002yz}
C.~Anastasiou and K.~Melnikov, {\it {Higgs boson production at hadron colliders
  in NNLO QCD}},  {\em Nucl. Phys. B} {\bf 646} (2002) 220--256,
  [\href{http://arxiv.org/abs/hep-ph/0207004}{{\tt hep-ph/0207004}}].

\bibitem{Maitre:2005uu}
D.~Maitre, {\it {HPL, a mathematica implementation of the harmonic
  polylogarithms}},  {\em Comput. Phys. Commun.} {\bf 174} (2006) 222--240,
  [\href{http://arxiv.org/abs/hep-ph/0507152}{{\tt hep-ph/0507152}}].

\bibitem{Gehrmann-DeRidder:2003pne}
A.~Gehrmann-De~Ridder, T.~Gehrmann, and G.~Heinrich, {\it {Four particle phase
  space integrals in massless QCD}},  {\em Nucl. Phys. B} {\bf 682} (2004)
  265--288, [\href{http://arxiv.org/abs/hep-ph/0311276}{{\tt hep-ph/0311276}}].

\bibitem{Magerya:2019cvz}
V.~Magerya and A.~Pikelner, {\it {Cutting massless four-loop propagators}},
  {\em JHEP} {\bf 12} (2019) 026, [\href{http://arxiv.org/abs/1910.07522}{{\tt
  arXiv:1910.07522}}].

\bibitem{Korchemsky:1999kt}
G.~P. Korchemsky and G.~F. Sterman, {\it {Power corrections to event shapes and
  factorization}},  {\em Nucl. Phys. B} {\bf 555} (1999) 335--351,
  [\href{http://arxiv.org/abs/hep-ph/9902341}{{\tt hep-ph/9902341}}].

\bibitem{Dokshitzer:1999sh}
Y.~L. Dokshitzer, G.~Marchesini, and B.~R. Webber, {\it {Nonperturbative
  effects in the energy energy correlation}},  {\em JHEP} {\bf 07} (1999) 012,
  [\href{http://arxiv.org/abs/hep-ph/9905339}{{\tt hep-ph/9905339}}].

\bibitem{Schindler:2023cww}
S.~T. Schindler, I.~W. Stewart, and Z.~Sun, {\it {Renormalons in the
  energy-energy correlator}},  \href{http://arxiv.org/abs/2305.19311}{{\tt
  arXiv:2305.19311}}.

\bibitem{Sjostrand:2014zea}
T.~Sj\"ostrand, S.~Ask, J.~R. Christiansen, R.~Corke, N.~Desai, P.~Ilten,
  S.~Mrenna, S.~Prestel, C.~O. Rasmussen, and P.~Z. Skands, {\it {An
  introduction to PYTHIA 8.2}},  {\em Comput. Phys. Commun.} {\bf 191} (2015)
  159--177, [\href{http://arxiv.org/abs/1410.3012}{{\tt arXiv:1410.3012}}].

\bibitem{Tulipant:2017ybb}
Z.~Tulip\'ant, A.~Kardos, and G.~Somogyi, {\it {Energy\textendash{}energy
  correlation in electron\textendash{}positron annihilation at NNLL + NNLO
  accuracy}},  {\em Eur. Phys. J. C} {\bf 77} (2017), no.~11 749,
  [\href{http://arxiv.org/abs/1708.04093}{{\tt arXiv:1708.04093}}].

\bibitem{Kardos:2018kqj}
A.~Kardos, S.~Kluth, G.~Somogyi, Z.~Tulip\'ant, and A.~Verbytskyi, {\it
  {Precise determination of $\alpha _{S}(M_Z)$ from a global fit of
  energy\textendash{}energy correlation to NNLO+NNLL predictions}},  {\em Eur.
  Phys. J. C} {\bf 78} (2018), no.~6 498,
  [\href{http://arxiv.org/abs/1804.09146}{{\tt arXiv:1804.09146}}].

\bibitem{Cao:2023oef}
H.~Cao, X.~Liu, and H.~X. Zhu, {\it {Toward precision measurements of nucleon
  energy correlators in lepton-nucleon collisions}},  {\em Phys. Rev. D} {\bf
  107} (2023), no.~11 114008, [\href{http://arxiv.org/abs/2303.01530}{{\tt
  arXiv:2303.01530}}].

\bibitem{Tarasov:1980au}
O.~V. Tarasov, A.~A. Vladimirov, and A.~Y. Zharkov, {\it {The Gell-Mann-Low
  Function of QCD in the Three Loop Approximation}},  {\em Phys. Lett. B} {\bf
  93} (1980) 429--432.

\bibitem{Larin:1993tp}
S.~A. Larin and J.~A.~M. Vermaseren, {\it {The Three loop QCD Beta function and
  anomalous dimensions}},  {\em Phys. Lett. B} {\bf 303} (1993) 334--336,
  [\href{http://arxiv.org/abs/hep-ph/9302208}{{\tt hep-ph/9302208}}].

\bibitem{vanRitbergen:1997va}
T.~van Ritbergen, J.~A.~M. Vermaseren, and S.~A. Larin, {\it {The Four loop
  beta function in quantum chromodynamics}},  {\em Phys. Lett. B} {\bf 400}
  (1997) 379--384, [\href{http://arxiv.org/abs/hep-ph/9701390}{{\tt
  hep-ph/9701390}}].

\bibitem{Czakon:2004bu}
M.~Czakon, {\it {The Four-loop QCD beta-function and anomalous dimensions}},
  {\em Nucl. Phys. B} {\bf 710} (2005) 485--498,
  [\href{http://arxiv.org/abs/hep-ph/0411261}{{\tt hep-ph/0411261}}].

\bibitem{Gehrmann:2010ue}
T.~Gehrmann, E.~W.~N. Glover, T.~Huber, N.~Ikizlerli, and C.~Studerus, {\it
  {Calculation of the quark and gluon form factors to three loops in QCD}},
  {\em JHEP} {\bf 06} (2010) 094, [\href{http://arxiv.org/abs/1004.3653}{{\tt
  arXiv:1004.3653}}].

\end{thebibliography}\endgroup
\bibliographystyle{JHEP}
\end{document}